%% file: ms.tex
\documentclass[nanomaterials,article,accept,moreauthors,pdftex]{Definitions/mdpi}
\usepackage[utf8]{inputenc} 
\usepackage{geometry}			
\usepackage{hyperref} 		
\usepackage{epsfig}
\usepackage{graphicx}
\usepackage{tabularx}
\usepackage{color}				
\usepackage{amsthm}
\usepackage{amsfonts}
\usepackage{mathtools}
\usepackage{adjustbox}
\usepackage{pgfplots}			
\usepackage{tikz}					
\usetikzlibrary{shapes}
\usetikzlibrary{calc}
\usetikzlibrary{positioning}
\usepackage{tikzscale}		
\usepackage{latexsym} 		
\usepackage{float}				
\usepackage{fancyhdr} 		
\usepackage{titlesec} 		
\usepackage{listings}
\usepackage{marginnote}
\usepackage[normalem]{ulem}
\usepackage{etoolbox}
\usepackage{appendix}
\usepackage{soul}
\usepackage{dcolumn}				
\usepackage{bm}						
\usepackage{multirow}
\usepackage{upgreek}

\ifdefined\ExtComp
\usepgfplotslibrary{external}
\fi

\definecolor{naranja_1}{rgb}{0.70,0.25,0}
\definecolor{gris-azul}{rgb}{0.27,0.33,0.415}  
\definecolor{azul-gris}{rgb}{0.192,0.278,0.4}  
\definecolor{verde-chulo}{rgb}{0.25,0.5,0.5}  
\definecolor{verde-chulo2}{rgb}{0,0.5,0.4} 
\definecolor{beige}{rgb}{0.96, 0.96, 0.86}

\definecolor{orcidlogocol}{HTML}{A6CE39}

\definecolor{color1}{rgb}{1, 0, 0}%
\definecolor{color2}{rgb}{0, 0, 1}%
\definecolor{color3}{rgb}{0, 1, 0}%
\definecolor{color4}{rgb}{1, 0, 1}%
\definecolor{color5}{rgb}{0, 1, 1}%

\definecolor{mycolor1}{rgb}{0.00000,0.44700,0.74100}%
\definecolor{mycolor2}{rgb}{0.85000,0.32500,0.09800}%
\definecolor{mycolor3}{rgb}{0.46600,0.67400,0.18800}%
\definecolor{mycolor4}{rgb}{0.49400,0.18400,0.55600}%
\definecolor{mycolor5}{rgb}{0.92900,0.69400,0.12500}%
\definecolor{mycolor6}{rgb}{0.30100,0.74500,0.93300}%
\definecolor{mycolor7}{rgb}{0.63500,0.07800,0.18400}%

\pgfplotsset{
	colormap={parula}{
		rgb255=(53,42,135)
		rgb255=(15,92,221)
		rgb255=(18,125,216)
		rgb255=(7,156,207)
		rgb255=(21,177,180)
		rgb255=(89,189,140)
		rgb255=(165,190,107)
		rgb255=(225,185,82)
		rgb255=(252,206,46)
		rgb255=(249,251,14)
	},
}

\pgfplotsset{compat=1.5}
\pgfplotsset{minor grid style={dashed}}

\newcommand{\secref}[1]{\hyperref[#1]{Section \ref*{#1}}}		
\newcommand{\subsecref}[1]{\hyperref[#1]{Subsection \ref*{#1}}}		
\newcommand{\appref}[1]{\hyperref[#1]{Appendix \ref*{#1}}}		
\newcommand{\chref}[1]{\hyperref[#1]{Chapter \ref*{#1}}}		
\newcommand{\figref}[1]{\hyperref[#1]{Figure \ref*{#1}}}		
\newcommand{\sfigref}[2]{\hyperref[#1]{Figure \ref*{#1}#2}}		
\newcommand{\sfigrefN}[2]{\hyperref[#1]{\ref*{#1}#2}}			
\newcommand{\tabref}[1]{\hyperref[#1]{Table \ref*{#1}}}			

\newcommand{\RemoveTickExp}{
	yticklabel style={
		/pgf/number format/fixed,
		/pgf/number format/precision=5
	},
	scaled y ticks=false
	}

\newcommand{\SubfigLetter}[3]{\node[fill=white] at (rel axis cs: #2,#3) {\textbf{#1}}}

\pgfkeys{/tikz/savenumber/.code 2 args={\global\edef#1{#2}}}

\newcolumntype{L}[1]{>{\raggedright\arraybackslash}p{#1}}
\newcolumntype{C}[1]{>{\centering\arraybackslash}p{#1}}
\newcolumntype{R}[1]{>{\raggedleft\arraybackslash}p{#1}}
\firstpage{1} 
\pubvolume{9}
\issuenum{7}
\articlenumber{1027}
\pubyear{2019}
\copyrightyear{2019}
\doinum{10.3390/nano9071027}
\history{Received: 23 June 2019; Revised: 12 July 2019; Accepted: 15 July 2019; Published: 18 July 2019 }

\pdfoutput=1

\Title{GFET Asymmetric Transfer Response Analysis through Access Region Resistances}

\newcommand{\commentsColor}{black}

\Author{{Alejandro Toral-Lopez} $^{1,}$*\orcidA{}, Enrique G. Marin $^{1,2}$\orcidB{}, Francisco Pasadas $^{3}$\orcidC{}, \mbox{Jose Maria Gonzalez-Medina $^{1}$\orcidD{}}, Francisco G. Ruiz $^{1,4}$\orcidE{}, David Jiménez $^{3}$\orcidG{} and \mbox{Andres Godoy $^{1,4,}$*\orcidH{}}}

\AuthorNames{Alejandro Toral-Lopez, Enrique G. Marin, Francisco Pasadas, Jose Maria Gonzalez-Medina, Francisco G.Ruiz, David Jiménez and Andres Godoy}

\address{%
$^{1}$ \quad Dpto. Electrónica, Fac. Ciencias, Univ. Granada, 18071 Granada, Spain\\
$^{2}$ \quad Dipartimento di Ingegneria dell’Informazione, Università di Pisa, 56122 Pisa, Italy\\
$^{3}$ \quad Dept. d'Enginyeria Electrònica, Escola d'Enginyeria, Univ. Autònoma de Barcelona, 08193 Bellaterra, Spain \\
$^{4}$ \quad Pervasive Electronics Advanced Research Laboratory, CITIC, Univ. Granada, 18017 Granada, Spain 
}

\corres{{Correspondence: atoral@ugr.es (A.T.-L.); agodoy@ugr.es (A.G.)}}

\abstract{Graphene-based devices are planned to augment the functionality of Si and III-V based technology in radio-frequency (RF) electronics. The expectations in designing graphene {field-effect} transistors (GFETs) with enhanced RF performance have attracted significant experimental efforts, mainly concentrated on achieving high mobility samples. However, little attention has been paid, so far, to the role of the access regions in these devices. \mbox{Here, we analyse} in detail, via numerical simulations, how the GFET transfer response is severely impacted by these regions, showing that they play a significant role in the asymmetric saturated behaviour commonly observed in GFETs. We also investigate how the modulation of the access region conductivity (i.e., by the influence of a back gate) and the presence of imperfections in the graphene layer (e.g., charge puddles) affects the transfer response. The analysis is extended to assess the application of GFETs for RF applications, by~evaluating their cut-off frequency.}


\begin{document}

\section{Introduction}\label{sec:intro}
Two-dimensional materials (2DMs) have awakened the great interest of the nanotechnology community during the last decade \cite{Fiori2014}. Their striking physical properties, intrinsically different from their 3D counterparts, open a vast field of opportunities only partially exploited so far. Among these alternatives, 2DMs find a~natural spot in electronics, where their monoatomic thickness makes them especially attractive to overcome the hurdles related to the transistor scaling-down \cite{Lee_2014}.

Graphene is not only the pioneer, but also the most singular member of the  2DM family \cite{Neto2009}. It~is characterized by a gapless Dirac-cone bandstructure, where electrons and holes have symmetric dispersion relationships. The literature is abundant in Graphene Field-Effect Transistors (GFETs) {\color{\commentsColor} \cite{Guerriero2017, Lin2009, Meric2008}}, where this particular band structure is manifested in an ambipolar behaviour {\color{\commentsColor} and a poor $I_{\rm ON}/I_{\rm OFF}$ ratio (direct consequence of the easiness to switch the carrier transport from electrons to holes and vice versa)}. This issue jeopardizes the use of GFETs in digital electronics{\color{\commentsColor}, although a successful demonstration has been achieved in \cite{Rizzi2012}.} In~radio-frequency (RF), however, graphene has revealed itself as an interesting candidate \cite{Pandey2018}, and devices with cut-off frequencies of hundreds of GHz have already been demonstrated \cite{Wu2016, Cheng2012}{\color{\commentsColor}, even reaching wafer scale integration \cite{Lin2010}, or~being applied for flexible electronics \cite{Georgiou2012, Wang2019}. The main strategies to boost GFETs performance have consisted of the scaling-down of the gate oxide thickness \cite{Guerriero2017, Liao2010a}, the encapsulation in hexagonal boron nitride \cite{Mayorov2011} or the improvement in the quality of the graphene-insulator stack \cite{Rizzi2012, Farmer2010}. In particular, clean self-aligned fabrication, based in pre-deposited gold, has been proposed in \cite{Feng2014}; while the self-aligned transfer of the gate stack (processed in a sacrificial substrate) has been detailed in \cite{Fiori2013a}.}

The transfer characteristic of experimental GFETs is V-shaped, but very often shows an asymmetry with respect to the Dirac voltage {\color{\commentsColor} \cite{Bartolomeo2015}}, usually associated with different electron and hole mobilities. {\color{\commentsColor} These~mobility dissimilarities are the common path to handle the device response asymmetry, leaving out of the spot the relevance of the gate underlapped areas \cite{Mayorov2011, Jain2015, Al-Amin2016}. These access regions (intended to minimize the capacitance coupling between the gate and the source and drain) impact, \mbox{however, strongly on} the GFET electrical behaviour, as they constitute a noticeable resistance pathway for carrier transport. Partial attempts on the modelling of this issue have 	 been discussed from an analytical resistance-based perspective in \cite{Jain2015, Wang2011}, but a comprehensive study of their impact in the GFET performance is still lacking \cite{Fiori2013a}}. In this work, we direct our attention to this asymmetric response of GFETs and, by means of detailed numerical simulations, we explain such effect studying the impact of the access regions in the transfer characteristic as well as in the RF performance of such devices.

The rest of the document is organized as follows. Section \ref{sbsec:methods} presents the numerical model employed for this study. To check and validate it we compare, in Section \ref{sbsec:val}, the simulated transfer response of two GFETs against the corresponding experimental measurements. Section \ref{sbsec:AcRAnalysis} contains a thorough analysis of the access resistances and a discussion of its influence on the cut-off frequency, $f_{\rm T}$. Finally, the main conclusions are drawn in Section \ref{sec:concl}.

\section{Results}
\unskip
\subsection{Device Simulation}\label{sbsec:methods}

\begin{figure}[H]
	\centering
	\includegraphics[width=0.6\linewidth]{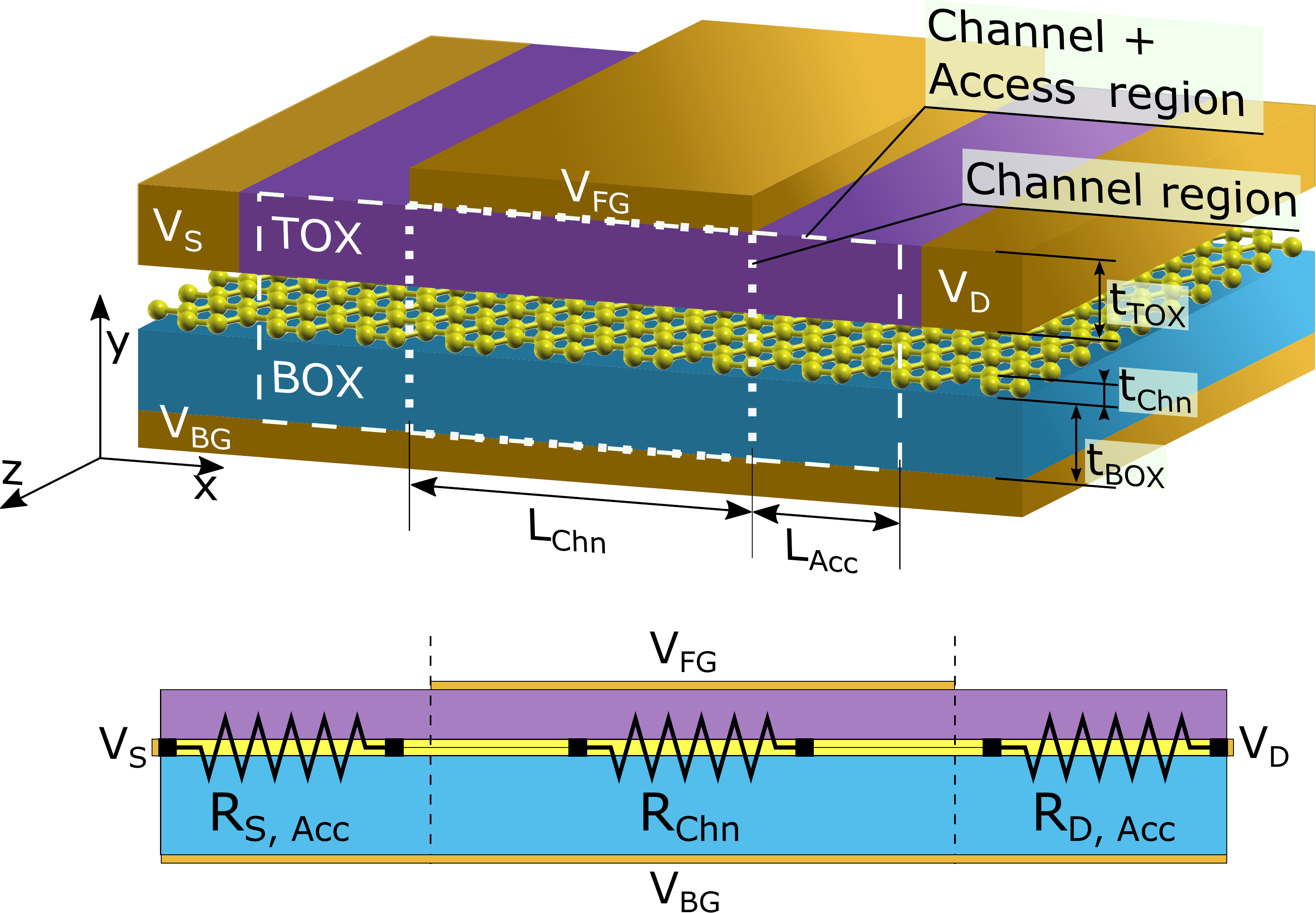}
	\caption{Schematic of the simulated GFET and the characteristic resistances of the device. The dashed and dotted rectangles indicate the regions used for the different simulations. While the dotted rectangle only encompasses the channel region, the dashed one includes the access regions.}
	\label{fig:graphenemosfet}
\end{figure}

A schematic depiction of the physical structure of the simulated GFET is shown in Figure~\ref{fig:graphenemosfet}. The~graphene flake is sandwiched in between a top insulator layer, with thickness $t_{\rm TOX}$ and dielectric permittivity $\varepsilon_{\rm TOX}$, \mbox{and an insulating} substrate, with thickness $t_{\rm BOX}$ and dielectric permittivity $\varepsilon_{\rm BOX}$. Both oxides are assumed thick enough as to neglect any tunnelling current through them. A~four-terminal device is considered, with~a~front gate extending over a length $L_{\text{Chn}}$ (the device channel length), giving rise to two under-lapped regions of length $L_{\text{Acc}}$ (the access region length) that connect it with the source and drain terminals. The back gate, when~considered, extends all along the structure including the channel as well as the access regions. \mbox{$V_{\rm FG}$, $V_{\rm BG}$, and $V_{\rm D}$} stand for the front gate, back gate, and drain terminal biases respectively, while the source terminal, $V_{\rm S}$, is assumed to be grounded. The total resistance of this structure, $R_{\rm T}$, can be schematically split into the series combination of three resistances corresponding to the source access region ($R_{\rm S, Acc}$), the channel region ($R_{\rm Chn}$) and the drain access region ($R_{\rm D, Acc}$).

To determine the I-V response of GFET devices, we have self-consistently solved the coupled Poisson, Drift-Diffusion and continuity equations \cite{Ancona_2010, Curatola_2005}. For the device modelling, we have considered a~longitudinal $x-y$ section of the GFET, assuming invariance along the device width ($z$). The resulting 2D Poisson equation is given by:
\begin{equation}
\nabla\left(\varepsilon\left(x,y\right) \nabla V\left(x,y\right)\right) = -\rho\left(x,y\right)
\label{eq:pois_eq}
\end{equation}
where $V$ is the electrostatic potential; $\rho$ is the net charge density in the structure, that comprises the mobile (electrons and holes) and fixed (dopants) charges; and $\varepsilon$ is the dielectric permittivity.

The Drift-Diffusion transport equation is formulated in terms of the pseudo-Fermi level ($E_{\rm F}$) as proposed in \cite{Feijoo_2016}:
\begin{equation}
J(x) = q\left[\mu_{\rm n} n_{\rm 1D}(x) + \mu_{\rm p} p_{\rm 1D}(x)\right]\frac{dV_{E_{\rm F}}}{dx}
\label{eq:J_DD_Grph}
\end{equation}
where $V_{E_{F}}$ is the potential associated with this level and $n_{\rm 1D}$ ($p_{\rm 1D}$) is the graphene electron (hole) 1D density profile. Here, $\mu_{\rm n}$ ($\mu_{\rm p}$) stands for the electron (hole) mobility. Due to the extreme confinement, the carriers are supposed to move only along the transport direction ($x$). $J$ must comply with the continuity equation that, under steady-state conditions, is formulated as: $\nabla \cdot \mathbf{J} = 0$. Ohmic contacts are assumed at the source and drain terminals, with the Fermi level at the source grounded, $E_{\rm F,S} = 0$, and at the drain given by $E_{\rm F,D}=-qV_{\rm DS}$. The equation system is then iteratively solved for each set of terminal biases, until a convergence threshold is achieved for the potential and charge concentrations. 

In addition to the mobile charge and dopants in the graphene layer, we account for the existence of puddles~\cite{Wilmart2016, Martin2007}. Their associated charge density, $N_{\rm p}$, is assumed constant and added to both electron and hole charge densities \cite{Fregonese2013}. {\color{\commentsColor} In this way, puddles impact on} the overall graphene layer conductivity while conserving a neutral net charge character.

\subsection{Validation}\label{sbsec:val}

\begin{figure}[H]
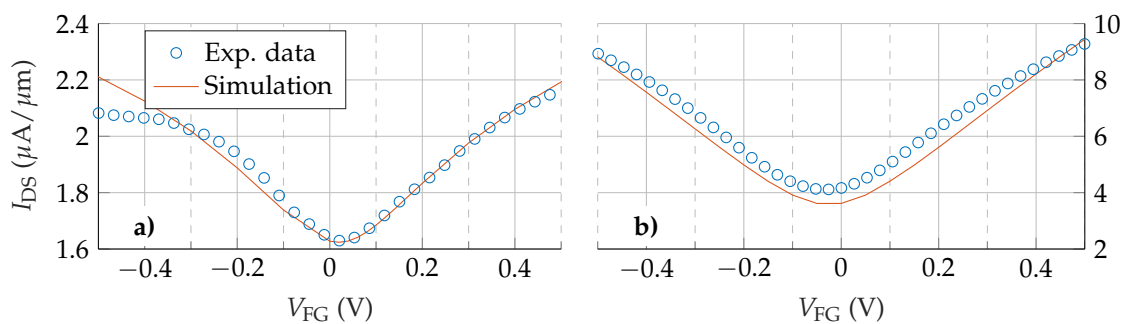

	\centering
	\begin{tikzpicture}
	\input{Figuras/GFET_WangComp}
	\input{Figuras/GFET_ZhangComp}
	\end{tikzpicture}
	\caption{Comparison between the simulation results and the experimental data extracted from \cite{Wang2010} (\textbf{a}) and~\cite{Zhang2012}~(\textbf{b}).}
	\label{fig:validation}
\end{figure}

To assess the capability of the numerical simulator to reproduce and explain the experimental results, we have first validated it against the devices fabricated in \cite{Wang2010, Zhang2012}. Both~are GFETs based on monolayer graphene embedded between a SiO$_2$ layer, which acts as a substrate, \mbox{and a Y$_2$O$_3$} layer, which acts as a front gate dielectric. In both cases, this Y$_2$O$_3$ layer is 5 nm thick while the substrate is 300 nm thick in \cite{Wang2010}, and~286 nm thick in \cite{Zhang2012}. For the device presented in~\cite{Wang2010}, the~distance between the source and drain contacts is 1.5 $\upmu$m and the front gate length is 600 nm, while~in \cite{Zhang2012} the device is 8.2 $\upmu$m long and its front gate is 7 $\upmu$m long. In other words, in both experimental devices the gate contact does not cover the whole region between source and drain contacts, thus creating two symmetrical under-lapped regions at both channel edges; namely, the~device access regions. To reproduce the data reported in \cite{Wang2010}, the same mobility is assumed for both types of carriers, electrons and holes ($\mu = \mu_{\rm n} = \mu_{\rm p}$) with a value of 90 cm$^{2}/$Vs, and a puddle charge density of \mbox{7$\cdot 10^{11}$ cm$^{-2}$} is considered. N-type chemical doping of 10$^{12}$ cm$^{-2}$ is defined for the graphene layer. To~account for the graphene-metal contact resistances, which are in series with the total resistance of the structure, \mbox{$R_{\rm T}$, we include} two additional 100 nm long N-type doped regions (5$\cdot 10^{10}$ cm$^{-2}$) in both source and drain ends~\cite{Venica2016}. The back gate is grounded and $V_{\rm DS}$ is set to 0.1V. To fit the data presented in~\cite{Zhang2012}, the values used are $\mu = $1091 cm$^{2}/$Vs, $N_{\rm p} = 8\cdot 10^{11}$ cm$^{-2}$ and the graphene layer chemical doping is set to 10$^{11}$ cm$^{-2}$. The back gate is also grounded and $V_{\rm DS}$ is set to 0.05 V. The experimental and simulated transfer characteristics are shown in Figure \ref{fig:validation}a \cite{Wang2010} and Figure \ref{fig:validation}{b} \cite{Zhang2012}. The~simulated I-V characteristics match very accurately with the experimental results in the whole range of biases and are able to catch the transfer response of the electron and hole branches, especially in Figure \ref{fig:validation}{b}.

\subsection{Access Region Analysis}\label{sbsec:AcRAnalysis}

As mentioned in Section \ref{sec:intro}, the existence of access regions and puddles is a very common scenario in the experimental realization of GFETs due to the difficulties to precisely control the fabrication process in this early stage of the technology. They modify the behaviour of the transistors, in many cases determining their performance, and therefore deserving a particular attention that is usually obliterated. Hence, once the numerical simulator has been validated, we now proceed to analyse the effect of the access regions. 

\subsubsection{Including the Access Regions}

To begin with, we have considered a test structure where the front gate covers the whole device length (i.e.,~suppressing the access regions) and compared the results with those obtained later when access regions are included. {\color{\commentsColor} These scenarios are illustrated in Figure \ref{fig:graphenemosfet} by the dotted and dashed frames respectively}. The~material stack comprises a monolayer graphene sandwiched between a 3 nm thick HfO$_2$ layer (front gate insulator) and a 27 nm thick SiO$_2$ layer (back gate insulator). The front gate, which determines the channel length ($L_{\rm Chn}$), is~100 nm long and both access regions are 35 nm long ($L_{\rm Acc}$). Electron and hole mobilities are equal (\mbox{$\mu = 1500$ cm$^{2}/$Vs}) and no chemical doping or puddle charge density is considered in the graphene layer.

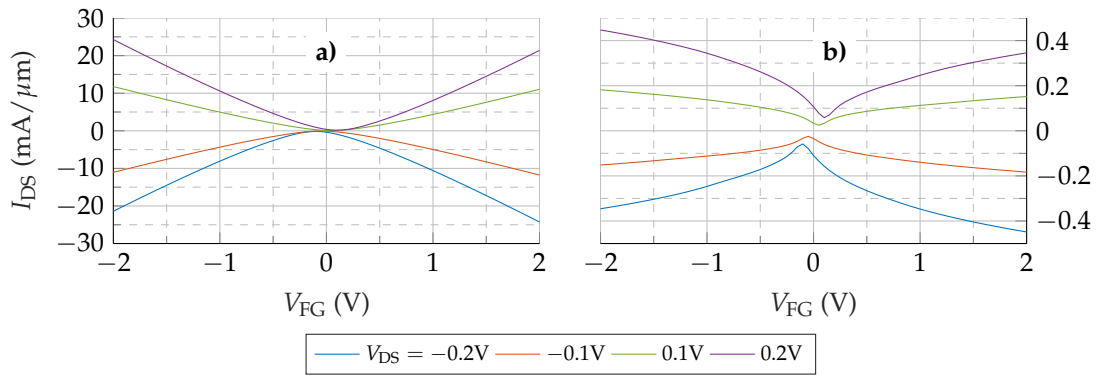
\begin{figure}[H]
	\centering
	\begin{tikzpicture}
	\input{Figuras/GFET_IdsVgs_EqMob}
	\input{Figuras/GFET_IdsVgs_Acc}
	\end{tikzpicture}
	\caption{$I_{\rm DS}-V_{\rm FG}$ curves of the device without (\textbf{a}) and with (\textbf{b}) access regions.}
	\label{fig:gfetidsvgsdiffmob}
\end{figure}

The transfer characteristic of the device without access regions is depicted in Figure \ref{fig:gfetidsvgsdiffmob}{a} for different values of $V_{\rm DS}$. As can be observed, the device exhibits the ambipolar V-shaped $I-V$ response of an~ideal GFET. The~minimum of the $I-V$ curve defines the Dirac voltage ($V_{\rm Dirac}$) that is shifted to larger $V_{\text{FG}}$ when $V_{\rm DS}$ increases. The behaviour is perfectly symmetric {\color{\commentsColor} with respect to} $V_{\rm Dirac}$, reflecting the symmetry between electron and hole properties.

Next, the GFET including the access regions is investigated. The resulting transfer characteristic is shown in Figure \ref{fig:gfetidsvgsdiffmob}{b}. Comparing Figure \ref{fig:gfetidsvgsdiffmob}{b} and Figure \ref{fig:gfetidsvgsdiffmob}{a}, a marked variation of the GFET response is observed. First, there is a notable decrease in the values of $I_{\text{DS}}$, around a factor x100. Second, the transfer characteristic shows a saturation trend for high $|V_{\text{FG}}|$ which resembles much better the experimental response. \mbox{Third, and more} important, the $I-V$ characteristic is no longer symmetric with respect to $V_{\text{Dirac}}$, though the mobility is identical for both kinds of carriers.

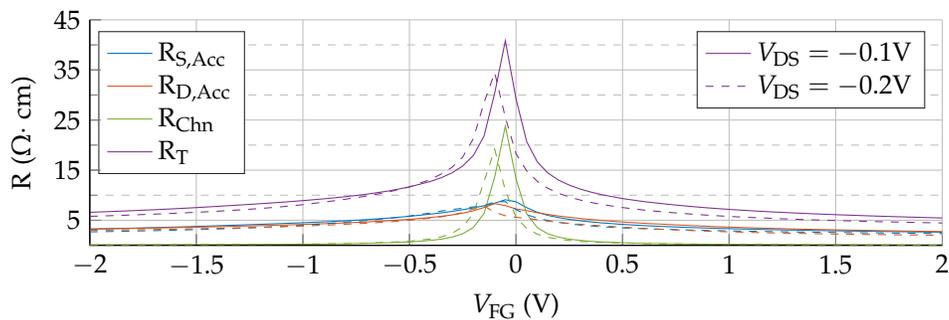
\begin{figure}[H]
	\centering
	\input{Figuras/GFET_NoVbg_Rdev_NoPuddles_VDS-0.1V}
	\caption{Resistance of the three device regions (channel, source and drain access regions) compared with the total resistance as a function of the gate potential, for two $V_{\rm DS}$ values: $-$0.1 V (solid) and \mbox{$-$0.2 V (dashed)}.}
	\label{fig:gfetnovbgrdevnopuddles}
\end{figure}

To provide insights into these changes, the resistance of the different regions of the device are calculated. Figure \ref{fig:gfetnovbgrdevnopuddles} shows their values for $V_{\rm DS}=-$0.1 V and $V_{\rm DS}=-$0.2 V. Mirror symmetric behaviour is observed for positive $V_{\rm DS}$. The access region resistances, $R_{\rm S, Acc}$ and $R_{\rm D, Acc}$, show values comparable with the channel resistance, $R_{\rm Chn}$. At the Dirac voltage, where the channel resistivity is the highest, $R_{\rm Chn}$ commands the series association, but still the access regions have a noticeable contribution. For $|V_{\rm FG} - V_{\rm Dirac}| > 0.1$ V the total resistance is mainly determined by $R_{\rm S, Acc}$ and $R_{\rm D, Acc}$. Consequently, the total resistance ($R_{\rm T}$) is not controlled just by the channel conductivity and, therefore, by the gate terminal. The weak dependence of $R_{\rm S, Acc}$ and $R_{\rm D, Acc}$ on $V_{\rm FG}$ is reflected in the $I_{\rm DS}$ trend to saturation. As the values of $R_{\rm S, Acc}$ and $R_{\rm D, Acc}$ are higher than the channel resistance, a larger fraction of $V_{\rm DS}$ drops in the access regions. This fact reduces the potential at the channel edges with respect to the no-access-regions scenario, reducing the output current. In addition, the $R_{\rm Acc}-V_{\rm FG}$ dependence is not symmetric, so neither are the access region potential drops, resulting into a~non-symmetric reduction of the output current, that is, an asymmetric $I_{\rm DS} - V_{\rm FG}$ curve shown in Figure \ref{fig:gfetidsvgsdiffmob}{b}. {\color{\commentsColor} This lack of equivalence between the source and drain access regions is explored in detail in the following~section.}

\begin{figure}[H]
	\centering
	\input{Figuras/RIds_Dev.tex}
	\caption{\color{\commentsColor} Transfer response (\textbf{a}, \textbf{b}) and structure resistances (\textbf{c}, \textbf{d}) as a function of the gate bias. These~results are obtained reducing the length of either the source (\textbf{a}, \textbf{c}, solid lines) or drain access region (\textbf{b}, \textbf{d}, dashed lines) down to 17.5 nm, and increasing the length of either the source (\textbf{a}, \textbf{c}, solid~lines) or the drain access region (\textbf{b}, \textbf{d}, dashed lines) up to 70 nm.}
	\label{fig:RIds_dev}
\end{figure}
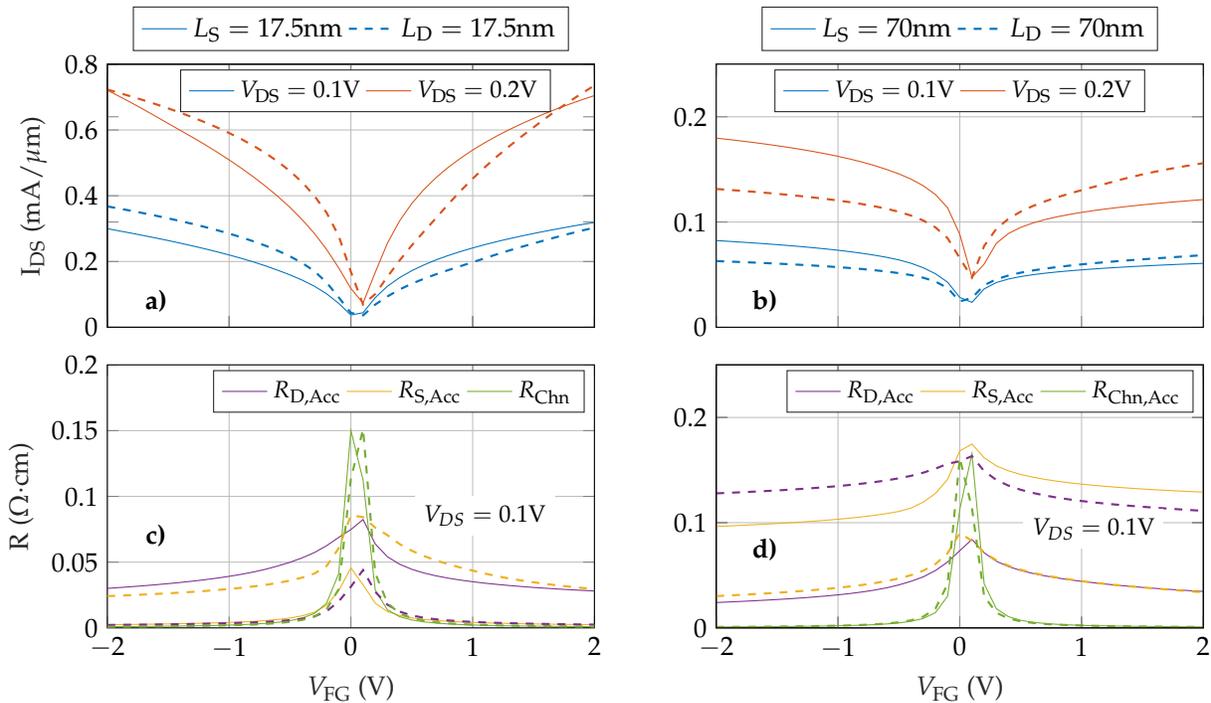

{\color{\commentsColor}
\subsubsection{Gate Misalignment}

In the previous section, we assumed that the gate is perfectly aligned in the middle of the channel leading to identical source and drain access regions ($L_{\rm S} = L_{\rm D} = L_{\rm Acc}$) at both ends. A more realistic scenario should consider the impact of having non-equal $L_{\rm S}$ and $L_{\rm D}$, enabling us to test the non-equivalent role of $R_{\rm S, Acc}$ and $R_{\rm D, Acc}$ on the GFET response. For this purpose, we have analysed GFETs where the top gate contact is not placed in the centre of the structure,  resulting in access regions of different length. In particular, we have kept $L_{\rm S}$ (or $L_{\rm D}$) equal to 35 nm while $L_{\rm D}$ (or $L_{\rm S}$) is modified. Specifically, we considered four scenarios: (i) short source, (ii)~short drain, (iii) long source and (iv) long drain. The length of the short and long regions is set to 17.5 nm and 70 nm, respectively. The~$I_{\rm DS}-V_{\rm FG}$ curves, along with the resistances $R_{\rm S, Acc}$, $R_{\rm D, Acc}$ and $R_{\rm Chn}$ obtained in each case, are~depicted in Figure~\ref{fig:RIds_dev}.}

As expected, there are significant differences between devices. Shortening either the source or the drain access regions results in a higher output current (Figure \ref{fig:RIds_dev}{a}) and reduces both its saturation and its asymmetry with respect to the elongated scenario (Figure \ref{fig:RIds_dev}{b}). When comparing the shorter regions (Figure \ref{fig:RIds_dev}{a}) it is clearly observable that the $L_{\rm S}$ = 17.5 nm device (solid lines) has a more symmetric response than the $L_{\rm D}$ = 17.5 nm (dashed lines). This is more evident for $V_{\rm DS}$ = 0.1 V and emphasizes the role of the source access region with respect to the drain access region. An equivalent conclusion can be achieved from the elongated devices (Figure~\ref{fig:RIds_dev}{b}). The longer $L_{\rm S}$ results in an increased asymmetry between both branches. These results can be explained by analysing the resistances of the structure. Figure \ref{fig:RIds_dev}{c},{d} show $R_{\rm S, Acc}$, $R_{\rm D, Acc}$ and $R_{\rm Chn}$ as a function of $V_{\rm FG}$ for $V_{\rm DS}$ = 0.1 V. When any access region is shortened (Figure \ref{fig:RIds_dev}{c}), its resistance is similar or lower than the channel resistance regardless $V_{\rm FG}$. The longer region resistance controls the total current (except for $V_{\rm FG}$ close to zero). When one of the regions is enlarged this effect is emphasized. The transfer responses in Figure \ref{fig:RIds_dev}{b} are clearly saturated due to the dominant role in the total conductivity of the longer access region.

\subsubsection{Impact of Electrostatic Doping and Puddles}

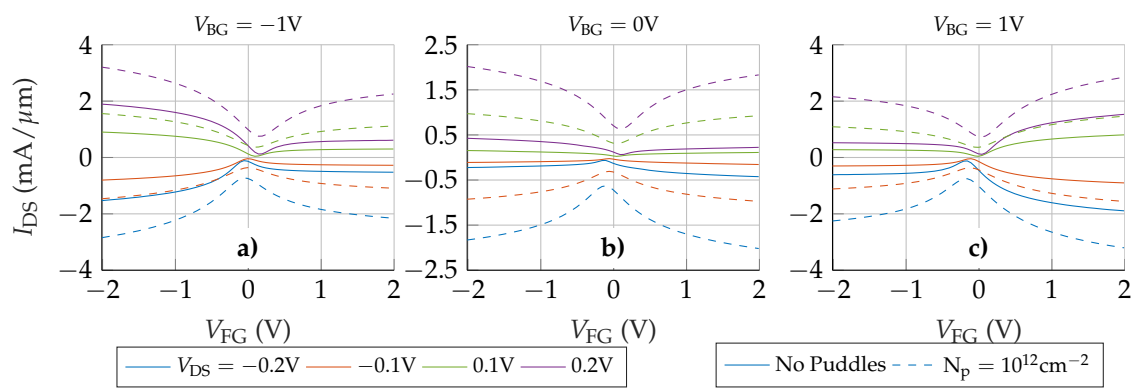
\begin{figure}[H]
	\centering
	\begin{tikzpicture}
	\input{Figuras/GFET_Vbg-1V_Np1e12_Comp}
	\input{Figuras/GFET_Vbg0V_Np1e12_Comp}
	\input{Figuras/GFET_Vbg1V_Np1e12_Comp}
	\end{tikzpicture}
	\caption{$I_{\rm DS}-V_{\rm FG}$ characteristics of the complete structure when three different back gate potentials are used ($-$1 V (\textbf{a}),  0 V (\textbf{b}) and 1 V (\textbf{c})). Solid lines correspond to the device without puddles and dashed lines to the device with $N_{\rm p} = 10^{12}$ cm$^{-2}$.}
	\label{fig:gfetidsvgsaccbg-1v}
\end{figure}

To reduce the impact of the access regions in the overall device performance, it is possible to increase their conductivity by means of an electrostatic doping using the back-gate terminal. In the following we analyse how the back gate influences the GFET behaviour. Figure \ref{fig:gfetidsvgsaccbg-1v} shows the transfer characteristic for three different values of $V_{\rm BG}$: $-$1 V, 0 V and 1 V (solid lines). For $V_{\rm BG} = $ 0 V the results are quite similar to the scenario without back gate. In the other two cases, depending on the polarity of $V_{\rm BG}$, electrons or holes are accumulated in the graphene layer. As a result, the P-type (N-type) branch is enhanced for $V_{\rm BG} = -$1 V (\mbox{$V_{\rm BG} =$ 1 V}), regardless the value of $V_{\rm DS}$. As in the previous scenario, the~origin of this behaviour can be traced back to the resistance associated with the access regions.

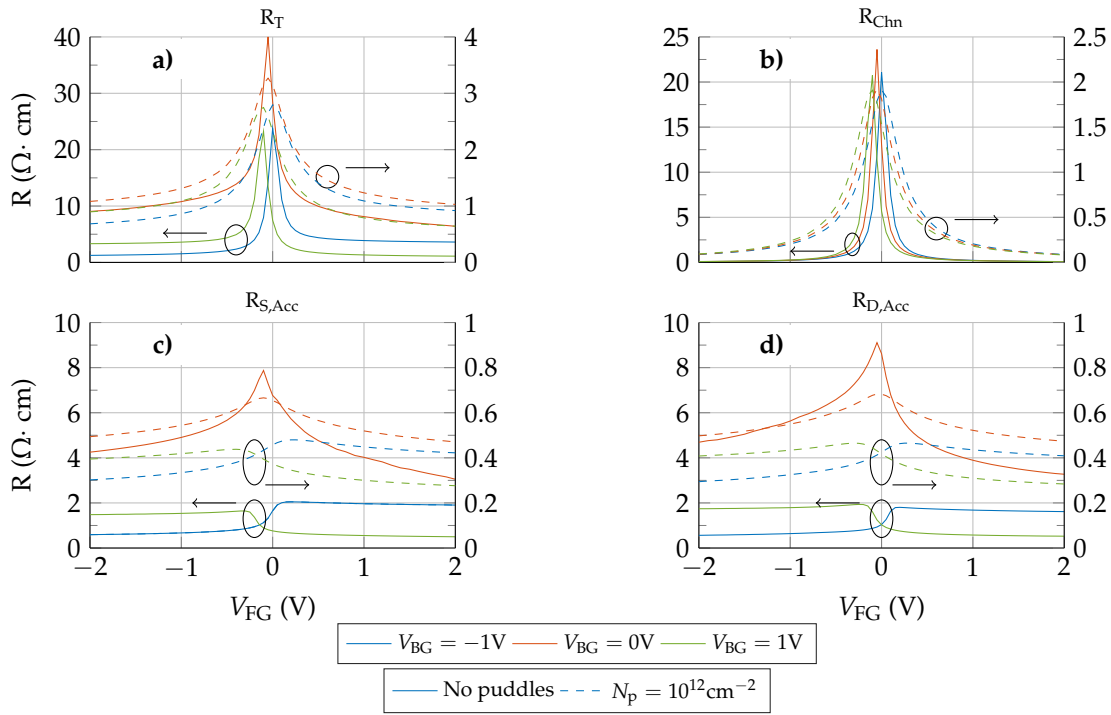
\begin{figure}[H]
	\centering
	\input{Figuras/GFET_Vbg1V_Rdev_NoPuddles}
	\caption{Total (\textbf{a}), channel (\textbf{b}), source (\textbf{c}) and drain (\textbf{d}) resistances for different back gate biases and \mbox{$V_{\rm DS}=-$0.1 V}. Solid lines (referred to the left axis) correspond to the no puddles scenario while dashed lines (referred to the right axis) depict the values obtained when a puddle concentration of \mbox{$N_{\rm p} = 10^{12}$ cm$^{-2}$} is considered.}
	\label{fig:gfetvbg1vrdevnopuddles}
\end{figure}

Figure \ref{fig:gfetvbg1vrdevnopuddles} depicts the device resistances for different $V_{\rm BG}$ and $V_{\rm DS} = -$0.1 V (without puddles, solid~lines). For $|V_{\rm BG}| =$ 1 V the total resistance near the Dirac voltage is dominated by $R_{\rm Chn}$. When $V_{\rm FG}$ is increased above $V_{\rm Dirac}$, the symmetry of $R_{\rm Chn}$ is kept since it is mostly controlled by the front gate, while the asymmetry of $R_{\rm S, Acc}$ and $R_{\rm D, Acc}$ is exacerbated due to the electrostatic doping, giving~rise to the large asymmetry observed in the transfer response, in Figure \ref{fig:gfetidsvgsaccbg-1v}. In particular, the asymmetric step-like dependence of the access resistances on $V_{\rm FG}$ (for $V_{\rm BG} \neq 0$ V) is the result of the electrostatic competition between the front and back gates to control the access regions closer to the channel. When~$V_{\rm FG}$ and $V_{\rm BG}$ have the same polarity, they add their electric forces to increase the carrier density in the aforementioned zones, increasing the conductivity and therefore lowering the whole access resistance. However, if $V_{\rm FG}$ is opposite to $V_{\rm BG}$, both gates compete to accumulate different types of charges, resulting in a depleted region close to the channel edges that decreases the conductivity and increases the overall access region resistances. An equivalent conclusion was achieved in \cite{Wilmart2016} where a~strong modulation of the total resistance by two additional gates is observed, as in Figure \ref{fig:gfetvbg1vrdevnopuddles}.

An additional aspect that cannot be overlooked is the effect of the presence of puddles in the graphene layer \cite{Martin2007, Zhang2009}. To shed light on this issue Figure \ref{fig:gfetidsvgsaccbg-1v} includes the $I_{\rm DS}-V_{\rm FG}$ response when a puddle charge density of $N_{\rm p} = 10^{12}$ cm$^{-2}$ is considered (dashed lines). Two major changes are observed after including the puddles: (i) the total current is increased, and (ii) the asymmetry is clearly reduced. These changes derive from the equal contribution of puddles to the conductivity of both electrons and holes, and explain why the $I-V$ curves of some experimental devices are reasonably symmetric close to the Dirac voltage, where the conductivity of puddles is dominant. In this situation, the conductivity of the whole graphene layer is increased for electrons and holes, in contrast with the electrostatic doping generated by the back gate. This non-selective improvement of the conductivity is translated into the resistances of the device: Figure \ref{fig:gfetvbg1vrdevnopuddles} includes the $R-V_{\rm FG}$ relation for \mbox{$N_{\rm p} = 10^{12}$ cm$^{-2}$} (dashed lines). The step-like behaviour of $R_{\rm S, Acc}$ and $R_{\rm D, Acc}$ is softened when the puddles are included, resembling the $V_{\rm BG} = 0$ V case.

\subsubsection{RF Performance}

Finally, to determine the impact of the access regions in the RF performance, we evaluate the cut-off frequency, $f_{\rm T}$, as a RF figure of merit (FoM). The value of $f_{\rm T}$ is calculated as in \cite{Marian2017, Schwierz2013}:

\begin{equation}
f_{\rm T} = \frac{1}{2\pi}\frac{g_{\rm m}}{C_{\rm fg}}
\end{equation}
where $g_{\rm m}$ is the transconductance and $C_{\rm fg}$ the front gate capacitance.

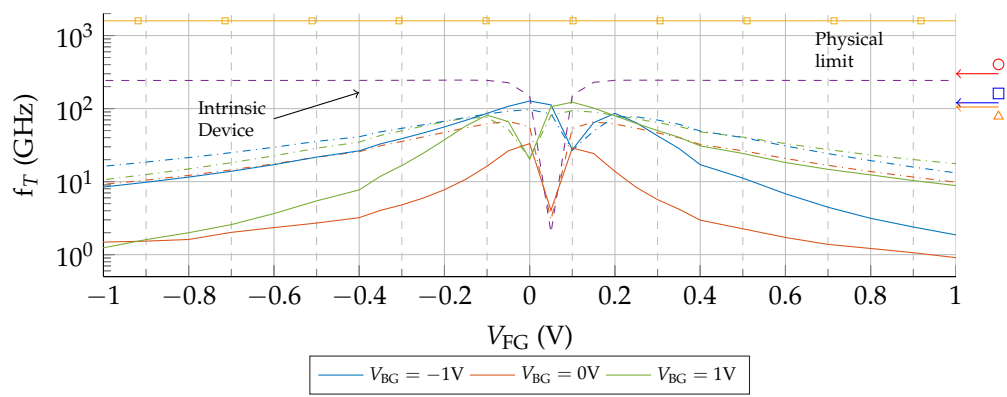
\begin{figure}[H]
	\centering
	\begin{tikzpicture}
	\input{Figuras/GFET_Puddles_VBG_fT_Comp}
	\end{tikzpicture}
	\caption{$f_{\rm T}$ of the back-gated device with access regions under two scenarios: no puddles (solid lines) and $N_{\rm p} = 10^{12}$ cm$^{-2}$ (dash-dotted lines). The values obtained for the intrinsic device are depicted by the purple dashed line. The arrows labelled by marks on the right side axis indicate the values of $f_{\rm T}$ extracted from \cite{Liao2010} (circle) and \cite{Wu2012} (square and triangle). The yellow line indicates the physical limit for graphene $v_{\rm F}/2\pi L$, determined by the transit time $L/v_{\rm F}$, with the Fermi velocity \mbox{$v_{\rm F} \approx 10^8$} cm/s and $L = $100 nm (squares).}
	\label{fig:gfetnopuddlesvbg0vftcomp}
\end{figure}

Figure \ref{fig:gfetnopuddlesvbg0vftcomp} shows $f_{\rm T}$ as a function of $V_{\rm FG}$ under two scenarios: no puddles (solid lines) and \mbox{$N_{\rm p} = 10^{12}$cm$^{-2}$} (dash-dotted lines). To assess the impact of the access regions, the performance of the intrinsic device (structure indicated by the dotted rectangle in Figure \ref{fig:graphenemosfet}) is depicted too (dashed lines). In addition, to evaluate the magnitude of the calculated values, the experimental measurements of $f_{\rm T}$ reported in \cite{Liao2010} and \cite{Wu2012} are indicated by the arrows on the right side axis of Figure \ref{fig:gfetnopuddlesvbg0vftcomp}. Despite~the device structure and the bias conditions are different, the channel lengths of these experimental devices are similar to the ones simulated here (144 nm \cite{Liao2010} and 140~nm~\cite{Wu2012}), and therefore constitute a~good reference. Importantly, a de-embedding procedure was carried out for the RF measurements of these experimental devices by using specific ``short'' and ``open'' structures with identical layouts in order to remove the effects of the parasitics associated with the pads and connections, but~not the contact and access region resistances.

Including the access regions results in a quite different response compared with the intrinsic device, as the associated parasitic resistances provoke a bias dependent decay of $f_{\rm T}$. Considering the scenario without puddles, when the back gate is properly biased, $f_{\rm T}$ is considerably improved. If we analyse Figure \ref{fig:gfetnopuddlesvbg0vftcomp} in combination with Figure \ref{fig:gfetvbg1vrdevnopuddles}, those combinations of $V_{\rm FG}$, $V_{\rm BG}$ for which the \mbox{$R_{\rm S} - V_{\rm FG}\,(R_{\rm D} - V_{\rm FG})$} curve shows its minimum values, are those for which $f_{\rm T}$ shows a greater improvement. When $R_{\rm S}\,(R_{\rm D})$ is higher, $f_{\rm T}$ is spoiled with respect to the $V_{\rm BG} = 0$ V case. This~relation between the access region conductivity and the improvement of the RF performance was experimentally observed in \cite{Al-Amin2016} where a higher $f_{\rm T}$ was demonstrated when a GFET with two additional electrodes was properly biased to control such conductivity. When puddles are included, the channel conductivity increases, what reduces the control of the back-gate bias, and simultaneously results in a~more symmetric $f_{\rm T}-V_{\rm FG}$ dependence.

\section{Conclusions}\label{sec:concl}

GFETs have been thoroughly studied in order to assess the impact of the access regions in the device performance. The validation of our approach against two experimental devices spotlights the importance of these regions as well as the presence of puddles to reproduce the state-of-the-art technology. When the access regions are considered, the transfer response reveals a lower, saturated~and asymmetric $I_{\rm DS} - V_{\rm FG}$ characteristic that is not observed in their absence. To explore the impact of a variable conductivity of these regions we have included a back gate in the structure able to introduce an electrostatic doping. The back gate increases the output current as well as the asymmetry of the transfer characteristic. The latter effect is explained in terms of the competition of the back and front gates that results in a depletion of the amount of carriers close to the channel edges when both biases have an opposite polarity. The influence of puddles is also theoretically investigated, observing that they reduce the asymmetry of $I_{\rm DS} - V_{\rm FG}$.

The analysis of the impact of the access regions and puddles have been extended to the prediction of the cut-off frequency to assess the properties of GFETs for potential RF applications. Our results reveal an important degradation of the $f_{\rm T}-V_{\rm FG}$ relation due to access regions. The application of an appropriate back gate bias can tune the access region conductivity generating a remarkable improvement in the RF performance. The presence of puddles also mitigates this degradation, but~neglects the possibility of tuning the access regions conductivity.

\vspace{12pt} 

\authorcontributions{A.T.-L., E.G.M., F.G.R. and A.G. conceived the work. A.T.-L., F.P., and J.M.G.-M. performed the numerical simulations under the supervision of E.G.M., F.G.R., D.J. and A.G. All authors analysed the results, contributed~to the discussion and wrote the manuscript.}


\funding{This research was founded by Spanish government grant numbers TEC2017-89955-P (MINECO/AEI/FEDER, UE), TEC2015-67462-C2-1-R (MINECO), IJCI-2017-32297 (MINECO/AEI), FPU16/04043 and FPU14/02579, and the European Union's Horizon 2020 Research and Innovation Program under Grant GrapheneCore2 785219.}


\conflictsofinterest{The authors declare no conflict of interest.} 

\abbreviations{The following abbreviations are used in this manuscript:\\

\noindent 
\begin{tabular}{@{}ll}
2DM & Two-dimensional material \\
MOSFET & Metal-Oxide-Semiconductor Field-Effect Transistor \\
GFET & Graphene Field-Effect Transistors \\
RF & Radio-Frequency
\end{tabular}}

%

\reftitle{References}


\externalbibliography{yes}





\end{document}

%% file: Figuras/GFET_IdsVgs_EqMob.tex
%
%
%

\begin{axis}[%
width=0.35\linewidth,
height=3cm,
at={(0\linewidth,0cm)},
scale only axis,
xmin=-2,
xmax=2,
minor x tick num=1,
xtick={-2,-1,...,2},
xlabel style={font=\color{white!15!black}},
xlabel={$V_{\rm FG}$ (V)},
ymin=-30,
ymax=30,
ytick={-30,-20,...,30},
minor y tick num=1,
ylabel style={font=\color{white!15!black}},
ylabel={$I_{\rm DS}$ (mA/$\mu$m)},
axis background/.style={fill=white},
axis x line*=bottom,
axis y line*=left,
xmajorgrids,
xminorgrids,
ymajorgrids,
yminorgrids,
legend style={at={(1.05,-0.4)}, anchor=north, legend cell align=left, align=left, draw=white!15!black, legend columns=4, font=\footnotesize}
]
\addplot [color=mycolor1]
table[row sep=crcr, y expr=1e-3*\thisrowno{1}*0.1]{%
	-2	-214055.305810551\\
	-1.9	-199965.634492867\\
	-1.8	-186006.741040873\\
	-1.7	-172190.602207411\\
	-1.6	-158528.267929907\\
	-1.5	-145034.691944375\\
	-1.4	-131727.204289703\\
	-1.3	-118625.280129862\\
	-1.2	-105752.724963415\\
	-1.1	-93138.4585005085\\
	-1	-80818.5728951754\\
	-0.9	-68836.4231860499\\
	-0.8	-57251.4928128322\\
	-0.7	-46150.9815203101\\
	-0.6	-35679.5366316721\\
	-0.5	-25993.9378932943\\
	-0.45	-21413.0136498882\\
	-0.4	-17088.5624434106\\
	-0.35	-13050.0786475505\\
	-0.3	-9373.91707365926\\
	-0.25	-6262.03461128325\\
	-0.2	-3801.87515118326\\
	-0.15	-2194.66778474902\\
	-0.1	-1622.40861447003\\
	-0.05	-2172.30397462508\\
	0	-3760.54257294782\\
	0.05	-6201.59058142594\\
	0.1	-9300.06464500767\\
	0.15	-12893.7027298439\\
	0.2	-16841.886052844\\
	0.25	-21140.3151279026\\
	0.3	-25716.7596424271\\
	0.35	-30541.0996856746\\
	0.4	-35606.9487709044\\
	0.45	-40839.167905543\\
	0.5	-46217.0373872851\\
	0.6	-57360.7844784924\\
	0.7	-68957.4416761027\\
	0.8	-80943.8255861216\\
	0.9	-93265.8129111815\\
	1	-105881.410428119\\
	1.1	-118755.361076021\\
	1.2	-131858.488655898\\
	1.3	-145167.081954673\\
	1.4	-158661.676530084\\
	1.5	-172324.92487357\\
	1.6	-186141.935851298\\
	1.7	-200101.705250051\\
	1.8	-214192.050165516\\
	1.9	-228402.997072509\\
	2	-242727.008901837\\
};
\addlegendentry{$V_{\rm DS} = -0.2$V}

\addplot [color=mycolor2]
table[row sep=crcr, y expr=1e-3*\thisrowno{1}*0.1]{%
	-2	-110564.650344689\\
	-1.9	-103488.635371564\\
	-1.8	-96474.9504021847\\
	-1.7	-89529.6496496005\\
	-1.6	-82658.8617709251\\
	-1.5	-75867.2505544954\\
	-1.4	-69165.2625019615\\
	-1.3	-62559.7332918546\\
	-1.2	-56063.3241421914\\
	-1.1	-49687.0815906959\\
	-1	-43449.3347763176\\
	-0.9	-37367.9910102368\\
	-0.8	-31469.2592199905\\
	-0.7	-25788.5652485399\\
	-0.6	-20373.1880105673\\
	-0.5	-15286.5575257006\\
	-0.45	-12893.4495382432\\
	-0.4	-10617.6698624238\\
	-0.35	-8450.19810700475\\
	-0.3	-6435.9994910765\\
	-0.25	-4609.99414182275\\
	-0.2	-3011.85504621291\\
	-0.15	-1698.45373431165\\
	-0.1	-821.226546804956\\
	-0.05	-497.635857746491\\
	0	-807.012871721023\\
	0.05	-1667.96264564067\\
	0.1	-2942.81143278993\\
	0.15	-4516.39442103525\\
	0.2	-6313.18007647211\\
	0.25	-8315.31327486712\\
	0.3	-10478.3524367735\\
	0.35	-12774.4950332832\\
	0.4	-15184.8443426447\\
	0.45	-17696.2433954118\\
	0.5	-20300.2919717659\\
	0.6	-25759.4254995728\\
	0.7	-31469.3436605933\\
	0.8	-37388.0712027365\\
	0.9	-43483.8719374981\\
	1	-49730.6831072937\\
	1.1	-56113.4628159291\\
	1.2	-62614.6110643863\\
	1.3	-69222.9155443557\\
	1.4	-75927.084917001\\
	1.5	-82720.1370937199\\
	1.6	-89591.9605833674\\
	1.7	-96538.250978493\\
	1.8	-103552.560867016\\
	1.9	-110629.127518446\\
	2	-117763.831651267\\
};
\addlegendentry{$-0.1$V}

\addplot [color=mycolor3]
table[row sep=crcr, y expr=1e-3*\thisrowno{1}*0.1]{%
	-2	117763.833249485\\
	-1.9	110629.128538983\\
	-1.8	103552.562733713\\
	-1.7	96538.2534942416\\
	-1.6	89591.9654136485\\
	-1.5	82720.1422992887\\
	-1.4	75927.0922420942\\
	-1.3	69222.9253291249\\
	-1.2	62614.6252912117\\
	-1.1	56113.4791183975\\
	-1	49730.7071945926\\
	-0.9	43483.9037374521\\
	-0.8	37388.1158833802\\
	-0.7	31469.3767444436\\
	-0.6	25759.5082150904\\
	-0.5	20300.3434315904\\
	-0.45	17696.2445214807\\
	-0.4	15184.8460512941\\
	-0.35	12774.4966279464\\
	-0.3	10478.3536270289\\
	-0.25	8315.31614787691\\
	-0.2	6313.18246058355\\
	-0.15	4516.35519578859\\
	-0.1	2942.81046458787\\
	-0.05	1667.96416300225\\
	0	806.379459222743\\
	0.05	497.635035032106\\
	0.1	821.226827999879\\
	0.15	1698.45825976077\\
	0.2	3011.86252089962\\
	0.25	4610.00312463682\\
	0.3	6435.97984382553\\
	0.35	8450.20660141677\\
	0.4	10617.6369657369\\
	0.45	12893.4195339495\\
	0.5	15286.5088946612\\
	0.6	20373.1576116862\\
	0.7	25788.5483335502\\
	0.8	31469.2530165151\\
	0.9	37367.989735251\\
	1	43449.3352972985\\
	1.1	49687.0801679818\\
	1.2	56063.3227502455\\
	1.3	62559.7367934593\\
	1.4	69165.2617673216\\
	1.5	75867.2545508282\\
	1.6	82658.8654459927\\
	1.7	89529.6499235704\\
	1.8	96474.9508102456\\
	1.9	103488.636076975\\
	2	110564.650527867\\
};
\addlegendentry{$0.1$V}

\addplot [color=mycolor4]
table[row sep=crcr, y expr=1e-3*\thisrowno{1}*0.1]{%
	-2	242727.007253567\\
	-1.9	228403.001928303\\
	-1.8	214192.04822826\\
	-1.7	200101.710627019\\
	-1.6	186141.942481556\\
	-1.5	172324.930935451\\
	-1.4	158661.676074298\\
	-1.3	145167.081416463\\
	-1.2	131858.488908599\\
	-1.1	118755.361111878\\
	-1	105881.412518966\\
	-0.9	93265.815297185\\
	-0.8	80943.8235480024\\
	-0.7	68957.4489224424\\
	-0.6	57360.7995227202\\
	-0.5	46217.081232354\\
	-0.45	40839.2208600688\\
	-0.4	35606.9893403111\\
	-0.35	30541.2171528357\\
	-0.3	25716.747752604\\
	-0.25	21140.3118074698\\
	-0.2	16841.8876345973\\
	-0.15	12893.7043268691\\
	-0.1	9300.06596554712\\
	-0.05	6201.59805534909\\
	0	3758.15713606586\\
	0.05	2172.30560167149\\
	0.1	1622.40897182322\\
	0.15	2194.66719136023\\
	0.2	3801.87699780625\\
	0.25	6262.0325267196\\
	0.3	9373.92114433591\\
	0.35	13050.101701525\\
	0.4	17088.5387432867\\
	0.45	21412.9806375925\\
	0.5	25993.9908897849\\
	0.6	35679.4659712681\\
	0.7	46150.9600294586\\
	0.8	57251.4857932842\\
	0.9	68836.4206037327\\
	1	80818.5715279688\\
	1.1	93138.4593607042\\
	1.2	105752.725886633\\
	1.3	118625.28191186\\
	1.4	131727.198116611\\
	1.5	145034.694347144\\
	1.6	158528.261795876\\
	1.7	172190.597173854\\
	1.8	186006.736158289\\
	1.9	199965.62995576\\
	2	214055.300670812\\
};
\addlegendentry{$0.2$V}

\SubfigLetter{a)}{0.5}{0.85};


\end{axis}

%% file: Figuras/GFET_IdsVgs_Acc.tex
%
%
%

\begin{axis}[%
width=0.35\linewidth,
height=3cm,
at={(0.4\linewidth,0cm)},
scale only axis,
xmin=-2,
xmax=2,
xtick={-2,-1,...,2},
minor x tick num=1,
xlabel style={font=\color{white!15!black}},
xlabel={$V_{\rm FG}$ (V)},
ymin=-0.5,
ymax=0.5,
ytick={-0.8, -0.6, -0.4, -0.2, 0, 0.2, 0.4, 0.6, 0.8},
minor y tick num=1,
ylabel style={font=\color{white!15!black}},
axis background/.style={fill=white},
axis x line*=bottom,
axis y line*=right,
xmajorgrids,
xminorgrids,
ymajorgrids,
yminorgrids,
]
\addplot [color=mycolor1]
table[row sep=crcr, y expr=1e-3*\thisrowno{1}*0.1]{%
	-2	-3459.22425605042\\
	-1.9	-3381.15571376395\\
	-1.8	-3300.6152968537\\
	-1.7	-3216.57853512986\\
	-1.6	-3129.14253790442\\
	-1.5	-3036.98604933154\\
	-1.4	-2939.34179782031\\
	-1.3	-2835.30571766167\\
	-1.2	-2723.32465329705\\
	-1.1	-2601.4548486522\\
	-1	-2467.05287817161\\
	-0.9	-2319.1716297734\\
	-0.8	-2175.9266980747\\
	-0.7	-2033.9933219024\\
	-0.6	-1883.44275149411\\
	-0.5	-1716.6665876892\\
	-0.45	-1623.54371586778\\
	-0.4	-1521.03352443261\\
	-0.35	-1405.75578214302\\
	-0.3	-1286.2573767626\\
	-0.25	-1137.86147902582\\
	-0.2	-943.207205860088\\
	-0.15	-705.996063618551\\
	-0.1	-583.676823428313\\
	-0.05	-788.527063873543\\
	0	-1097.2803500694\\
	0.05	-1348.43061731146\\
	0.1	-1571.87775906122\\
	0.15	-1763.2128990559\\
	0.2	-1929.58581006358\\
	0.25	-2076.86706185446\\
	0.3	-2212.22340028983\\
	0.35	-2335.91050577831\\
	0.4	-2449.86040294623\\
	0.45	-2555.91583398471\\
	0.5	-2657.36796774137\\
	0.6	-2844.19500628601\\
	0.7	-3014.96293471339\\
	0.8	-3174.81232251231\\
	0.9	-3329.15811464859\\
	1	-3470.35593501947\\
	1.1	-3600.30022601092\\
	1.2	-3721.25341669807\\
	1.3	-3834.55825997932\\
	1.4	-3941.37368551957\\
	1.5	-4042.75526929558\\
	1.6	-4139.24428610587\\
	1.7	-4231.45517660123\\
	1.8	-4319.90084112586\\
	1.9	-4404.82998207662\\
	2	-4486.80430147776\\
};

\addplot [color=mycolor2]
table[row sep=crcr, y expr=1e-3*\thisrowno{1}*0.1]{%
	-2	-1517.41997709509\\
	-1.9	-1482.65590346374\\
	-1.8	-1446.62651572823\\
	-1.7	-1409.23796253878\\
	-1.6	-1370.03317503784\\
	-1.5	-1328.89197770154\\
	-1.4	-1285.94102462154\\
	-1.3	-1245.07458341577\\
	-1.2	-1204.61409795826\\
	-1.1	-1163.65715612689\\
	-1	-1121.49267504004\\
	-0.9	-1077.59889194911\\
	-0.8	-1031.07757515801\\
	-0.7	-980.351770022321\\
	-0.6	-923.896563958471\\
	-0.5	-859.190018035055\\
	-0.45	-822.502542117376\\
	-0.4	-783.837662166859\\
	-0.35	-741.373903791388\\
	-0.3	-692.10956947809\\
	-0.25	-633.045337909517\\
	-0.2	-559.213288491704\\
	-0.15	-460.400288492877\\
	-0.1	-331.062667690021\\
	-0.05	-246.317018318334\\
	0	-343.643097626381\\
	0.05	-484.910769215204\\
	0.1	-602.632210263322\\
	0.15	-695.061403507392\\
	0.2	-769.985578353213\\
	0.25	-833.320172532061\\
	0.3	-889.952175248628\\
	0.35	-940.955566589766\\
	0.4	-988.182153611337\\
	0.45	-1032.2815922316\\
	0.5	-1073.08248372817\\
	0.6	-1146.89924571965\\
	0.7	-1212.70154557199\\
	0.8	-1277.55518418373\\
	0.9	-1338.96979406041\\
	1	-1395.88789290782\\
	1.1	-1449.14708431628\\
	1.2	-1499.47975272734\\
	1.3	-1546.90541934578\\
	1.4	-1592.30547878175\\
	1.5	-1635.6911458677\\
	1.6	-1677.47928359954\\
	1.7	-1718.17374720225\\
	1.8	-1757.36482145149\\
	1.9	-1795.8863061764\\
	2	-1833.47892446508\\
};

\addplot [color=mycolor3]
table[row sep=crcr, y expr=1e-3*\thisrowno{1}*0.1]{%
	-2	1818.3263039986\\
	-1.9	1780.11627255561\\
	-1.8	1741.17318859259\\
	-1.7	1701.44213305721\\
	-1.6	1660.2899959253\\
	-1.5	1618.06354860682\\
	-1.4	1574.01247732202\\
	-1.3	1527.95531144562\\
	-1.2	1479.75967888429\\
	-1.1	1428.56812345526\\
	-1	1374.49113531614\\
	-0.9	1316.64835609826\\
	-0.8	1254.0680667001\\
	-0.7	1188.11660799674\\
	-0.6	1121.48528553636\\
	-0.5	1046.99635330138\\
	-0.45	1007.32392849943\\
	-0.4	965.270301572823\\
	-0.35	919.9928710352\\
	-0.3	870.701820105819\\
	-0.25	815.592339103785\\
	-0.2	752.683530163687\\
	-0.15	680.238967736539\\
	-0.1	590.791754531489\\
	-0.05	476.795058642145\\
	0	336.004339915463\\
	0.05	245.933928637613\\
	0.1	332.745795501573\\
	0.15	462.662149628937\\
	0.2	562.252647907846\\
	0.25	636.735235870672\\
	0.3	696.007184957833\\
	0.35	745.338685618598\\
	0.4	787.829293999008\\
	0.45	826.926260584568\\
	0.5	863.463494867547\\
	0.6	928.161231392938\\
	0.7	984.582018368366\\
	0.8	1035.2224416477\\
	0.9	1081.769891752\\
	1	1125.83273337917\\
	1.1	1168.04175805754\\
	1.2	1209.18123328528\\
	1.3	1249.80576647236\\
	1.4	1291.09809364761\\
	1.5	1333.85214381288\\
	1.6	1374.31751626285\\
	1.7	1412.89190698308\\
	1.8	1449.81569485817\\
	1.9	1485.31333949063\\
	2	1519.62823817815\\
};

\addplot [color=mycolor4]
table[row sep=crcr, y expr=1e-3*\thisrowno{1}*0.1]{%
	-2	4471.99247142833\\
	-1.9	4389.19525459154\\
	-1.8	4303.03913674722\\
	-1.7	4213.7429926242\\
	-1.6	4120.31640959921\\
	-1.5	4022.85171403208\\
	-1.4	3919.95891433068\\
	-1.3	3811.73614609257\\
	-1.2	3696.59511347488\\
	-1.1	3573.54476391452\\
	-1	3440.66999040679\\
	-0.9	3295.61012651733\\
	-0.8	3140.64440536947\\
	-0.7	2979.87618325068\\
	-0.6	2806.89634507141\\
	-0.5	2616.4310545747\\
	-0.45	2512.59397946102\\
	-0.4	2406.9303069689\\
	-0.35	2293.71462919686\\
	-0.3	2170.95060255623\\
	-0.25	2036.53451853548\\
	-0.2	1887.48058207704\\
	-0.15	1724.63827641068\\
	-0.1	1537.69334080508\\
	-0.05	1318.29201675577\\
	0	1062.0158350406\\
	0.05	779.301249289983\\
	0.1	582.940339030137\\
	0.15	709.450243536836\\
	0.2	947.975878157059\\
	0.25	1142.2496253937\\
	0.3	1290.48267397089\\
	0.35	1409.55790501796\\
	0.4	1524.17995235417\\
	0.45	1626.39325163826\\
	0.5	1719.0423847765\\
	0.6	1884.88222881474\\
	0.7	2034.32321298924\\
	0.8	2174.65232571539\\
	0.9	2315.00216596007\\
	1	2460.06100574272\\
	1.1	2595.19354881602\\
	1.2	2718.35693053149\\
	1.3	2831.2640453631\\
	1.4	2935.98842386624\\
	1.5	3034.1792805868\\
	1.6	3126.76864432505\\
	1.7	3214.60618275249\\
	1.8	3298.90728600251\\
	1.9	3379.68854198353\\
	2	3457.98107519749\\
};

\SubfigLetter{b)}{0.55}{0.85};

\end{axis}

%% file: Figuras/GFET_NoVbg_Rdev_NoPuddles_VDS-0.1V.tex
%
%
%
\begin{tikzpicture}

\begin{axis}[%
width=0.7\linewidth,
height=3cm,
at={(0\linewidth,0)},
scale only axis,
xmin=-2,
xmax=2,
xtick={-2,-1.5,...,2},
xlabel={$V_{\rm FG}$ (V)},
ymin=0,
ymax=45,
ytick={5,15,...,45},
minor y tick num=1,
ylabel={R ($\Omega\cdot$ cm)},
axis background/.style={fill=white},
axis x line*=bottom,
axis y line*=left,
xmajorgrids,
ymajorgrids,
yminorgrids,
clip mode=individual,
legend style={at={(0.01,0.95)}, anchor=north west ,legend cell align=left, align=left, draw=white!15!black, legend columns=1}
]
\addplot [color=mycolor1]
table[row sep=crcr, y expr=\thisrowno{1}*10]{%
	-2	0.324123005351235\\
	-1.9	0.332567041429697\\
	-1.8	0.341818844083769\\
	-1.7	0.352078393988623\\
	-1.6	0.363468490487053\\
	-1.5	0.376417520305468\\
	-1.4	0.390867824385881\\
	-1.3	0.404920458652117\\
	-1.2	0.419346108282266\\
	-1.1	0.434593876297958\\
	-1	0.45108972232593\\
	-0.9	0.469186035595865\\
	-0.8	0.489514936571144\\
	-0.7	0.513405681211237\\
	-0.6	0.542203055498683\\
	-0.5	0.5783295839383\\
	-0.4	0.623224004171384\\
	-0.35	0.647913401163434\\
	-0.3	0.676653714380504\\
	-0.25	0.710066528554265\\
	-0.2	0.747202132497488\\
	-0.15	0.789717134206943\\
	-0.1	0.840149660862515\\
	-0.05	0.909423471459724\\
	0	0.869044162323204\\
	0.05	0.751758314708158\\
	0.1	0.659605056030518\\
	0.15	0.597300522825207\\
	0.2	0.552630444236682\\
	0.25	0.518612884577343\\
	0.3	0.490757998041873\\
	0.35	0.467805186055642\\
	0.4	0.448436055491737\\
	0.5	0.417281545058728\\
	0.6	0.393225849445595\\
	0.7	0.373840860247765\\
	0.8	0.357847680852293\\
	0.9	0.344290805650801\\
	1	0.332538661893811\\
	1.1	0.322164503971019\\
	1.2	0.312935699437706\\
	1.3	0.304645893741859\\
	1.4	0.297077621228652\\
	1.5	0.290175394327596\\
	1.6	0.283803509425297\\
	1.7	0.277886709978691\\
	1.8	0.272402461332895\\
	1.9	0.267226917859496\\
	2	0.26241059866415\\
};
\addlegendentry{R$_{\rm S, Acc}$}

\addplot [color=mycolor2]
table[row sep=crcr, y expr=\thisrowno{1}*10]{%
	-2	0.325786270747447\\
	-1.9	0.332170284256765\\
	-1.8	0.339008090447885\\
	-1.7	0.346276583744479\\
	-1.6	0.354253832120118\\
	-1.5	0.362807147900702\\
	-1.4	0.372196746780494\\
	-1.3	0.382123096011345\\
	-1.2	0.392797409168966\\
	-1.1	0.404449980007199\\
	-1	0.41733198364213\\
	-0.9	0.431753290112132\\
	-0.8	0.448198424913122\\
	-0.7	0.467357087155443\\
	-0.6	0.490327190345677\\
	-0.5	0.518800175116963\\
	-0.4	0.55557856982599\\
	-0.35	0.578489475606468\\
	-0.3	0.606493455384787\\
	-0.25	0.642043093525377\\
	-0.2	0.689208461211978\\
	-0.15	0.754812471114483\\
	-0.1	0.832514053647143\\
	-0.05	0.794414259755383\\
	0	0.725458764139786\\
	0.05	0.692625797690614\\
	0.1	0.651451620253853\\
	0.15	0.615464342088887\\
	0.2	0.585074322690122\\
	0.25	0.559311297552319\\
	0.3	0.536113512259636\\
	0.35	0.515600571297723\\
	0.4	0.496813032032758\\
	0.5	0.464785051429119\\
	0.6	0.439412506714344\\
	0.7	0.418615385637898\\
	0.8	0.397852134620302\\
	0.9	0.379313348582895\\
	1	0.363544239123224\\
	1.1	0.349906433127576\\
	1.2	0.337847162429501\\
	1.3	0.327233915173473\\
	1.4	0.317661566386566\\
	1.5	0.308998998818035\\
	1.6	0.30107906729331\\
	1.7	0.293688272557843\\
	1.8	0.286901491900866\\
	1.9	0.280495420948072\\
	2	0.274447972696299\\
};
\addlegendentry{R$_{\rm D, Acc}$}

\addplot [color=mycolor3]
table[row sep=crcr, y expr=\thisrowno{1}*10]{%
	-2	0.00922784362739762\\
	-1.9	0.00986038945168063\\
	-1.8	0.0105786813033378\\
	-1.7	0.0114018047223874\\
	-1.6	0.0123533236150769\\
	-1.5	0.0134633957860031\\
	-1.4	0.0147762339004532\\
	-1.3	0.0163437032350603\\
	-1.2	0.0182470385713911\\
	-1.1	0.0205977605216751\\
	-1	0.0235704000272537\\
	-0.9	0.0274261803130263\\
	-0.8	0.0325933199012637\\
	-0.7	0.03982587174507\\
	-0.6	0.0505347588032801\\
	-0.5	0.0676826808377246\\
	-0.4	0.0983117111164567\\
	-0.35	0.124129844990283\\
	-0.3	0.163926647012288\\
	-0.25	0.230655019776782\\
	-0.2	0.356545254438628\\
	-0.15	0.635786021891761\\
	-0.1	1.36471105560601\\
	-0.05	2.37811968189512\\
	0	1.33170072249815\\
	0.05	0.6259949227336\\
	0.1	0.353008272792332\\
	0.15	0.229033861392574\\
	0.2	0.163225574936773\\
	0.25	0.123775985032499\\
	0.3	0.098122386644424\\
	0.35	0.080442853565801\\
	0.4	0.067635363821971\\
	0.5	0.0505212612739027\\
	0.6	0.0398248672585673\\
	0.7	0.0325971420865879\\
	0.8	0.0274222649803517\\
	0.9	0.0235628710175159\\
	1	0.020590281711079\\
	1.1	0.0182411443343181\\
	1.2	0.0163399776641594\\
	1.3	0.0147754189937477\\
	1.4	0.0134662751492844\\
	1.5	0.0123579181019356\\
	1.6	0.0114067986035843\\
	1.7	0.0105837723386756\\
	1.8	0.00986531638305281\\
	1.9	0.00923250029960928\\
	2	0.00867159727707867\\
};
\addlegendentry{R$_{\rm Chn}$}

\addplot [color=mycolor4]
table[row sep=crcr, y expr=\thisrowno{1}*10]{%
	-2	0.65913711972608\\
	-1.9	0.674597715138143\\
	-1.8	0.691405615834991\\
	-1.7	0.709756782455489\\
	-1.6	0.730075646222248\\
	-1.5	0.752688063992173\\
	-1.4	0.777840805066828\\
	-1.3	0.803387257898522\\
	-1.2	0.830390556022623\\
	-1.1	0.859641616826832\\
	-1	0.891992105995313\\
	-0.9	0.928365506021023\\
	-0.8	0.97030668138553\\
	-0.7	1.02058864011175\\
	-0.6	1.08306500464764\\
	-0.5	1.16481243989299\\
	-0.4	1.27711428511383\\
	-0.35	1.35053272176019\\
	-0.3	1.44707381677758\\
	-0.25	1.58276464185642\\
	-0.2	1.79295584814809\\
	-0.15	2.18031562721319\\
	-0.1	3.03737477011567\\
	-0.05	4.08195741311023\\
	0	2.92620364896114\\
	0.05	2.07037903513237\\
	0.1	1.6640649490767\\
	0.15	1.44179872630667\\
	0.2	1.30093034186358\\
	0.25	1.20170016716216\\
	0.3	1.12499389694593\\
	0.35	1.06384861091917\\
	0.4	1.01288445134647\\
	0.5	0.93258785776175\\
	0.6	0.872463223418507\\
	0.7	0.825053387972251\\
	0.8	0.783122080452947\\
	0.9	0.747167025251211\\
	1	0.716673182728114\\
	1.1	0.690312081432913\\
	1.2	0.667122839531367\\
	1.3	0.64665522790908\\
	1.4	0.628205462764502\\
	1.5	0.611532311247566\\
	1.6	0.596289375322191\\
	1.7	0.582158754875209\\
	1.8	0.569169269616814\\
	1.9	0.556954839107177\\
	2	0.545530168637528\\
};
\addlegendentry{R$_{\rm T}$}

\end{axis}

\begin{axis}[%
width=0.7\linewidth,
height=3cm,
at={(0\linewidth,0)},
scale only axis,
xmin=-2,
xmax=2,
xticklabels={,,},
ymin=0,
ymax=45,
ytick={5,15,...,45},
yticklabels={,,},
axis x line*=none,
axis y line*=none,
legend style={at={(0.98,0.95)}, anchor=north east,legend cell align=left, align=left, draw=white!15!black, legend columns=1}
]
\addplot [color=mycolor4]
table[row sep=crcr, x expr=\thisrowno{0}-4]{%
	-2	0.264631301546647\\
	-1.9	0.269659371550285\\
	-1.8	0.274975853868789\\
	-1.7	0.280668469803016\\
	-1.6	0.286782220774672\\
	-1.5	0.293353985427198\\
	-1.4	0.300521499467427\\
	-1.3	0.308318319591512\\
	-1.2	0.316946040683498\\
	-1.1	0.326564086676763\\
	-1	0.337324986127828\\
	-0.9	0.349564809812954\\
	-0.8	0.363725422372381\\
	-0.7	0.380446703437305\\
	-0.6	0.400515597497228\\
	-0.5	0.425497650412468\\
	-0.4	0.457324491869277\\
	-0.35	0.477088529418094\\
	-0.3	0.500535628454631\\
	-0.25	0.529116084531815\\
	-0.2	0.564966371659975\\
	-0.15	0.610297098589303\\
	-0.1	0.673239167315708\\
	-0.05	0.766069859299002\\
	0	0.889839296052859\\
	0.05	0.898988645833913\\
	0.1	0.827853402376215\\
	0.15	0.780836607307993\\
	0.2	0.739215707334586\\
	0.25	0.702442962030588\\
	0.3	0.670092094127425\\
	0.35	0.642374911047888\\
	0.4	0.618437373266701\\
	0.5	0.574334754519967\\
	0.6	0.53908932150136\\
	0.7	0.510841482999145\\
	0.8	0.487397719443746\\
	0.9	0.467299359306883\\
	1	0.449298987413889\\
	1.1	0.432931322049611\\
	1.2	0.417743685728306\\
	1.3	0.403380319863081\\
	1.4	0.389221006208871\\
	1.5	0.374995089148453\\
	1.6	0.362447405569578\\
	1.7	0.351350629806447\\
	1.8	0.341339327073552\\
	1.9	0.332280043809296\\
	2	0.324003387922139\\
};
\addlegendentry{$V_{\rm DS} = -0.1$V};

\addplot [color=mycolor1, dashed, forget plot]
table[row sep=crcr, y expr=\thisrowno{1}*10]{%
	-2	0.269148462048739\\
	-1.9	0.276453171759189\\
	-1.8	0.284362444281665\\
	-1.7	0.293046728284857\\
	-1.6	0.302643487693616\\
	-1.5	0.313405640056941\\
	-1.4	0.325582518845253\\
	-1.3	0.33963791930347\\
	-1.2	0.356114756035564\\
	-1.1	0.375927476057401\\
	-1	0.400339132198123\\
	-0.9	0.430802323503964\\
	-0.8	0.462772055440106\\
	-0.7	0.496859386866287\\
	-0.6	0.536344891700252\\
	-0.5	0.584381102689612\\
	-0.4	0.646251227783275\\
	-0.35	0.685042945996133\\
	-0.3	0.713450116118953\\
	-0.25	0.740622555254752\\
	-0.2	0.765924389669508\\
	-0.15	0.790781162709189\\
	-0.1	0.865410489727601\\
	-0.05	0.871445520640642\\
	0	0.712313427395384\\
	0.05	0.61242940518402\\
	0.1	0.542239815646571\\
	0.15	0.493253453774541\\
	0.2	0.457027231694274\\
	0.25	0.428898072648419\\
	0.3	0.40581647285681\\
	0.35	0.386746326943219\\
	0.4	0.370629753011579\\
	0.5	0.347598521111763\\
	0.6	0.331573387693797\\
	0.7	0.318143985015649\\
	0.8	0.306678447117958\\
	0.9	0.296623709483111\\
	1	0.288042961030982\\
	1.1	0.280483265664325\\
	1.2	0.273712968987453\\
	1.3	0.267612392782025\\
	1.4	0.262061775998612\\
	1.5	0.256977356896532\\
	1.6	0.252277635588471\\
	1.7	0.247926965611047\\
	1.8	0.243854299511814\\
	1.9	0.240076058231406\\
	2	0.23651310282895\\
};

\addplot [color=mycolor2, forget plot, dashed]
table[row sep=crcr, y expr=\thisrowno{1}*10]{%
	-2	0.299651884286031\\
	-1.9	0.305035749384358\\
	-1.8	0.310806443191298\\
	-1.7	0.317086406305394\\
	-1.6	0.323857322454865\\
	-1.5	0.331307543159427\\
	-1.4	0.339604286469689\\
	-1.3	0.348820648617048\\
	-1.2	0.359272577791525\\
	-1.1	0.371260161921873\\
	-1	0.385384912387139\\
	-0.9	0.402175145073946\\
	-0.8	0.420887107820228\\
	-0.7	0.442174679944617\\
	-0.6	0.467775267705206\\
	-0.5	0.500043831010188\\
	-0.4	0.543132763592035\\
	-0.35	0.570636242556898\\
	-0.3	0.605834850219374\\
	-0.25	0.655025727230691\\
	-0.2	0.727148827374046\\
	-0.15	0.785884302135077\\
	-0.1	0.659142353417445\\
	-0.05	0.589388579528511\\
	0	0.557510633500527\\
	0.05	0.540254451457934\\
	0.1	0.510344853878311\\
	0.15	0.48268324101087\\
	0.2	0.458702531709627\\
	0.25	0.438025005112536\\
	0.3	0.419341656167008\\
	0.35	0.403026215794602\\
	0.4	0.388740455724633\\
	0.5	0.361208567799275\\
	0.6	0.336424527185039\\
	0.7	0.316010727744886\\
	0.8	0.298443400699583\\
	0.9	0.282605639418089\\
	1	0.26932666475037\\
	1.1	0.25814802807648\\
	1.2	0.248546708369969\\
	1.3	0.24016190458535\\
	1.4	0.232753692620656\\
	1.5	0.226114691569516\\
	1.6	0.220145426935224\\
	1.7	0.214717616265264\\
	1.8	0.209771442738306\\
	1.9	0.205204481187456\\
	2	0.200989180958603\\
};

\addplot [color=mycolor3, forget plot, dashed]
table[row sep=crcr, y expr=\thisrowno{1}*10]{%
	-2	0.00949226181683884\\
	-1.9	0.0101623663243556\\
	-1.8	0.0109268431499362\\
	-1.7	0.0118057161771367\\
	-1.6	0.0128261177029919\\
	-1.5	0.01402505459021\\
	-1.4	0.0154476627334456\\
	-1.3	0.0171661631647075\\
	-1.2	0.0192713054196736\\
	-1.1	0.0219110173993954\\
	-1	0.02530388942418\\
	-0.9	0.0298046830471821\\
	-0.8	0.0359806880474954\\
	-0.7	0.0448661645857458\\
	-0.6	0.0585642912639579\\
	-0.5	0.0817369035731621\\
	-0.4	0.127238588544708\\
	-0.35	0.169338570452464\\
	-0.3	0.238833521630325\\
	-0.25	0.36698746722075\\
	-0.2	0.636138148502849\\
	-0.15	1.27315315335013\\
	-0.1	1.91734414457782\\
	-0.05	1.09090818087776\\
	0	0.56047328546077\\
	0.05	0.334971545305852\\
	0.1	0.222753434270953\\
	0.15	0.160508884607764\\
	0.2	0.122414116493865\\
	0.25	0.0973867356829418\\
	0.3	0.0799980258645271\\
	0.35	0.0673431683937614\\
	0.4	0.0577930692766045\\
	0.5	0.0444269598540377\\
	0.6	0.0356796662625375\\
	0.7	0.0296117524296675\\
	0.8	0.0251844751602723\\
	0.9	0.0218250762341321\\
	1	0.0192062982968519\\
	1.1	0.0171154300860388\\
	1.2	0.0154077769078058\\
	1.3	0.0139928578707212\\
	1.4	0.0128001035335463\\
	1.5	0.0117843547286078\\
	1.6	0.0109091608472962\\
	1.7	0.010147788961356\\
	1.8	0.00948036078135352\\
	1.9	0.00889086404849787\\
	2	0.00836636980259694\\
};

\addplot [color=mycolor4, dashed]
table[row sep=crcr, y expr=\thisrowno{1}*10]{%
	-2	0.578292608151609\\
	-1.9	0.591651287467903\\
	-1.8	0.606095730622899\\
	-1.7	0.621938850767387\\
	-1.6	0.639326927851473\\
	-1.5	0.658738237806578\\
	-1.4	0.680634468048388\\
	-1.3	0.705624731085226\\
	-1.2	0.734658639246762\\
	-1.1	0.76909865537867\\
	-1	0.811027934009441\\
	-0.9	0.862782151625092\\
	-0.8	0.919639851307829\\
	-0.7	0.983900231396649\\
	-0.6	1.06268445066942\\
	-0.5	1.16616183727296\\
	-0.4	1.31662257992002\\
	-0.35	1.4250177590055\\
	-0.3	1.55811848796865\\
	-0.25	1.76263574970619\\
	-0.2	2.1292113655464\\
	-0.15	2.8498186181944\\
	-0.1	3.44189698772286\\
	-0.05	2.55174228104691\\
	0	1.83029734635668\\
	0.05	1.48765540194781\\
	0.1	1.27533810379584\\
	0.15	1.13644557939318\\
	0.2	1.03814387989777\\
	0.25	0.964309813443897\\
	0.3	0.905156154888345\\
	0.35	0.857115711131582\\
	0.4	0.817163278012817\\
	0.5	0.753234048765075\\
	0.6	0.703677581141374\\
	0.7	0.663766465190203\\
	0.8	0.630306322977813\\
	0.9	0.601054425135332\\
	1	0.576575924078205\\
	1.1	0.555746723826844\\
	1.2	0.537667454265228\\
	1.3	0.521767155238096\\
	1.4	0.507615572152814\\
	1.5	0.494876403194656\\
	1.6	0.483332223370991\\
	1.7	0.472792370837666\\
	1.8	0.463106103031473\\
	1.9	0.454171403467359\\
	2	0.44586865359015\\
};
\addlegendentry{$V_{\rm DS} = -0.2$V};


\end{axis}

\end{tikzpicture}%

%% file: Figuras/RIds_Dev.tex
\begin{tikzpicture}
	{\begin{axis}[%
	width=0.4\textwidth,
	height=3.5cm,
	at={(0, 0)},
	scale only axis,
	xmin=-2,
	xmax=2,
	xticklabels={,,},
	xlabel style={font=\color{white!15!black}},
	ymin=0,
	ymax=0.8,
	ylabel style={font=\color{white!15!black}},
	ylabel={$\text{I}_{\text{DS}}\text{ (}\text{mA/}\mu\text{m)}$},
	title style={font=\bfseries},
	xmajorgrids,
	ymajorgrids,
	legend style={legend cell align=left, align=left, draw=white!15!black, anchor=south, at={(0.5, 1.05)}, legend columns=2},
	]
		\addplot [color=mycolor1]
		table[row sep=crcr, y expr=\thisrowno{1}*1e-4]{%
			-2	2999.27919370532\\
			-1.9	2928.73760803504\\
			-1.8	2859.35733399715\\
			-1.7	2787.84099292561\\
			-1.6	2713.55130332533\\
			-1.5	2637.14642652702\\
			-1.4	2557.23018919976\\
			-1.3	2474.39123824281\\
			-1.2	2387.52844553635\\
			-1.1	2295.72937174279\\
			-1	2199.18145868845\\
			-0.9	2096.38972101205\\
			-0.8	1986.13765738214\\
			-0.7	1866.07237355847\\
			-0.6	1733.92728802142\\
			-0.5	1593.18263850208\\
			-0.4	1433.92108224579\\
			-0.3	1250.04560167251\\
			-0.2	1019.68786621041\\
			-0.0999999999999999	726.41686938867\\
			0	372.003305702433\\
			0.0999999999999999	442.073164486649\\
			0.2	900.964136684944\\
			0.3	1247.5902751295\\
			0.4	1509.0945395483\\
			0.5	1720.63338854377\\
			0.6	1896.00835798557\\
			0.7	2049.32515700435\\
			0.8	2183.26071147061\\
			0.9	2303.11339347327\\
			1	2412.02831250852\\
			1.1	2513.28059145589\\
			1.2	2606.51392602625\\
			1.3	2693.16219565584\\
			1.4	2774.62274006444\\
			1.5	2852.00129376099\\
			1.6	2924.82943247781\\
			1.7	2994.54817912615\\
			1.8	3062.00427190985\\
			1.9	3126.41915204138\\
			2	3185.86152977507\\
		};
		\addlegendentry{$L_{\rm S} = 17.5$nm};
		\addplot [color=mycolor2, forget plot]
		table[row sep=crcr, y expr=\thisrowno{1}*1e-4]{%
			-2	7210.16859649048\\
			-1.9	7018.60381039027\\
			-1.8	6822.81371557991\\
			-1.7	6618.87398156174\\
			-1.6	6409.50266251784\\
			-1.5	6197.42764613484\\
			-1.4	5986.03074079556\\
			-1.3	5776.80269836902\\
			-1.2	5558.49423284002\\
			-1.1	5330.15143512535\\
			-1	5091.27495230887\\
			-0.9	4839.1887414029\\
			-0.8	4572.41428147106\\
			-0.7	4287.89844554001\\
			-0.6	3980.64954584818\\
			-0.5	3645.55988775659\\
			-0.4	3271.67943246596\\
			-0.3	2856.08949838106\\
			-0.2	2370.67464616421\\
			-0.0999999999999999	1807.06896368927\\
			0	1179.48221113623\\
			0.0999999999999999	735.43671309097\\
			0.2	1668.04290295091\\
			0.3	2557.92413787002\\
			0.4	3235.33172183455\\
			0.5	3763.11574452699\\
			0.6	4192.29376708125\\
			0.7	4553.47026716994\\
			0.8	4865.19469461774\\
			0.9	5139.6105709604\\
			1	5389.23008368632\\
			1.1	5613.60002235437\\
			1.2	5819.76527262481\\
			1.3	6009.98785601101\\
			1.4	6185.19637469498\\
			1.5	6351.83369613143\\
			1.6	6510.05224305029\\
			1.7	6655.84076369007\\
			1.8	6792.18972181475\\
			1.9	6922.62034999618\\
			2	7046.51242444407\\
		};
		
		\addplot [color=mycolor1, dashed, line width=0.8pt]
		table[row sep=crcr, y expr=\thisrowno{1}*1e-4]{%
			-2	3675.75841177125\\
			-1.9	3608.48873841444\\
			-1.8	3539.38850729202\\
			-1.7	3467.00646531452\\
			-1.6	3391.58831636612\\
			-1.5	3312.62080272287\\
			-1.4	3229.67898003898\\
			-1.3	3142.99783869745\\
			-1.2	3051.425231415\\
			-1.1	2953.68051276823\\
			-1	2848.83378103408\\
			-0.9	2736.17811790869\\
			-0.8	2613.43706372861\\
			-0.7	2479.01759062919\\
			-0.6	2328.35431223331\\
			-0.5	2157.09738653234\\
			-0.4	1954.59756475677\\
			-0.3	1709.09432065287\\
			-0.2	1363.09660837512\\
			-0.0999999999999999	910.337657621132\\
			0	437.055531799942\\
			0.0999999999999999	360.374505999482\\
			0.2	690.616198778768\\
			0.3	936.464074889376\\
			0.4	1149.04613673703\\
			0.5	1324.85922192799\\
			0.6	1476.20383744543\\
			0.7	1612.46188056651\\
			0.8	1738.79353416558\\
			0.9	1859.31057303205\\
			1	1977.41661356942\\
			1.1	2099.61864095026\\
			1.2	2220.59361794108\\
			1.3	2334.83285463949\\
			1.4	2443.97713095857\\
			1.5	2549.63272572293\\
			1.6	2653.39565620178\\
			1.7	2753.97928001694\\
			1.8	2850.1697843511\\
			1.9	2942.48461884038\\
			2	3031.46315947969\\
		};
		\addlegendentry{$L_{\rm D} = 17.5$nm};
		\addplot [color=mycolor2, dashed, line width=0.8pt, forget plot]
		table[row sep=crcr, y expr=\thisrowno{1}*1e-4]{%
			-2	7229.34030119683\\
			-1.9	7124.33689145149\\
			-1.8	7015.00037126476\\
			-1.7	6901.79317204684\\
			-1.6	6782.65399361136\\
			-1.5	6657.64962476618\\
			-1.4	6525.21966426637\\
			-1.3	6385.56871803261\\
			-1.2	6236.58558468275\\
			-1.1	6078.52814214291\\
			-1	5908.31261892824\\
			-0.9	5723.10258231935\\
			-0.8	5519.84139845431\\
			-0.7	5293.82436961573\\
			-0.6	5038.95075544453\\
			-0.5	4749.0665254858\\
			-0.4	4409.13045295149\\
			-0.3	4000.03987400663\\
			-0.2	3486.37544253915\\
			-0.0999999999999999	2777.73659253881\\
			0	1687.54407281019\\
			0.0999999999999999	691.043597276022\\
			0.2	1011.17534303579\\
			0.3	1536.03920769708\\
			0.4	2031.43757098922\\
			0.5	2490.24158495807\\
			0.6	2912.59808339968\\
			0.7	3324.16031125484\\
			0.8	3745.64515304611\\
			0.9	4143.77110814323\\
			1	4519.08504324732\\
			1.1	4868.44624526847\\
			1.2	5194.56466708076\\
			1.3	5503.32484596856\\
			1.4	5798.05137683895\\
			1.5	6081.57558618897\\
			1.6	6360.14681309621\\
			1.7	6626.91086246253\\
			1.8	6880.13429835455\\
			1.9	7122.48759025529\\
			2	7353.96035332204\\
		};
		
		\SubfigLetter{a)}{0.1}{0.1};
	\end{axis}}

	\begin{axis}[%
	width=0.4\textwidth,
	height=3.5cm,
	at={(0\textwidth, 0)},
	scale only axis,
	xmin=-2,
	xmax=2,
	xticklabels={,,},
	ymin=0,
	ymax=0.25,
	yticklabels={,,},
	legend style={legend cell align=left, align=left, draw=white!15!black, anchor=north, at={(0.53, 0.98)}, legend columns=2},
	]
	\addplot [color=mycolor1]
	table[row sep=crcr, y expr=\thisrowno{1}*1e-4]{%
		-2	823.995811101145\\
	};
	\addlegendentry{\small $V_{\rm DS} = 0.1$V};
	\addplot [color=mycolor2]
	table[row sep=crcr, y expr=\thisrowno{1}*1e-4]{%
		-2	823.995811101145\\
	};
	\addlegendentry{\small $V_{\rm DS} = 0.2$V};
	\end{axis}

	\begin{axis}[%
	width=0.4\textwidth,
	height=3.5cm,
	at={(0.5\textwidth, 0)},
	scale only axis,
	xmin=-2,
	xmax=2,
	xticklabels={,,},
	ymin=0,
	ymax=0.25,
	yticklabels={,,},
	legend style={legend cell align=left, align=left, draw=white!15!black, anchor=south, at={(0.5, 1.05)}, legend columns=2},
	]
	\addplot [color=mycolor1]
	table[row sep=crcr, y expr=\thisrowno{1}*1e-4]{%
		-2	823.995811101145\\
	};
	\addlegendentry{$L_{\rm S} = 70$nm};
	\addplot [color=mycolor1, dashed, line width=0.8pt]
	table[row sep=crcr, y expr=\thisrowno{1}*1e-4]{%
		-2	823.995811101145\\
	};
	\addlegendentry{$L_{\rm D} = 70$nm};
	\end{axis}
	
	\begin{axis}[%
	width=0.4\textwidth,
	height=3.5cm,
	at={(0.5\textwidth, 0)},
	scale only axis,
	xmin=-2,
	xmax=2,
	xticklabels={,,},
	xlabel style={font=\color{white!15!black}},
	ymin=0,
	ymax=0.25,
	axis background/.style={fill=white},
	xmajorgrids,
	ymajorgrids,
	legend style={legend cell align=left, align=left, draw=white!15!black, anchor=north, at={(0.5, 0.98)}, legend columns=2},
	]
		\addplot [color=mycolor1]
		table[row sep=crcr, y expr=\thisrowno{1}*1e-4]{%
			-2	823.995811101145\\
			-1.9	816.808110763357\\
			-1.8	809.397025791444\\
			-1.7	801.703534588255\\
			-1.6	793.641681667374\\
			-1.5	785.074057889456\\
			-1.4	775.974552501817\\
			-1.3	766.188231886439\\
			-1.2	755.697814101601\\
			-1.1	744.287343584004\\
			-1	731.799205898566\\
			-0.9	718.002286318601\\
			-0.8	702.566878250108\\
			-0.7	684.999475037339\\
			-0.6	664.586182787591\\
			-0.5	640.139548465785\\
			-0.4	608.232259146606\\
			-0.3	565.348130993448\\
			-0.2	508.762578645561\\
			-0.0999999999999999	428.203747160138\\
			0	283.349402311991\\
			0.0999999999999999	236.74220097857\\
			0.2	364.399451324825\\
			0.3	424.762262756749\\
			0.4	458.518691448031\\
			0.5	481.72892246243\\
			0.6	499.353939471268\\
			0.7	513.667707473269\\
			0.8	525.777043722452\\
			0.9	536.282510097072\\
			1	545.623824239464\\
			1.1	554.098130009728\\
			1.2	561.878802713983\\
			1.3	569.065160116375\\
			1.4	575.681568038506\\
			1.5	581.823598221279\\
			1.6	587.594361161119\\
			1.7	592.953665849835\\
			1.8	598.054902053846\\
			1.9	602.822216135944\\
			2	607.410770064748\\
		};
		\addlegendentry{\small$V_{\rm DS} = 0.1$V};
		\addplot [color=mycolor2]
		table[row sep=crcr, y expr=\thisrowno{1}*1e-4]{%
			-2	1797.53971032283\\
			-1.9	1784.62046287745\\
			-1.8	1771.01498573322\\
			-1.7	1756.75434496141\\
			-1.6	1741.52728667568\\
			-1.5	1725.42996377211\\
			-1.4	1708.15851468361\\
			-1.3	1689.70170693646\\
			-1.2	1669.75398786709\\
			-1.1	1648.09775136771\\
			-1	1624.46490093668\\
			-0.9	1598.3792381802\\
			-0.8	1569.29748078555\\
			-0.7	1536.46302184291\\
			-0.6	1499.10317779805\\
			-0.5	1457.59371643866\\
			-0.4	1407.5446955697\\
			-0.3	1344.75665163084\\
			-0.2	1261.46960678621\\
			-0.0999999999999999	1136.09576585496\\
			0	882.773210012258\\
			0.0999999999999999	474.169256104705\\
			0.2	600.23041374352\\
			0.3	797.676218759615\\
			0.4	890.978184069387\\
			0.5	949.140854763318\\
			0.6	990.95784486956\\
			0.7	1023.39187446535\\
			0.8	1049.88906609843\\
			0.9	1072.18567657292\\
			1	1091.66070720412\\
			1.1	1108.88411586339\\
			1.2	1124.33820678574\\
			1.3	1138.38409214911\\
			1.4	1151.33835782365\\
			1.5	1163.36918913607\\
			1.6	1174.49064951697\\
			1.7	1185.05690320959\\
			1.8	1194.95967596615\\
			1.9	1204.35033048169\\
			2	1213.31040980626\\
		};
		\addlegendentry{\small$V_{\rm DS} = 0.2$V};
		\addplot [color=mycolor1, dashed, line width=0.8pt, forget plot]
		table[row sep=crcr, y expr=\thisrowno{1}*1e-4]{%
			-2	629.024044011885\\
			-1.9	624.674422632994\\
			-1.8	620.086971516362\\
			-1.7	615.309158052495\\
			-1.6	610.211974594234\\
			-1.5	604.876850927345\\
			-1.4	599.143966666227\\
			-1.3	593.029808465116\\
			-1.2	586.450602609883\\
			-1.1	579.2944910684\\
			-1	571.473005427259\\
			-0.9	562.828873007127\\
			-0.8	553.157162496101\\
			-0.7	542.155779403126\\
			-0.6	529.354923457894\\
			-0.5	514.273518597373\\
			-0.4	495.601571479708\\
			-0.3	470.798010783334\\
			-0.2	434.791921724662\\
			-0.0999999999999999	372.947518955711\\
			0	245.084720704198\\
			0.0999999999999999	279.009896709139\\
			0.2	399.188472959609\\
			0.3	456.148913911459\\
			0.4	492.263622045686\\
			0.5	518.766803248551\\
			0.6	539.777424071545\\
			0.7	557.261549731219\\
			0.8	572.285868064642\\
			0.9	585.553629447596\\
			1	597.782252482416\\
			1.1	608.87593992443\\
			1.2	619.09318926722\\
			1.3	628.535958127721\\
			1.4	637.372972786344\\
			1.5	645.734630583993\\
			1.6	653.770018325507\\
			1.7	661.655268949729\\
			1.8	669.634345395494\\
			1.9	677.534913428433\\
			2	685.007893631382\\
		};
		
		\addplot [color=mycolor2, dashed, line width=0.8pt, forget plot]
		table[row sep=crcr, y expr=\thisrowno{1}*1e-4]{%
			-2	1314.31938248176\\
			-1.9	1305.95193445767\\
			-1.8	1297.27094381301\\
			-1.7	1288.10031245397\\
			-1.6	1278.42222022144\\
			-1.5	1268.15855318592\\
			-1.4	1257.28996230534\\
			-1.3	1245.54092621729\\
			-1.2	1233.07481658385\\
			-1.1	1219.48037996781\\
			-1	1204.65270696347\\
			-0.9	1188.31536747476\\
			-0.8	1170.13399917775\\
			-0.7	1149.6173886365\\
			-0.6	1126.05962479571\\
			-0.5	1097.10338731785\\
			-0.4	1061.2536861088\\
			-0.3	1014.71030772337\\
			-0.2	949.575129928005\\
			-0.0999999999999999	846.642586874558\\
			0	655.903767146764\\
			0.0999999999999999	472.662547713084\\
			0.2	772.510039359262\\
			0.3	935.373324927408\\
			0.4	1030.84331525433\\
			0.5	1098.89826915395\\
			0.6	1152.19469589791\\
			0.7	1196.63939400005\\
			0.8	1235.27510363553\\
			0.9	1269.92283796111\\
			1	1302.89860266904\\
			1.1	1334.11444891282\\
			1.2	1364.55851705847\\
			1.3	1396.29132457309\\
			1.4	1425.73645069332\\
			1.5	1452.54033430282\\
			1.6	1477.2045721704\\
			1.7	1500.11301156745\\
			1.8	1521.23126167877\\
			1.9	1540.90369406396\\
			2	1559.23196191459\\
		};

		\SubfigLetter{b)}{0.1}{0.1};
	\end{axis}
	
	\begin{axis}[%
width=0.4\textwidth,
height=3.5cm,
at={(0\textwidth, -4cm)},
scale only axis,
xmin=-2,
xmax=2,
xlabel style={font=\color{white!15!black}},
xlabel = {$V_{\rm FG}$ (V)},
ymin=0,
ymax=0.2,
\RemoveTickExp,
ylabel style={font=\color{white!15!black}},
ylabel={R ($\Omega \cdot$cm)},
axis background/.style={fill=white},
xmajorgrids,
ymajorgrids,
legend style={legend cell align=left, align=left, draw=white!15!black, anchor=north east, at={(0.98, 0.98)}, legend columns=3},
]
		\addplot [color=mycolor4]
		table[row sep=crcr, y expr=\thisrowno{1}*10]{%
			-2	0.00300391870138985\\
			-1.9	0.00306894395702591\\
			-1.8	0.00313477724003335\\
			-1.7	0.00320550720363579\\
			-1.6	0.00328230174413527\\
			-1.5	0.0033647376312692\\
			-1.4	0.00345523397671177\\
			-1.3	0.00355360140840821\\
			-1.2	0.00366220884617118\\
			-1.1	0.00378381379157662\\
			-1	0.00391957245244943\\
			-0.9	0.00407378855754606\\
			-0.8	0.00425087335207881\\
			-0.7	0.00445878216066545\\
			-0.6	0.00470874416024247\\
			-0.5	0.00498907362527564\\
			-0.4	0.00532416103588789\\
			-0.3	0.00572666271950323\\
			-0.2	0.0062472841685855\\
			-0.0999999999999999	0.00687909411956733\\
			0	0.00747782286468843\\
			0.0999999999999999	0.00823852332866973\\
			0.2	0.00638587864639944\\
			0.3	0.00540894789551025\\
			0.4	0.00484173560446671\\
			0.5	0.00445585204794427\\
			0.6	0.00417601056901629\\
			0.7	0.0039566670303247\\
			0.8	0.00378091883264373\\
			0.9	0.00363475266091361\\
			1	0.0035104833361124\\
			1.1	0.00340152141725264\\
			1.2	0.00330647767675759\\
			1.3	0.00322223279322905\\
			1.4	0.00314634385208971\\
			1.5	0.00307700854904257\\
			1.6	0.00301439173697837\\
			1.7	0.0029565634417007\\
			1.8	0.0029023581770862\\
			1.9	0.00285222692470331\\
			2	0.00280796373896373\\
		};
		\addlegendentry{\small$R_{\rm D, Acc}$};
		\addplot [color=mycolor5]
		table[row sep=crcr, y expr=\thisrowno{1}*10]{%
			-2	0.000243287727758909\\
			-1.9	0.000252823613273765\\
			-1.8	0.000263344425010699\\
			-1.7	0.000274930190464866\\
			-1.6	0.000287836978669793\\
			-1.5	0.00030233009328615\\
			-1.4	0.000318806044094275\\
			-1.3	0.000337700619359066\\
			-1.2	0.000359722996358972\\
			-1.1	0.000385586745567197\\
			-1	0.000416144707254947\\
			-0.9	0.000453229190227156\\
			-0.8	0.00049950808234173\\
			-0.7	0.000559437897709595\\
			-0.6	0.000638427566587804\\
			-0.5	0.000748467367851791\\
			-0.4	0.000917122086639704\\
			-0.3	0.0011881014066925\\
			-0.2	0.00171238238899618\\
			-0.0999999999999999	0.00282165673046256\\
			0	0.00456593182434564\\
			0.0999999999999999	0.00328228834418498\\
			0.2	0.00188276472088517\\
			0.3	0.00124800066392751\\
			0.4	0.000941779985632373\\
			0.5	0.000761782701416846\\
			0.6	0.000646765963059172\\
			0.7	0.00056276782044582\\
			0.8	0.000501764295630356\\
			0.9	0.000454905625049215\\
			1	0.000417275002625789\\
			1.1	0.000385752300603021\\
			1.2	0.000359630417357811\\
			1.3	0.000337668540075167\\
			1.4	0.000318797605655196\\
			1.5	0.000302335083233159\\
			1.6	0.000287841042882249\\
			1.7	0.000274836470824993\\
			1.8	0.000263095509542851\\
			1.9	0.000252611483410573\\
			2	0.000243081474501913\\
		};
		\addlegendentry{\small$R_{\rm S, Acc}$};
		\addplot [color=mycolor3]
		table[row sep=crcr, y expr=\thisrowno{1}*10]{%
			-2	8.81228039848383e-05\\
			-1.9	9.39484971618303e-05\\
			-1.8	0.000100534584341333\\
			-1.7	0.000108039211187831\\
			-1.6	0.000116659821236351\\
			-1.5	0.000126638949268824\\
			-1.4	0.000138332448219146\\
			-1.3	0.000152179119578431\\
			-1.2	0.000168813132351191\\
			-1.1	0.000189106671876353\\
			-1	0.000214371955186603\\
			-0.9	0.000246472597471334\\
			-0.8	0.000288480323432598\\
			-0.7	0.000345373364476146\\
			-0.6	0.000425936478399849\\
			-0.5	0.000546714440198273\\
			-0.4	0.000742787243775304\\
			-0.3	0.00109997980006706\\
			-0.2	0.00187273963367647\\
			-0.0999999999999999	0.00412084094109615\\
			0	0.0150176213252287\\
			0.0999999999999999	0.011241448110546\\
			0.2	0.00286812683522927\\
			0.3	0.00137684658299594\\
			0.4	0.000854505455798221\\
			0.5	0.00060237424991923\\
			0.6	0.000457725684790385\\
			0.7	0.000365242199710393\\
			0.8	0.00030178591041641\\
			0.9	0.000255830759570309\\
			1	0.000221198425446516\\
			1.1	0.000194290599979105\\
			1.2	0.000172840193829881\\
			1.3	0.000155372642961465\\
			1.4	0.000140919902576752\\
			1.5	0.000128766432062237\\
			1.6	0.000118427599048628\\
			1.7	0.000109536826808362\\
			1.8	0.000101808829014105\\
			1.9	9.5043177835097e-05\\
			2	8.90741652198753e-05\\
		};
		\addlegendentry{\small$R_{\rm Chn}$};
		\addplot [color=mycolor4, dashed, line width=0.8pt]
		table[row sep=crcr, y expr=\thisrowno{1}*10]{%
			-2	0.000223118819718495\\
			-1.9	0.000230995848390903\\
			-1.8	0.000239603956995401\\
			-1.7	0.000249123143588656\\
			-1.6	0.000259659400939339\\
			-1.5	0.000271454624316925\\
			-1.4	0.000284713464870655\\
			-1.3	0.00029952780235559\\
			-1.2	0.000316404580679349\\
			-1.1	0.000335978597491568\\
			-1	0.000358983779220353\\
			-0.9	0.000386493685929269\\
			-0.8	0.000420164171787826\\
			-0.7	0.000461505832759725\\
			-0.6	0.000514954335654446\\
			-0.5	0.000587805118892175\\
			-0.4	0.000696306971717636\\
			-0.3	0.000873002027781692\\
			-0.2	0.00125705584408434\\
			-0.0999999999999999	0.00201151409880755\\
			0	0.00313133755394353\\
			0.0999999999999999	0.00440002188327277\\
			0.2	0.00274654382395224\\
			0.3	0.00176638917987158\\
			0.4	0.00123793150679105\\
			0.5	0.000947263662242756\\
			0.6	0.000768292534652047\\
			0.7	0.000650579791998652\\
			0.8	0.000567346781580839\\
			0.9	0.000504257365485838\\
			1	0.00045580301868088\\
			1.1	0.000417283228137496\\
			1.2	0.000385766371362003\\
			1.3	0.000359236195541471\\
			1.4	0.000336711791324147\\
			1.5	0.000317480969254232\\
			1.6	0.000300775490410893\\
			1.7	0.000286122144848592\\
			1.8	0.000273125625891015\\
			1.9	0.000261490275504814\\
			2	0.000250939562307605\\
		};
		
		\addplot [color=mycolor5, dashed, line width=0.8pt]
		table[row sep=crcr, y expr=\thisrowno{1}*10]{%
			-2	0.00241479237558211\\
			-1.9	0.00245239727991393\\
			-1.8	0.00249200108261404\\
			-1.7	0.00253478990240466\\
			-1.6	0.00258076335831573\\
			-1.5	0.00263047860307908\\
			-1.4	0.00268453694047107\\
			-1.3	0.00274311400304123\\
			-1.2	0.00280744466657309\\
			-1.1	0.00287906623110287\\
			-1	0.00295947042876342\\
			-0.9	0.0030498962665915\\
			-0.8	0.00315363287286967\\
			-0.7	0.00327431038217899\\
			-0.6	0.00341908087850167\\
			-0.5	0.00359560922586775\\
			-0.4	0.00382371525735769\\
			-0.3	0.0041299823542998\\
			-0.2	0.00469575421315435\\
			-0.0999999999999999	0.00606070584185141\\
			0	0.00855604749717939\\
			0.0999999999999999	0.0084268932062804\\
			0.2	0.00762215313301572\\
			0.3	0.00705008692245616\\
			0.4	0.00637553401287069\\
			0.5	0.00586602022637759\\
			0.6	0.00546552129901156\\
			0.7	0.00513036729861254\\
			0.8	0.00484269745630708\\
			0.9	0.0045892791566382\\
			1	0.00435803400484622\\
			1.1	0.00413396172914867\\
			1.2	0.00393098109284775\\
			1.3	0.00375721375593054\\
			1.4	0.00360488826409246\\
			1.5	0.0034682260976243\\
			1.6	0.00334309303824068\\
			1.7	0.00322994515152586\\
			1.8	0.00312889798637561\\
			1.9	0.00303785992914854\\
			2	0.00295515348628358\\
		};
		
		\addplot [color=mycolor3, dashed, line width=0.8pt]
		table[row sep=crcr, y expr=\thisrowno{1}*10]{%
			-2	8.37999179800732e-05\\
			-1.9	8.91080213253368e-05\\
			-1.8	9.50822243242137e-05\\
			-1.7	0.000101854290489179\\
			-1.6	0.000109589892440998\\
			-1.5	0.000118489756638212\\
			-1.4	0.000128840884579029\\
			-1.3	0.000141009864471935\\
			-1.2	0.000155483216271338\\
			-1.1	0.000172977191717376\\
			-1	0.000194464304905173\\
			-0.9	0.000221421517729741\\
			-0.8	0.000256133193171529\\
			-0.7	0.000302215516114869\\
			-0.6	0.000365877465469486\\
			-0.5	0.000458721186044505\\
			-0.4	0.000604331006248781\\
			-0.3	0.000859631484831624\\
			-0.2	0.00140214531292477\\
			-0.0999999999999999	0.00295165780370446\\
			0	0.0113354406297338\\
			0.0999999999999999	0.0151000768966083\\
			0.2	0.00416656526905469\\
			0.3	0.00188755613951394\\
			0.4	0.00110444740883763\\
			0.5	0.000744866123828414\\
			0.6	0.000547820898863949\\
			0.7	0.00042659660392148\\
			0.8	0.000345809267543192\\
			0.9	0.000288758437813309\\
			1	0.000246638482973585\\
			1.1	0.000214446154172008\\
			1.2	0.000189129110667845\\
			1.3	0.000168811311818575\\
			1.4	0.000152163577291173\\
			1.5	0.000138308732239879\\
			1.6	0.000126610182890719\\
			1.7	0.000116628136751162\\
			1.8	0.00010800699314624\\
			1.9	0.000100502819932602\\
			2	9.39178629192828e-05\\
		};
		
		\SubfigLetter{c)}{0.1}{0.35};,
		\node[fill=white, anchor=south west] at (rel axis cs: 0.63, 0.35) {\small $V_{DS} = 0.1$V};
	\end{axis}
	
	\begin{axis}[%
width=0.4\textwidth,
height=3.5cm,
at={(0.5\textwidth, -4cm)},
scale only axis,
xmin=-2,
xmax=2,
xlabel style={font=\color{white!15!black}},
xlabel = {$V_{\rm FG}$ (V)},
ymin=0,
ymax=0.25,
\RemoveTickExp,
xmajorgrids,
ymajorgrids,
]
\addplot [color=mycolor4]
table[row sep=crcr, y expr=\thisrowno{1}*10]{%
	-2	0.00241053556662731\\
	-1.9	0.00246216533060665\\
	-1.8	0.00251584678074208\\
	-1.7	0.0025721028679212\\
	-1.6	0.00263241593934607\\
	-1.5	0.00269758603347978\\
	-1.4	0.00276884197939857\\
	-1.3	0.0028471742162497\\
	-1.2	0.00293399942009943\\
	-1.1	0.00303145148067459\\
	-1	0.00314189701555844\\
	-0.9	0.00326852525154781\\
	-0.8	0.00341685333930725\\
	-0.7	0.00359392331969362\\
	-0.6	0.0038108005294462\\
	-0.5	0.00408644697914102\\
	-0.4	0.00442922100873105\\
	-0.3	0.00485213030923027\\
	-0.2	0.00545052689360071\\
	-0.0999999999999999	0.00626394542122649\\
	0	0.00733823439284818\\
	0.0999999999999999	0.00841249884928476\\
	0.2	0.00713651267913803\\
	0.3	0.00633552776426894\\
	0.4	0.00582753780405176\\
	0.5	0.0054598196344325\\
	0.6	0.00517501327654171\\
	0.7	0.00494297298866681\\
	0.8	0.00474656075029396\\
	0.9	0.00457675060085906\\
	1	0.00442581811695672\\
	1.1	0.00428918576182231\\
	1.2	0.00416420421881484\\
	1.3	0.0040494221424544\\
	1.4	0.00394482809225647\\
	1.5	0.00384945445731741\\
	1.6	0.00376142480601291\\
	1.7	0.00368012412121659\\
	1.8	0.00360457836172467\\
	1.9	0.0035340859014982\\
	2	0.00346763139730422\\
};

\addplot [color=mycolor5]
table[row sep=crcr, y expr=\thisrowno{1}*10]{%
	-2	0.00964177243846093\\
	-1.9	0.00969160055505208\\
	-1.8	0.00974400671934093\\
	-1.7	0.0097994944302815\\
	-1.6	0.0098580951281576\\
	-1.5	0.00992145623026901\\
	-1.4	0.00998911543584736\\
	-1.3	0.0100630916608981\\
	-1.2	0.0101427892184444\\
	-1.1	0.0102304783299251\\
	-1	0.0103274737937125\\
	-0.9	0.0104359984663685\\
	-0.8	0.0105582541655636\\
	-0.7	0.0106990465178481\\
	-0.6	0.0108651337398356\\
	-0.5	0.0110683144117194\\
	-0.4	0.0113941352558027\\
	-0.3	0.0119539242073072\\
	-0.2	0.0127746874637315\\
	-0.0999999999999999	0.014116553188982\\
	0	0.0168200436217629\\
	0.0999999999999999	0.0174761255956974\\
	0.2	0.0161697035262479\\
	0.3	0.0154007196119009\\
	0.4	0.0149274072381659\\
	0.5	0.0145855698101748\\
	0.6	0.0143239982921584\\
	0.7	0.0141129147172049\\
	0.8	0.0139378234867237\\
	0.9	0.0137895414664558\\
	1	0.013661578797435\\
	1.1	0.0135488205293474\\
	1.2	0.0134482497353433\\
	1.3	0.0133579165798882\\
	1.4	0.0132766565388909\\
	1.5	0.0132021083935229\\
	1.6	0.0131327056096777\\
	1.7	0.0130699068049094\\
	1.8	0.0130100041683235\\
	1.9	0.0129555846019896\\
	2	0.0129031575938414\\
};

\addplot [color=mycolor3]
table[row sep=crcr, y expr=\thisrowno{1}*10]{%
	-2	8.49384618602421e-05\\
	-1.9	9.03519277730526e-05\\
	-1.8	9.64503033028161e-05\\
	-1.7	0.000103368349953146\\
	-1.6	0.000111273544689025\\
	-1.5	0.000120379507267847\\
	-1.4	0.000130984355669459\\
	-1.3	0.000143456197893907\\
	-1.2	0.000158324143522941\\
	-1.1	0.000176306910823002\\
	-1	0.000198459019536188\\
	-0.9	0.000226279883523481\\
	-0.8	0.000262189927665807\\
	-0.7	0.000310026705817116\\
	-0.6	0.000376393775612113\\
	-0.5	0.000473547856433761\\
	-0.4	0.00062655733622157\\
	-0.3	0.000894652440595252\\
	-0.2	0.00145033910334297\\
	-0.0999999999999999	0.00301345639604634\\
	0	0.011271095711075\\
	0.0999999999999999	0.0165422485382035\\
	0.2	0.00419312870154024\\
	0.3	0.00183199026360015\\
	0.4	0.00106952254185987\\
	0.5	0.000723457119813262\\
	0.6	0.000534490883637492\\
	0.7	0.000417935060750348\\
	0.8	0.000339964321559821\\
	0.9	0.000284689409831904\\
	1	0.000243758119511206\\
	1.1	0.000212400038096365\\
	1.2	0.000187684508402345\\
	1.3	0.000167761745856124\\
	1.4	0.000151414912195839\\
	1.5	0.000137768357918615\\
	1.6	0.000126236923819404\\
	1.7	0.000116375691418904\\
	1.8	0.000107849505062575\\
	1.9	0.000100419847975275\\
	2	9.38922801510366e-05\\
};

\addplot [color=mycolor4, dashed, line width=0.8pt]
table[row sep=crcr, y expr=\thisrowno{1}*10]{%
	-2	0.0127873551983812\\
	-1.9	0.0128362633930547\\
	-1.8	0.0128885365838554\\
	-1.7	0.0129435101518639\\
	-1.6	0.013003054139766\\
	-1.5	0.0130661079914324\\
	-1.4	0.0131349070003557\\
	-1.3	0.0132093469954872\\
	-1.2	0.0132903469149143\\
	-1.1	0.0133796333833761\\
	-1	0.0134784551148379\\
	-0.9	0.0135889448800731\\
	-0.8	0.0137137516606336\\
	-0.7	0.0138569937966269\\
	-0.6	0.0140247606396989\\
	-0.5	0.0142185666139183\\
	-0.4	0.0144525130711192\\
	-0.3	0.0147501232874026\\
	-0.2	0.0151354830669523\\
	-0.0999999999999999	0.0155946601106239\\
	0	0.0158440513117831\\
	0.0999999999999999	0.0163261445206655\\
	0.2	0.0149035569708036\\
	0.3	0.014064157746736\\
	0.4	0.0135200801328627\\
	0.5	0.01312887959104\\
	0.6	0.0128274920948355\\
	0.7	0.0125844785751112\\
	0.8	0.0123815532032215\\
	0.9	0.0122079452444263\\
	1	0.0120586449320283\\
	1.1	0.0119264185743286\\
	1.2	0.0118077220338053\\
	1.3	0.011700523712747\\
	1.4	0.0116026284822808\\
	1.5	0.0115119650737662\\
	1.6	0.0114274754396811\\
	1.7	0.0113473725101981\\
	1.8	0.0112704547620334\\
	1.9	0.0111964593208549\\
	2	0.0111273314233476\\
};

\addplot [color=mycolor5, dashed, line width=0.8pt]
table[row sep=crcr, y expr=\thisrowno{1}*10]{%
	-2	0.00302393941951767\\
	-1.9	0.00308005579573351\\
	-1.8	0.0031398108156065\\
	-1.7	0.0032027699924408\\
	-1.6	0.00327061382071818\\
	-1.5	0.00334242357709344\\
	-1.4	0.00342049159885401\\
	-1.3	0.00350472832591255\\
	-1.2	0.00359684255619308\\
	-1.1	0.00369861619897123\\
	-1	0.00381174973499807\\
	-0.9	0.00393916272914172\\
	-0.8	0.00408472905090507\\
	-0.7	0.00425390214087364\\
	-0.6	0.00445535434876882\\
	-0.5	0.00470065171483081\\
	-0.4	0.00501295427429072\\
	-0.3	0.00543703953597779\\
	-0.2	0.00605985075573392\\
	-0.0999999999999999	0.00710407720648687\\
	0	0.00900635293987076\\
	0.0999999999999999	0.00827625563225984\\
	0.2	0.00717785519449136\\
	0.3	0.00643023849902273\\
	0.4	0.00591274694388468\\
	0.5	0.00553042325451699\\
	0.6	0.00523213326899622\\
	0.7	0.00498962981164791\\
	0.8	0.00478691753450616\\
	0.9	0.00461180788325318\\
	1	0.00444720309669433\\
	1.1	0.00430209477400965\\
	1.2	0.0041715988731422\\
	1.3	0.00405385833281713\\
	1.4	0.00394582309029476\\
	1.5	0.00384561048583942\\
	1.6	0.00375020730428403\\
	1.7	0.00365699072040947\\
	1.8	0.00356160251173271\\
	1.9	0.00346827656184579\\
	2	0.00338239083679492\\
};

\addplot [color=mycolor3, dashed, line width=0.8pt]
table[row sep=crcr, y expr=\thisrowno{1}*10]{%
	-2	8.75541295953498e-05\\
	-1.9	9.33052732470246e-05\\
	-1.8	9.97991642251066e-05\\
	-1.7	0.000107192095856353\\
	-1.6	0.000115677440692422\\
	-1.5	0.000125493636269132\\
	-1.4	0.000136975451974384\\
	-1.3	0.000150566811814325\\
	-1.2	0.000166855014393567\\
	-1.1	0.000186713091650287\\
	-1	0.000211362263128267\\
	-0.9	0.00024264271027063\\
	-0.8	0.000283489681169491\\
	-0.7	0.000338672390760928\\
	-0.6	0.000416559626750869\\
	-0.5	0.000533040127179136\\
	-0.4	0.000721963989083882\\
	-0.3	0.00106798683204822\\
	-0.2	0.00182904815340171\\
	-0.0999999999999999	0.00416987088227289\\
	0	0.0161395394091115\\
	0.0999999999999999	0.0113761539349259\\
	0.2	0.00300906381486954\\
	0.3	0.00144777521770651\\
	0.4	0.000893640928630056\\
	0.5	0.000625742231984451\\
	0.6	0.000473026076211068\\
	0.7	0.000375963693029187\\
	0.8	0.000309583891077491\\
	0.9	0.000261717762656166\\
	1	0.000225771890093682\\
	1.1	0.000197931340203465\\
	1.2	0.000175767610659794\\
	1.3	0.000157789577875072\\
	1.4	0.000142925848526602\\
	1.5	0.000130465973986792\\
	1.6	0.000119870328234887\\
	1.7	0.000110773214586169\\
	1.8	0.000102875008197538\\
	1.9	9.59660215018833e-05\\
	2	8.9881005110844e-05\\
};

\node[fill=white, anchor=south west] at (rel axis cs: 0.63, 0.3) {\small $V_{DS} = 0.1$V};
\SubfigLetter{d)}{0.1}{0.3};
\end{axis}

	\begin{axis}[%
width=0.4\textwidth,
height=3.5cm,
at={(0.5\textwidth, -4cm)},
scale only axis,
xmin=-2,
xmax=2,
xticklabels={,,},
ymin=0,
ymax=0.25,
yticklabels={,,},
legend style={legend cell align=left, align=left, draw=white!15!black, anchor=north east, at={(0.98, 0.98)}, legend columns=3},
]
\addplot [color=mycolor4]
table[row sep=crcr, y expr=\thisrowno{1}*1e-4]{%
	-2	823.995811101145\\
};
		\addlegendentry{\small$R_{\rm D, Acc}$};
\addplot [color=mycolor5]
table[row sep=crcr, y expr=\thisrowno{1}*1e-4]{%
	-2	823.995811101145\\
};
		\addlegendentry{\small$R_{\rm S, Acc}$};
\addplot [color=mycolor3]
table[row sep=crcr, y expr=\thisrowno{1}*1e-4]{%
	-2	823.995811101145\\
};
	\addlegendentry{\small$R_{\rm Chn, Acc}$};
\end{axis}
\end{tikzpicture}

%% file: Figuras/GFET_Vbg-1V_Np1e12_Comp.tex
%
%
%

\begin{axis}[%
width=0.24\textwidth,
height=3cm,
at={(0\textwidth,0cm)},
scale only axis,
xmin=-2.01,
xmax=2,
xlabel style={font=\color{white!15!black}},
xlabel={$V_{\rm FG}$ (V)},
ymin=-4,
ymax=4,
ylabel style={font=\color{white!15!black}},
ylabel={$I_{\rm DS}$ (mA/$\mu$m)},
axis background/.style={fill=white},
axis x line*=bottom,
axis y line*=left,
xmajorgrids,
ymajorgrids,
clip mode=individual,
legend style={at={(0.05,-0.52)}, anchor=south west, legend cell align=left, align=left, draw=white!15!black, legend columns = 4, font=\footnotesize}
]
\addplot [color=mycolor1]
  table[row sep=crcr, y expr=1e-3*\thisrowno{1}*0.1]{%
-2.01	-15301.2524239672\\
-1.91	-15064.7140929436\\
-1.81	-14838.4711963865\\
-1.71	-14625.4804707619\\
-1.61	-14336.5014707586\\
-1.51	-14085.9932891591\\
-1.41	-13751.713604418\\
-1.31	-13451.9500739063\\
-1.21	-13079.3672886368\\
-1.11	-12718.4884667782\\
-1.01	-12285.5535338266\\
-0.91	-11791.6603165797\\
-0.81	-11218.8967009311\\
-0.71	-10543.2373372474\\
-0.61	-9749.53258512516\\
-0.51	-8804.04897356553\\
-0.41	-7575.43449248362\\
-0.36	-6817.78593317244\\
-0.31	-5941.64370688141\\
-0.26	-4934.78274405982\\
-0.21	-3808.84238910839\\
-0.16	-2638.12255445205\\
-0.11	-1619.39772336459\\
-0.06	-1129.04639009306\\
-0.01	-1306.01111247524\\
0.04	-1868.56611380372\\
0.09	-2488.07690476897\\
0.14	-3007.68201760706\\
0.19	-3404.44638826693\\
0.24	-3702.03069992161\\
0.29	-3926.81434741879\\
0.34	-4098.50834540438\\
0.39	-4240.59087104164\\
0.49	-4450.61924044881\\
0.59	-4597.76275943474\\
0.69	-4708.22838189087\\
0.79	-4795.29795200144\\
0.89	-4864.81598157777\\
0.99	-4914.1735873366\\
1.09	-4969.327909192\\
1.19	-5016.69770960923\\
1.29	-5047.34353836072\\
1.39	-5086.89570789161\\
1.49	-5110.99197337422\\
1.59	-5145.3802983874\\
1.69	-5164.98556219279\\
1.79	-5189.22041113385\\
1.89	-5211.91160845129\\
1.99	-5226.74259916433\\
};
\addlegendentry{$V_{\rm DS} = -0.2$V}

\addplot [color=mycolor2]
  table[row sep=crcr, y expr=1e-3*\thisrowno{1}*0.1]{%
-2.01	-8026.4723703299\\
-1.91	-7908.57379821277\\
-1.81	-7796.64962603187\\
-1.71	-7689.90492239222\\
-1.61	-7549.82869380603\\
-1.51	-7425.0340088292\\
-1.41	-7264.94919387524\\
-1.31	-7115.07871882897\\
-1.21	-6927.32625816332\\
-1.11	-6743.15759770127\\
-1.01	-6526.14489526572\\
-0.91	-6280.74836911618\\
-0.81	-5998.25047738769\\
-0.71	-5668.63763677052\\
-0.61	-5297.50045173754\\
-0.51	-4840.69009543812\\
-0.41	-4261.22288689707\\
-0.36	-3909.05458280567\\
-0.31	-3504.21532522195\\
-0.26	-3037.31271642384\\
-0.21	-2501.84178452563\\
-0.16	-1900.36720178213\\
-0.11	-1266.66446103058\\
-0.06	-705.499358100875\\
-0.01	-423.55430728761\\
0.04	-558.773763546488\\
0.09	-917.137369829803\\
0.14	-1276.87819620753\\
0.19	-1564.78460965658\\
0.24	-1779.30539121647\\
0.29	-1938.05117992709\\
0.34	-2058.48963110442\\
0.39	-2152.75266506993\\
0.49	-2285.68761085961\\
0.59	-2380.42735681627\\
0.69	-2448.12557068358\\
0.79	-2501.29572004504\\
0.89	-2545.6490027837\\
0.99	-2582.43292348463\\
1.09	-2610.29161118844\\
1.19	-2635.86829668066\\
1.29	-2661.63262102051\\
1.39	-2679.37470657594\\
1.49	-2700.60744536278\\
1.59	-2714.10088894625\\
1.69	-2731.83602247406\\
1.79	-2742.7235588184\\
1.89	-2755.47619587592\\
1.99	-2769.35372779678\\
};
\addlegendentry{$-0.1$V}

\addplot [color=mycolor3]
  table[row sep=crcr, y expr=1e-3*\thisrowno{1}*0.1]{%
-2.01	9017.28469721228\\
-1.91	8921.67571314467\\
-1.81	8812.73054485655\\
-1.71	8690.09156069372\\
-1.61	8570.65641086066\\
-1.51	8430.42930338283\\
-1.41	8288.94453120463\\
-1.31	8125.50290712922\\
-1.21	7954.85193516687\\
-1.11	7759.51576625231\\
-1.01	7547.32959734635\\
-0.91	7307.02985976782\\
-0.81	7033.32008755635\\
-0.71	6716.96507042756\\
-0.61	6345.5102138813\\
-0.51	5900.79293344459\\
-0.41	5355.93331849411\\
-0.36	5035.03430438838\\
-0.31	4674.12142760351\\
-0.26	4266.13637584967\\
-0.21	3802.86115155746\\
-0.16	3276.61020555682\\
-0.11	2682.00054213271\\
-0.06	2026.94287349195\\
-0.01	1347.95235935534\\
0.04	749.83941643014\\
0.09	433.309869558666\\
0.14	548.516599916512\\
0.19	912.562234365878\\
0.24	1296.46552334175\\
0.29	1612.42368036814\\
0.34	1855.619047455\\
0.39	2041.85233664599\\
0.49	2298.21663360886\\
0.59	2466.49088201811\\
0.69	2582.26889592637\\
0.79	2665.0787252831\\
0.89	2729.56889846675\\
0.99	2781.77683137704\\
1.09	2823.61440283141\\
1.19	2854.56195053315\\
1.29	2883.89541253036\\
1.39	2912.43964554827\\
1.49	2928.80500376032\\
1.59	2953.77495780048\\
1.69	2965.4857059629\\
1.79	2985.11014617239\\
1.89	2998.7593954275\\
1.99	3008.57199275243\\
};
\addlegendentry{$0.1$V}

\addplot [color=mycolor4]
  table[row sep=crcr, y expr=1e-3*\thisrowno{1}*0.1]{%
-2.01	18924.4010927324\\
-1.91	18751.1311962084\\
-1.81	18534.2803668922\\
-1.71	18275.5825931033\\
-1.61	18053.8122420165\\
-1.51	17761.1353224842\\
-1.41	17495.1180650845\\
-1.31	17158.3308222574\\
-1.21	16833.9118867424\\
-1.11	16435.8461074757\\
-1.01	16016.8721404259\\
-0.91	15546.8289489644\\
-0.81	15014.7094298429\\
-0.71	14404.2218846498\\
-0.61	13694.7408537906\\
-0.51	12850.7492213279\\
-0.41	11801.5996581257\\
-0.36	11194.1540681008\\
-0.31	10518.8590744921\\
-0.26	9763.90858066222\\
-0.21	8916.03327976311\\
-0.16	7960.95650879354\\
-0.11	6885.06286937069\\
-0.06	5683.49007469383\\
-0.01	4377.53900619533\\
0.04	3048.45577476989\\
0.09	1884.33742818149\\
0.14	1225.3153961861\\
0.19	1293.50950347796\\
0.24	1857.83624039062\\
0.29	2559.01362759139\\
0.34	3192.42978089351\\
0.39	3701.30412188809\\
0.49	4400.87417999815\\
0.59	4850.49873616317\\
0.69	5139.41924733028\\
0.79	5355.28830717145\\
0.89	5512.9834950929\\
0.99	5630.85089578465\\
1.09	5732.09069165261\\
1.19	5800.69388444219\\
1.29	5838.9830929487\\
1.39	5920.53684736668\\
1.49	5944.69936603267\\
1.59	6001.55148687053\\
1.69	6017.89189937573\\
1.79	6071.66919287574\\
1.89	6067.63503691069\\
1.99	6129.68485767083\\
};
\addlegendentry{$0.2$V}

\addplot [color=mycolor1, dashed, forget plot]
  table[row sep=crcr, y expr=1e-3*\thisrowno{1}*0.1]{%
-2.01	-28459.7727122332\\
-1.91	-28012.2650293548\\
-1.81	-27564.446755715\\
-1.71	-27111.535845613\\
-1.61	-26568.7276622834\\
-1.51	-26036.2962050499\\
-1.41	-25405.7134172968\\
-1.31	-24767.1211840527\\
-1.21	-24017.7885130741\\
-1.11	-23235.0368937955\\
-1.01	-22339.9188543786\\
-0.91	-21334.9201203811\\
-0.81	-20233.5197451902\\
-0.71	-18972.9252658384\\
-0.61	-17517.2973323014\\
-0.51	-15829.3832885755\\
-0.41	-13880.6997767282\\
-0.36	-12809.182363909\\
-0.31	-11685.8759621343\\
-0.26	-10535.893562978\\
-0.21	-9407.82521919014\\
-0.16	-8390.17581866939\\
-0.11	-7624.11857346673\\
-0.06	-7273.08458378078\\
-0.01	-7411.35129481601\\
0.04	-7957.5254032058\\
0.09	-8740.55525579117\\
0.14	-9617.98785194302\\
0.19	-10502.7087523166\\
0.24	-11350.829666532\\
0.29	-12143.0765841814\\
0.34	-12873.0143472519\\
0.39	-13541.1704771904\\
0.49	-14713.6884534009\\
0.59	-15695.0457815804\\
0.69	-16525.1334979065\\
0.79	-17235.1332165238\\
0.89	-17849.0766864253\\
0.99	-18385.6843750214\\
1.09	-18859.6400687397\\
1.19	-19282.4566764414\\
1.29	-19661.6675069397\\
1.39	-20005.3832318456\\
1.49	-20317.7424143746\\
1.59	-20604.7926672397\\
1.69	-20867.7028638215\\
1.79	-21112.6498802109\\
1.89	-21339.1792519061\\
1.99	-21540.2523412951\\
};
\addplot [color=mycolor2, dashed, forget plot]
  table[row sep=crcr, y expr=1e-3*\thisrowno{1}*0.1]{%
-2.01	-14643.1316977595\\
-1.91	-14422.5795418716\\
-1.81	-14198.9341090929\\
-1.71	-13970.45941653\\
-1.61	-13702.9514608624\\
-1.51	-13436.3405589566\\
-1.41	-13126.5219982999\\
-1.31	-12809.2451217897\\
-1.21	-12442.8410333351\\
-1.11	-12057.8515587662\\
-1.01	-11621.971655648\\
-0.91	-11134.9368703919\\
-0.81	-10599.3752331854\\
-0.71	-9991.65591749147\\
-0.61	-9293.70105213896\\
-0.51	-8487.35394301751\\
-0.41	-7554.89253737965\\
-0.36	-7038.10243024084\\
-0.31	-6489.08541890249\\
-0.26	-5912.72361201288\\
-0.21	-5322.27146072291\\
-0.16	-4740.36297832393\\
-0.11	-4209.75990888493\\
-0.06	-3799.5889286011\\
-0.01	-3601.05607684367\\
0.04	-3670.9459946259\\
0.09	-3960.51764542923\\
0.14	-4369.95926915655\\
0.19	-4825.24737044103\\
0.24	-5281.66173107848\\
0.29	-5718.98151364103\\
0.34	-6126.86026818191\\
0.39	-6502.39384912248\\
0.49	-7159.29290533373\\
0.59	-7706.81754862177\\
0.69	-8168.69953013594\\
0.79	-8562.16948782621\\
0.89	-8900.58900993959\\
0.99	-9194.60756941863\\
1.09	-9452.60949234107\\
1.19	-9682.18331848883\\
1.29	-9887.32521362564\\
1.39	-10071.9856671308\\
1.49	-10239.9537488006\\
1.59	-10392.9400356577\\
1.69	-10533.6436111794\\
1.79	-10663.4550607719\\
1.89	-10783.033708558\\
1.99	-10891.0701194554\\
};
\addplot [color=mycolor3, dashed, forget plot]
  table[row sep=crcr, y expr=1e-3*\thisrowno{1}*0.1]{%
-2.01	15583.6535711442\\
-1.91	15386.400197233\\
-1.81	15170.3219847416\\
-1.71	14935.7014652031\\
-1.61	14691.3448541746\\
-1.51	14421.517769655\\
-1.41	14135.6368539358\\
-1.31	13820.316259578\\
-1.21	13480.2899148973\\
-1.11	13104.6207493271\\
-1.01	12692.5344202552\\
-0.91	12234.5946360077\\
-0.81	11721.5536326378\\
-0.71	11142.3706383853\\
-0.61	10483.9831443747\\
-0.51	9729.70178438161\\
-0.41	8862.07365946096\\
-0.36	8380.58979345751\\
-0.31	7865.09141972966\\
-0.26	7315.21953182159\\
-0.21	6733.05834063898\\
-0.16	6123.9408489749\\
-0.11	5500.13604556211\\
-0.06	4886.54406983495\\
-0.01	4325.42639572037\\
0.04	3886.17865045035\\
0.09	3663.64521212425\\
0.14	3717.78993646933\\
0.19	4002.4773607343\\
0.24	4416.93754443595\\
0.29	4885.58891277799\\
0.34	5361.33653699214\\
0.39	5820.74728097946\\
0.49	6650.98738502349\\
0.59	7352.46470250321\\
0.69	7938.41551047274\\
0.79	8431.30705144748\\
0.89	8848.98996163569\\
0.99	9204.64705180489\\
1.09	9513.27604561511\\
1.19	9784.47592012638\\
1.29	10022.655464186\\
1.39	10235.7195690612\\
1.49	10427.6037952301\\
1.59	10600.1832205246\\
1.69	10757.9660219216\\
1.79	10902.2817106963\\
1.89	11034.2433017953\\
1.99	11160.9901074681\\
};
\addplot [color=mycolor4, dashed, forget plot]
  table[row sep=crcr, y expr=1e-3*\thisrowno{1}*0.1]{%
-2.01	32051.7944757977\\
-1.91	31676.6821993137\\
-1.81	31252.0525619288\\
-1.71	30777.589360872\\
-1.61	30313.2050321553\\
-1.51	29771.2252372342\\
-1.41	29226.8318324535\\
-1.31	28599.1637363592\\
-1.21	27950.7766154595\\
-1.11	27209.0230763475\\
-1.01	26411.2672783745\\
-0.91	25527.3957635199\\
-0.81	24530.2663117604\\
-0.71	23400.7665616851\\
-0.61	22126.3569965266\\
-0.51	20677.9719799705\\
-0.41	19022.2291090203\\
-0.36	18105.4757616486\\
-0.31	17123.9929520909\\
-0.26	16074.9413384977\\
-0.21	14958.1904982975\\
-0.16	13777.3337104554\\
-0.11	12544.3516179031\\
-0.06	11283.4746832173\\
-0.01	10044.610887757\\
0.04	8911.20138826161\\
0.09	8020.23508069134\\
0.14	7545.51040803826\\
0.19	7587.31140769919\\
0.24	8087.2657940409\\
0.29	8880.14233265839\\
0.34	9805.72573027342\\
0.39	10762.8386206937\\
0.49	12575.393340515\\
0.59	14145.8330998061\\
0.69	15469.2692312554\\
0.79	16579.7776940146\\
0.89	17520.2292309144\\
0.99	18313.2145399193\\
1.09	18986.0150391817\\
1.19	19572.884502347\\
1.29	20092.5422298721\\
1.39	20550.3108387699\\
1.49	20961.3797654664\\
1.59	21332.0511049782\\
1.69	21667.3881068633\\
1.79	21973.0714835107\\
1.89	22255.460639925\\
1.99	22521.1409548144\\
};

\node[anchor=south, font=\footnotesize] at (rel axis cs: 0.5, 1) {$V_{\rm BG}=-1$V};
\SubfigLetter{a)}{0.5}{0.1};

\end{axis}

%% file: Figuras/GFET_Vbg0V_Np1e12_Comp.tex
%
%
%

\begin{axis}[%
width=0.24\textwidth,
height=3cm,
at={(0.3\textwidth,0cm)},
scale only axis,
xmin=-2.01,
xmax=2,
xlabel style={font=\color{white!15!black}},
xlabel={$V_{\rm FG}$ (V)},
ymin=-2.5,
ymax=2.5,
ytick={-2.5,-1.5,...,2.5},
minor y tick num=0,
ylabel style={font=\color{white!15!black}},
axis background/.style={fill=white},
axis x line*=bottom,
axis y line*=left,
xmajorgrids,
ymajorgrids,
yminorgrids,
clip mode=individual,
legend style={at={(0.85,-0.52)}, anchor=south west, legend cell align=left, align=left, draw=white!15!black, font=\footnotesize, legend columns = 2}
]
\addplot [color=mycolor1]
  table[row sep=crcr, y expr=1e-3*\thisrowno{1}*0.1]{%
-2.01	-2233.71363848527\\
-1.91	-2204.49360021082\\
-1.81	-2186.57257642514\\
-1.71	-2154.04053943033\\
-1.61	-2120.50765149117\\
-1.51	-2084.84998951846\\
-1.41	-2038.43084978978\\
-1.31	-2003.22641308317\\
-1.21	-1966.73203758142\\
-1.11	-1931.26806141218\\
-1.01	-1889.21845038605\\
-0.91	-1841.81495574843\\
-0.81	-1792.0958473146\\
-0.71	-1732.85700637732\\
-0.61	-1661.54150558087\\
-0.51	-1572.24198221139\\
-0.41	-1449.93097876333\\
-0.36	-1367.2985574844\\
-0.31	-1261.214259944\\
-0.26	-1111.68311468919\\
-0.21	-907.464027191886\\
-0.16	-685.15876902808\\
-0.11	-650.467215740112\\
-0.06	-929.861729702597\\
-0.01	-1272.78031174215\\
0.04	-1548.20592928596\\
0.09	-1773.01896359962\\
0.14	-1957.33107160123\\
0.19	-2115.91224765757\\
0.24	-2248.73209800128\\
0.29	-2367.40063920641\\
0.34	-2478.26497795652\\
0.39	-2590.71557939566\\
0.49	-2840.95402661074\\
0.59	-3021.3236510103\\
0.69	-3186.30463430149\\
0.79	-3314.11097425354\\
0.89	-3437.93731001428\\
0.99	-3539.82047408851\\
1.09	-3636.74004798834\\
1.19	-3725.74855276643\\
1.29	-3808.25086043369\\
1.39	-3883.82697679264\\
1.49	-3956.01810427547\\
1.59	-4023.47446780002\\
1.69	-4087.16925828248\\
1.79	-4147.38412058758\\
1.89	-4204.86652562629\\
1.99	-4253.30978211322\\
};
\addlegendentry{No Puddles};

\addplot [color=mycolor2, forget plot]
  table[row sep=crcr, y expr=1e-3*\thisrowno{1}*0.1]{%
-2.01	-1106.52285629796\\
-1.91	-1091.70787169856\\
-1.81	-1080.56230567184\\
-1.71	-1064.58446945333\\
-1.61	-1048.15929689868\\
-1.51	-1030.90191509153\\
-1.41	-1009.42644885303\\
-1.31	-990.268764220394\\
-1.21	-970.155129089125\\
-1.11	-951.256281540414\\
-1.01	-927.595095008862\\
-0.91	-900.051712964886\\
-0.81	-874.873970274975\\
-0.71	-845.416810627075\\
-0.61	-811.356973486444\\
-0.51	-771.593777484543\\
-0.41	-721.313154442627\\
-0.36	-689.63606958847\\
-0.31	-651.450898389064\\
-0.26	-602.82945795433\\
-0.21	-537.449800505722\\
-0.16	-444.418658253583\\
-0.11	-321.910582330179\\
-0.06	-251.305214498382\\
-0.01	-353.373996566142\\
0.04	-492.511543727724\\
0.09	-602.914251370302\\
0.14	-687.364987339977\\
0.19	-755.683016273448\\
0.24	-810.746224498157\\
0.29	-858.751142015135\\
0.34	-902.760792958905\\
0.39	-940.675729565758\\
0.49	-1006.54393943983\\
0.59	-1059.69110147032\\
0.69	-1112.50511889891\\
0.79	-1152.85448368576\\
0.89	-1198.08927174531\\
0.99	-1230.80400710212\\
1.09	-1266.00046013533\\
1.19	-1299.16239034132\\
1.29	-1340.16688492931\\
1.39	-1376.87592128639\\
1.49	-1404.05814675134\\
1.59	-1436.56991065121\\
1.69	-1467.55961657565\\
1.79	-1497.3744343935\\
1.89	-1525.89373469943\\
1.99	-1560.07657727175\\
};

\addplot [color=mycolor3, forget plot]
  table[row sep=crcr, y expr=1e-3*\thisrowno{1}*0.1]{%
-2.01	1544.70351462274\\
-1.91	1517.02695731054\\
-1.81	1495.40516305124\\
-1.71	1465.12444626921\\
-1.61	1433.70071678769\\
-1.51	1400.40597147653\\
-1.41	1357.19555305538\\
-1.31	1319.67153067114\\
-1.21	1286.6084105742\\
-1.11	1259.05336989803\\
-1.01	1223.28130028567\\
-0.91	1180.20594725201\\
-0.81	1144.10871825958\\
-0.71	1099.25221272438\\
-0.61	1045.98298821134\\
-0.51	990.008953969193\\
-0.41	926.88293588915\\
-0.36	889.420395746436\\
-0.31	847.671220394363\\
-0.26	800.087790879427\\
-0.21	744.563257676984\\
-0.16	677.365816027553\\
-0.11	592.641441715389\\
-0.06	482.960294322051\\
-0.01	344.593588912217\\
0.04	248.230160393196\\
0.09	320.069684685515\\
0.14	441.588449265824\\
0.19	534.04348476354\\
0.24	599.338088600417\\
0.29	648.061733629421\\
0.34	686.84590661906\\
0.39	719.029645028452\\
0.49	770.547141364173\\
0.59	810.619198737824\\
0.69	845.62670807201\\
0.79	874.474470902311\\
0.89	904.380775041379\\
0.99	927.75268341703\\
1.09	951.25560338617\\
1.19	972.903284226429\\
1.29	996.011962619651\\
1.39	1015.30671858234\\
1.49	1030.10819184766\\
1.59	1047.18702760086\\
1.69	1063.52724682141\\
1.79	1079.28975057171\\
1.89	1094.34817357345\\
1.99	1113.2683076666\\
};

\addplot [color=mycolor4, forget plot]
  table[row sep=crcr, y expr=1e-3*\thisrowno{1}*0.1]{%
-2.01	4252.33295993808\\
-1.91	4197.20680128465\\
-1.81	4135.63911126316\\
-1.71	4075.93312757329\\
-1.61	4012.63209712217\\
-1.51	3945.54040655766\\
-1.41	3874.30372813834\\
-1.31	3797.10550462636\\
-1.21	3714.44150263708\\
-1.11	3627.79125491881\\
-1.01	3530.91270425812\\
-0.91	3418.96238109404\\
-0.81	3304.59438699888\\
-0.71	3168.63015259966\\
-0.61	2999.26725261759\\
-0.51	2800.33384587897\\
-0.41	2550.63035725561\\
-0.36	2453.253516771\\
-0.31	2345.88167077078\\
-0.26	2226.13498673666\\
-0.21	2090.24793516851\\
-0.16	1932.83612320806\\
-0.11	1745.30582508573\\
-0.06	1519.47706981324\\
-0.01	1239.94660605027\\
0.04	912.925767148859\\
0.09	644.234865406071\\
0.14	682.331077303813\\
0.19	903.403331635323\\
0.24	1105.83770794882\\
0.29	1254.43779967681\\
0.34	1361.12725027463\\
0.39	1444.36194736242\\
0.49	1567.6633169221\\
0.59	1657.66348965473\\
0.69	1729.75155045958\\
0.79	1788.26429714878\\
0.89	1841.41478149576\\
0.99	1885.13182249795\\
1.09	1926.83119567796\\
1.19	1965.6436955526\\
1.29	2008.9496754045\\
1.39	2047.5692138524\\
1.49	2075.30949785417\\
1.59	2110.48355005006\\
1.69	2143.12003574686\\
1.79	2174.8564418835\\
1.89	2205.24115508702\\
1.99	2248.03228510733\\
};

\addplot [color=mycolor1, dashed]
  table[row sep=crcr, y expr=1e-3*\thisrowno{1}*0.1]{%
-2.01	-18301.1852932602\\
-1.91	-18076.5541676531\\
-1.81	-17845.6235547117\\
-1.71	-17587.506832488\\
-1.61	-17309.9272807866\\
-1.51	-17008.4375086789\\
-1.41	-16673.1832940307\\
-1.31	-16312.7562292963\\
-1.21	-15914.1549479556\\
-1.11	-15474.2460310823\\
-1.01	-14973.0228962173\\
-0.91	-14398.9826584404\\
-0.81	-13746.598015934\\
-0.71	-12984.0451131275\\
-0.61	-12085.9110142653\\
-0.51	-11025.3547457835\\
-0.41	-9764.18469350672\\
-0.36	-9058.55356416944\\
-0.31	-8318.23876755351\\
-0.26	-7578.80836464814\\
-0.21	-6911.1516037473\\
-0.16	-6438.69829585511\\
-0.11	-6302.70711028969\\
-0.06	-6561.11196815674\\
-0.01	-7132.36710448458\\
0.04	-7864.51764220185\\
0.09	-8654.78072227202\\
0.14	-9439.59341375319\\
0.19	-10191.3830977636\\
0.24	-10892.6711761188\\
0.29	-11543.8178426753\\
0.34	-12146.7614716116\\
0.39	-12700.1411558611\\
0.49	-13681.2746237029\\
0.59	-14517.3455048175\\
0.69	-15248.489641835\\
0.79	-15879.7053382665\\
0.89	-16448.1245925546\\
0.99	-16944.7027693356\\
1.09	-17397.7173036054\\
1.19	-17809.022577087\\
1.29	-18191.9746063024\\
1.39	-18540.9163983014\\
1.49	-18857.8307048183\\
1.59	-19159.1640112887\\
1.69	-19439.6939936171\\
1.79	-19703.4493878238\\
1.89	-19951.4171548658\\
1.99	-20191.7403282599\\
};
\addlegendentry{N$_{\rm p} = 10^{12}$cm$^{-2}$};

\addplot [color=mycolor2, dashed, forget plot]
  table[row sep=crcr, y expr=1e-3*\thisrowno{1}*0.1]{%
-2.01	-9252.98733198174\\
-1.91	-9141.47863795287\\
-1.81	-9026.20806423925\\
-1.71	-8899.10993382832\\
-1.61	-8762.7455805174\\
-1.51	-8615.27457080256\\
-1.41	-8452.62005133064\\
-1.31	-8277.90278908117\\
-1.21	-8085.92878082072\\
-1.11	-7874.94466800377\\
-1.01	-7636.44488730005\\
-0.91	-7365.33272186101\\
-0.81	-7059.04049350988\\
-0.71	-6703.29946023913\\
-0.61	-6287.19204509855\\
-0.51	-5798.97570279026\\
-0.41	-5217.20616508342\\
-0.36	-4887.42496753078\\
-0.31	-4532.81693670391\\
-0.26	-4159.88023687299\\
-0.21	-3783.13788834539\\
-0.16	-3435.89124482441\\
-0.11	-3174.36875927438\\
-0.06	-3078.11599678061\\
-0.01	-3194.02157742183\\
0.04	-3473.31239407495\\
0.09	-3834.05361885516\\
0.14	-4222.30922934958\\
0.19	-4607.95176414112\\
0.24	-4974.65697843986\\
0.29	-5317.65767026443\\
0.34	-5635.80384185489\\
0.39	-5927.47451478181\\
0.49	-6442.53231922005\\
0.59	-6878.59995552914\\
0.69	-7255.97128108348\\
0.79	-7580.40521762847\\
0.89	-7869.31210379089\\
0.99	-8121.47151157415\\
1.09	-8349.68806578665\\
1.19	-8556.04258962173\\
1.29	-8746.6712414175\\
1.39	-8919.15019801056\\
1.49	-9075.45304466997\\
1.59	-9222.96675317145\\
1.69	-9360.5583033512\\
1.79	-9489.25580928789\\
1.89	-9610.46484433905\\
1.99	-9727.66148211983\\
};
\addplot [color=mycolor3, dashed, forget plot]
  table[row sep=crcr, y expr=1e-3*\thisrowno{1}*0.1]{%
-2.01	9718.33612710573\\
-1.91	9604.13137331386\\
-1.81	9486.05875951262\\
-1.71	9357.09060758185\\
-1.61	9219.31496717223\\
-1.51	9071.55663875842\\
-1.41	8909.1749529958\\
-1.31	8736.71416662934\\
-1.21	8548.73178740812\\
-1.11	8344.90315961483\\
-1.01	8116.53800408646\\
-0.91	7859.69417355035\\
-0.81	7575.12217970546\\
-0.71	7248.76033822899\\
-0.61	6871.56255274441\\
-0.51	6434.43872696032\\
-0.41	5921.37154596379\\
-0.36	5629.96977490527\\
-0.31	5313.11168966283\\
-0.26	4970.37240796971\\
-0.21	4603.33207625046\\
-0.16	4218.59037244879\\
-0.11	3829.93555233848\\
-0.06	3469.61488795192\\
-0.01	3190.29639704704\\
0.04	3075.58173086858\\
0.09	3173.24874148038\\
0.14	3435.38020097188\\
0.19	3782.95144987419\\
0.24	4159.91226016921\\
0.29	4533.04461297874\\
0.34	4887.84657853202\\
0.39	5217.79976296941\\
0.49	5800.02549333298\\
0.59	6288.79171677653\\
0.69	6705.10229201232\\
0.79	7059.63374465006\\
0.89	7369.0718894303\\
0.99	7636.74546572304\\
1.09	7875.10804124976\\
1.19	8088.32573136007\\
1.29	8282.880285871\\
1.39	8457.86806294582\\
1.49	8614.93018768338\\
1.59	8762.20046918436\\
1.69	8898.45385208461\\
1.79	9025.38011508388\\
1.89	9144.02543003668\\
1.99	9259.03666888193\\
};
\addplot [color=mycolor4, dashed, forget plot]
  table[row sep=crcr, y expr=1e-3*\thisrowno{1}*0.1]{%
-2.01	20172.2065927944\\
-1.91	19938.0315359238\\
-1.81	19696.7867426632\\
-1.71	19432.3990285428\\
-1.61	19151.355963237\\
-1.51	18849.4103789949\\
-1.41	18517.6677696614\\
-1.31	18169.5037838206\\
-1.21	17793.2868634025\\
-1.11	17388.8380868146\\
-1.01	16935.3774444287\\
-0.91	16426.5214308782\\
-0.81	15869.4529652333\\
-0.71	15233.1076919454\\
-0.61	14502.0007907096\\
-0.51	13663.0410027714\\
-0.41	12686.3820756165\\
-0.36	12133.2912536088\\
-0.31	11532.9489786942\\
-0.26	10882.0124842639\\
-0.21	10179.8025007528\\
-0.16	9429.5880967555\\
-0.11	8643.67354090855\\
-0.06	7853.65797486607\\
-0.01	7120.50215647112\\
0.04	6553.07150771372\\
0.09	6298.37657241759\\
0.14	6436.17451620653\\
0.19	6909.36982120588\\
0.24	7577.4564629953\\
0.29	8316.82415021892\\
0.34	9056.45003082709\\
0.39	9761.87931745124\\
0.49	11022.6015698891\\
0.59	12083.7345062839\\
0.69	12982.6791321981\\
0.79	13743.2066341068\\
0.89	14402.7811921995\\
0.99	14968.9672802348\\
1.09	15469.7161248546\\
1.19	15914.6297863187\\
1.29	16319.2505664316\\
1.39	16680.325483108\\
1.49	17002.0519513854\\
1.59	17303.0808841267\\
1.69	17580.3285699122\\
1.79	17837.98017334\\
1.89	18077.5141459283\\
1.99	18311.6452136774\\
};

\node[anchor=south, font=\footnotesize] at (rel axis cs: 0.5, 1) {$V_{\rm BG}=0$V};
\SubfigLetter{b)}{0.5}{0.1};

\end{axis}

%% file: Figuras/GFET_Vbg1V_Np1e12_Comp.tex
%
%
%

\begin{axis}[%
width=0.24\textwidth,
height=3cm,
at={(0.6\textwidth,0cm)},
scale only axis,
xmin=-2.01,
xmax=2,
xlabel style={font=\color{white!15!black}},
xlabel={$V_{\rm FG}$ (V)},
ymin=-4,
ymax=4,
ylabel style={font=\color{white!15!black}},
axis background/.style={fill=white},
axis x line*=bottom,
axis y line*=left,
xmajorgrids,
ymajorgrids,
clip mode=individual,
legend style={at={(0.03,0.97)}, anchor=north west, legend cell align=left, align=left, draw=white!15!black}
]
\addplot [color=mycolor1, forget plot]
  table[row sep=crcr, y expr=1e-3*\thisrowno{1}*0.1]{%
-2.01	-6132.38959962818\\
-1.91	-6070.03655380827\\
-1.81	-6073.21685148632\\
-1.71	-6018.96115131711\\
-1.61	-6003.28618951801\\
-1.51	-5945.82903175971\\
-1.41	-5922.27205143342\\
-1.31	-5840.73517590448\\
-1.21	-5802.06306345768\\
-1.11	-5733.92797529089\\
-1.01	-5632.9582230962\\
-0.91	-5515.88948604454\\
-0.81	-5357.87899016072\\
-0.71	-5141.31598164516\\
-0.61	-4852.2571404104\\
-0.51	-4402.10434665945\\
-0.41	-3702.02573457678\\
-0.36	-3192.93607422921\\
-0.31	-2559.30024133008\\
-0.26	-1857.96298131293\\
-0.21	-1293.57656999504\\
-0.16	-1225.35705157583\\
-0.11	-1884.38464346688\\
-0.06	-3048.50989494802\\
-0.01	-4377.58982464985\\
0.04	-5683.51580494728\\
0.09	-6885.05544720166\\
0.14	-7960.89014678054\\
0.19	-8915.92269429959\\
0.24	-9763.75687595044\\
0.29	-10518.5985229774\\
0.34	-11193.8542974441\\
0.39	-11801.2699041308\\
0.49	-12850.3177121626\\
0.59	-13694.3848139457\\
0.69	-14403.7654166346\\
0.79	-15014.2355635521\\
0.89	-15546.2733824026\\
0.99	-16016.2947705362\\
1.09	-16435.5187602231\\
1.19	-16834.3251871989\\
1.29	-17180.0463669751\\
1.39	-17472.7868502249\\
1.49	-17786.0360030072\\
1.59	-18028.6491956656\\
1.69	-18303.3968715742\\
1.79	-18535.1034093413\\
1.89	-18722.2671090956\\
1.99	-18924.9649011013\\
};

\addplot [color=mycolor2, forget plot]
  table[row sep=crcr, y expr=1e-3*\thisrowno{1}*0.1]{%
-2.01	-3011.3022132769\\
-1.91	-3001.02773765561\\
-1.81	-2986.48093697657\\
-1.71	-2967.02683494511\\
-1.61	-2955.34978673822\\
-1.51	-2930.44951717595\\
-1.41	-2914.06164883434\\
-1.31	-2885.82883085073\\
-1.21	-2856.25528981414\\
-1.11	-2825.48596689336\\
-1.01	-2783.70882723081\\
-0.91	-2731.41991531368\\
-0.81	-2667.42963718155\\
-0.71	-2584.31765257924\\
-0.61	-2468.07841351344\\
-0.51	-2299.57522278098\\
-0.41	-2042.68519510744\\
-0.36	-1856.25568557458\\
-0.31	-1612.8713574066\\
-0.26	-1296.72518137058\\
-0.21	-912.648488977343\\
-0.16	-548.539064590632\\
-0.11	-433.318350889322\\
-0.06	-749.853167303911\\
-0.01	-1347.96356346523\\
0.04	-2026.95223064642\\
0.09	-2681.98632177739\\
0.14	-3276.55603793931\\
0.19	-3802.7515852449\\
0.24	-4265.96709566403\\
0.29	-4673.87568366317\\
0.34	-5034.71856293713\\
0.39	-5355.55022769083\\
0.49	-5900.26983524261\\
0.59	-6344.90370560304\\
0.69	-6716.25801514664\\
0.79	-7032.5725884543\\
0.89	-7306.21921410308\\
0.99	-7546.47580187572\\
1.09	-7759.06711235664\\
1.19	-7955.15165026512\\
1.29	-8130.0358483981\\
1.39	-8284.96567386264\\
1.49	-8436.08290141389\\
1.59	-8565.85037448751\\
1.69	-8696.73003799048\\
1.79	-8813.50518397812\\
1.89	-8915.72610177634\\
1.99	-9017.9082553351\\
};

\addplot [color=mycolor3, forget plot]
  table[row sep=crcr, y expr=1e-3*\thisrowno{1}*0.1]{%
-2.01	2773.25366322594\\
-1.91	2758.45677506662\\
-1.81	2745.24681640628\\
-1.71	2734.18499570454\\
-1.61	2716.3713567729\\
-1.51	2702.85107931937\\
-1.41	2681.72262947776\\
-1.31	2663.81907253247\\
-1.21	2638.00024777435\\
-1.11	2612.39044463886\\
-1.01	2584.54247064271\\
-0.91	2547.87876194246\\
-0.81	2503.329386467\\
-0.71	2450.06919323487\\
-0.61	2382.04000720636\\
-0.51	2286.63193188887\\
-0.41	2153.59326707411\\
-0.36	2059.17537797464\\
-0.31	1938.59434878301\\
-0.26	1779.66491810452\\
-0.21	1564.97234746566\\
-0.16	1276.97753257925\\
-0.11	917.154580818602\\
-0.06	558.769743637842\\
-0.01	423.55648850388\\
0.04	705.50824307764\\
0.09	1266.66918744529\\
0.14	1900.34852065147\\
0.19	2501.79334809067\\
0.24	3037.22605578244\\
0.29	3504.07512194132\\
0.34	3908.87445111657\\
0.39	4260.99003465835\\
0.49	4840.3609606702\\
0.59	5297.08383106953\\
0.69	5668.35382949761\\
0.79	5997.94408366976\\
0.89	6280.40762979591\\
0.99	6525.77218728505\\
1.09	6743.02240485452\\
1.19	6927.5912611779\\
1.29	7104.06169089301\\
1.39	7277.31030856287\\
1.49	7413.58251211829\\
1.59	7562.93596360671\\
1.69	7678.1152217192\\
1.79	7797.36598163298\\
1.89	7922.01341539477\\
1.99	8027.23817617782\\
};

\addplot [color=mycolor4, forget plot]
  table[row sep=crcr, y expr=1e-3*\thisrowno{1}*0.1]{%
-2.01	5232.43112502247\\
-1.91	5216.40333108022\\
-1.81	5192.06499531965\\
-1.71	5168.26379594285\\
-1.61	5148.47627611949\\
-1.51	5114.06164883435\\
-1.41	5090.09477235379\\
-1.31	5050.3294027289\\
-1.21	5019.63040252827\\
-1.11	4972.2428740181\\
-1.01	4916.61975087721\\
-0.91	4867.35096038541\\
-0.81	4797.78820730564\\
-0.71	4710.49576939316\\
-0.61	4599.65612643809\\
-0.51	4452.30359869162\\
-0.41	4241.90706195333\\
-0.36	4099.59723493721\\
-0.31	3927.69296525453\\
-0.26	3702.6570339767\\
-0.21	3404.84701700212\\
-0.16	3007.93285846434\\
-0.11	2488.18576885583\\
-0.06	1868.59609050822\\
-0.01	1306.01855743974\\
0.04	1129.04836113102\\
0.09	1619.40536326084\\
0.14	2638.12391805856\\
0.19	3808.81400546153\\
0.24	4934.71291367286\\
0.29	5941.51002524782\\
0.34	6817.57719424355\\
0.39	7575.13810652776\\
0.49	8803.58329461269\\
0.59	9748.89934563291\\
0.69	10542.8820075539\\
0.79	11218.5903057275\\
0.89	11791.2544962589\\
0.99	12285.049425759\\
1.09	12718.3297213542\\
1.19	13079.7975189129\\
1.29	13426.2782331216\\
1.39	13781.8189333488\\
1.49	14057.3424918349\\
1.59	14367.2514547444\\
1.69	14597.4795092104\\
1.79	14839.6655277632\\
1.89	15093.296708682\\
1.99	15302.5431296692\\
};

\addplot [color=mycolor1, dashed, forget plot]
  table[row sep=crcr, y expr=1e-3*\thisrowno{1}*0.1]{%
-2.01	-22526.4513364504\\
-1.91	-22259.2681338733\\
-1.81	-21975.116730961\\
-1.71	-21669.4602160334\\
-1.61	-21334.0572844899\\
-1.51	-20963.4114707814\\
-1.41	-20552.2554440583\\
-1.31	-20094.5165026124\\
-1.21	-19574.7845755721\\
-1.11	-18987.7458867117\\
-1.01	-18314.8895355873\\
-0.91	-17522.9473735343\\
-0.81	-16582.0634654225\\
-0.71	-15471.161326815\\
-0.61	-14147.2455550386\\
-0.51	-12576.4050733882\\
-0.41	-10763.4703342545\\
-0.36	-9806.20166449492\\
-0.31	-8880.48059912849\\
-0.26	-8087.50994287192\\
-0.21	-7587.49621716585\\
-0.16	-7545.66559699949\\
-0.11	-8020.36324133852\\
-0.06	-8911.30420795407\\
-0.01	-10044.6810181864\\
0.04	-11283.5086275169\\
0.09	-12544.349974825\\
0.14	-13777.2859886978\\
0.19	-14958.0986229095\\
0.24	-16074.8288769676\\
0.29	-17123.8264563238\\
0.34	-18105.2621682688\\
0.39	-19021.9666218\\
0.49	-20677.656504731\\
0.59	-22125.9310828349\\
0.69	-23400.2743700535\\
0.79	-24529.7381247022\\
0.89	-25526.8043751344\\
0.99	-26410.5984607299\\
1.09	-27208.625935284\\
1.19	-27951.2170188143\\
1.29	-28616.8588365808\\
1.39	-29209.0581566676\\
1.49	-29791.4663873668\\
1.59	-30293.2882564967\\
1.69	-30799.9676063005\\
1.79	-31252.9691025037\\
1.89	-31653.9657465202\\
1.99	-32052.3724104307\\
};
\addplot [color=mycolor2, dashed, forget plot]
  table[row sep=crcr, y expr=1e-3*\thisrowno{1}*0.1]{%
-2.01	-11165.7565809134\\
-1.91	-11037.5653027615\\
-1.81	-10904.2678106473\\
-1.71	-10759.8972351609\\
-1.61	-10602.1676699385\\
-1.51	-10429.5701305748\\
-1.41	-10237.7236852604\\
-1.31	-10024.6104410802\\
-1.21	-9786.48483229008\\
-1.11	-9515.26214115854\\
-1.01	-9206.56661045907\\
-0.91	-8851.36197252498\\
-0.81	-8433.40751250957\\
-0.71	-7940.22299800505\\
-0.61	-7353.90869771604\\
-0.51	-6652.07923137162\\
-0.41	-5821.49645229559\\
-0.36	-5361.92428178056\\
-0.31	-4886.02956103342\\
-0.26	-4417.25122757905\\
-0.21	-4002.66397520836\\
-0.16	-3717.90530919661\\
-0.11	-3663.72261502073\\
-0.06	-3886.23218317892\\
-0.01	-4325.46068654823\\
0.04	-4886.55297735736\\
0.09	-5500.11690808119\\
0.14	-6123.88918446806\\
0.19	-6732.96789420312\\
0.24	-7315.0899484092\\
0.29	-7864.90157591006\\
0.34	-8380.35229066455\\
0.39	-8861.78354812934\\
0.49	-9729.30317542821\\
0.59	-10483.5091242864\\
0.69	-11141.8140985587\\
0.79	-11720.9506764887\\
0.89	-12233.9416669353\\
0.99	-12691.8390277094\\
1.09	-13104.1778094003\\
1.19	-13480.448939633\\
1.29	-13822.6115173717\\
1.39	-14133.7121740893\\
1.49	-14424.675719394\\
1.59	-14688.904921937\\
1.69	-14939.6870698436\\
1.79	-15170.9952987921\\
1.89	-15382.1658159013\\
1.99	-15584.2120949955\\
};
\addplot [color=mycolor3, dashed, forget plot]
  table[row sep=crcr, y expr=1e-3*\thisrowno{1}*0.1]{%
-2.01	10897.6094949767\\
-1.91	10787.8630306136\\
-1.81	10666.6869631756\\
-1.71	10536.7588475912\\
-1.61	10396.1346664037\\
-1.51	10243.1371772469\\
-1.41	10075.1554930513\\
-1.31	9890.3842768116\\
-1.21	9685.2449184072\\
-1.11	9455.4851219322\\
-1.01	9197.38765655367\\
-0.91	8903.18560955143\\
-0.81	8564.60894153634\\
-0.71	8170.85525361176\\
-0.61	7707.88183730741\\
-0.51	7160.13067953031\\
-0.41	6502.99174917113\\
-0.36	6127.33833784801\\
-0.31	5719.3420551371\\
-0.26	5281.91447927499\\
-0.21	4825.41852131736\\
-0.16	4370.06283455482\\
-0.11	3960.573176616\\
-0.06	3670.97182333534\\
-0.01	3601.07466247142\\
0.04	3799.59857738047\\
0.09	4209.75515443544\\
0.14	4740.33727284481\\
0.19	5322.22398184304\\
0.24	5912.65343331351\\
0.29	6488.98278934683\\
0.34	7037.97656903453\\
0.39	7554.73516178273\\
0.49	8487.13729682857\\
0.59	9293.42685655459\\
0.69	9991.33061837999\\
0.79	10599.018899074\\
0.89	11134.6521343605\\
0.99	11621.6680141418\\
1.09	12057.7319844216\\
1.19	12443.0191516718\\
1.29	12800.232665164\\
1.39	13136.7200019091\\
1.49	13426.6952628891\\
1.59	13713.8833241557\\
1.69	13960.364836259\\
1.79	14199.4650672643\\
1.89	14434.0877940569\\
1.99	14643.7099462519\\
};
\addplot [color=mycolor4, dashed, forget plot]
  table[row sep=crcr, y expr=1e-3*\thisrowno{1}*0.1]{%
-2.01	21551.0517174565\\
-1.91	21347.1823937646\\
-1.81	21117.9686476728\\
-1.71	20872.9620481756\\
-1.61	20610.0578861036\\
-1.51	20322.8840136019\\
-1.41	20010.4852008631\\
-1.31	19666.5442228817\\
-1.21	19287.3069196205\\
-1.11	18864.2248039382\\
-1.01	18390.0309628618\\
-0.91	17853.1032138856\\
-0.81	17238.8412005202\\
-0.71	16528.4375916751\\
-0.61	15697.8077901767\\
-0.51	14715.9317152365\\
-0.41	13542.8350538201\\
-0.36	12873.8006229702\\
-0.31	12143.6730969785\\
-0.26	11351.2657381817\\
-0.21	10503.0156488019\\
-0.16	9618.17320793898\\
-0.11	8740.65416622549\\
-0.06	7957.57299868725\\
-0.01	7411.37277510584\\
0.04	7273.09330775258\\
0.09	7624.11583617015\\
0.14	8390.1552104757\\
0.19	9407.78402384588\\
0.24	10535.8245498142\\
0.29	11685.7762599097\\
0.34	12809.0446187576\\
0.39	13880.5188031414\\
0.49	15829.1211606338\\
0.59	17516.9850499101\\
0.69	18972.5040359617\\
0.79	20233.0607561864\\
0.89	21334.6401691745\\
0.99	22339.6073932431\\
1.09	23234.9847038983\\
1.19	24018.0739358995\\
1.29	24744.678210324\\
1.39	25430.5167850228\\
1.49	26011.9961745226\\
1.59	26595.7460149463\\
1.69	27085.5909970972\\
1.79	27565.2742158922\\
1.89	28041.3245621931\\
1.99	28460.8395493138\\
};

\node[anchor=south, font=\footnotesize] at (rel axis cs: 0.5, 1) {$V_{\rm BG}=1$V};
\SubfigLetter{c)}{0.5}{0.1};

\end{axis}

%% file: Figuras/GFET_Vbg1V_Rdev_NoPuddles.tex
%
%
%
\begin{tikzpicture}

{\begin{axis}[%
width=0.3\linewidth,
height=3cm,
at={(0,0)},
scale only axis,
xmin=-2,
xmax=2,
xticklabels={,,},
ymin=0,
ymax=40,
ytick={0,10,...,50},
minor y tick num=1,
ylabel={R ($\Omega\cdot$ cm)},
axis background/.style={fill=white},
axis x line*=bottom,
axis y line*=left,
xmajorgrids,
ymajorgrids,
clip mode=individual,
legend style={at={(0.4\linewidth, -4.8cm)}, anchor=north ,legend cell align=left, align=left, draw=white!15!black, legend columns=4, font=\footnotesize}
]

\addplot [color=mycolor1]
table[row sep=crcr, y expr=\thisrowno{1}*10]{%
	-2	0.124712603883798\\
	-1.9	0.126578088643649\\
	-1.8	0.128402496802609\\
	-1.7	0.130193402858116\\
	-1.6	0.132618140382076\\
	-1.5	0.134858235848666\\
	-1.4	0.137842164279202\\
	-1.3	0.140760683601141\\
	-1.2	0.144592829303336\\
	-1.1	0.148563311883295\\
	-1	0.153529098268579\\
	-0.9	0.159559318409069\\
	-0.8	0.16711426330232\\
	-0.7	0.176883411695351\\
	-0.6	0.189347478125466\\
	-0.5	0.207316095340291\\
	-0.4	0.235654358330306\\
	-0.35	0.256980000210796\\
	-0.3	0.286787341418101\\
	-0.25	0.331020016405514\\
	-0.2	0.402051067852813\\
	-0.15	0.529522045830658\\
	-0.1	0.794647818792288\\
	-0.05	1.42679915532536\\
	0	2.3772665095426\\
	0.05	1.80933358407316\\
	0.1	1.10965585642\\
	0.15	0.793696900044358\\
	0.2	0.644846651101668\\
	0.25	0.565683189380771\\
	0.3	0.518544140456002\\
	0.35	0.4877071160282\\
	0.4	0.466021798482549\\
	0.5	0.438521294905129\\
	0.6	0.420843311523743\\
	0.7	0.409062766718536\\
	0.8	0.400270127628644\\
	0.9	0.393227134983497\\
	1	0.387574155912289\\
	1.1	0.383397439820231\\
	1.2	0.379645447739453\\
	1.3	0.375945079860884\\
	1.4	0.373434361594692\\
	1.5	0.370480804219682\\
	1.6	0.368623735368881\\
	1.7	0.366217830246854\\
	1.8	0.364752848738469\\
	1.9	0.363054910828503\\
	2	0.361227186947989\\
};
\addlegendentry{$V_{\rm BG} = -1$V};
\addplot [color=mycolor2]
table[row sep=crcr, y expr=\thisrowno{1}*10]{%
	-2	0.903862637883275\\
	-1.9	0.916135921959444\\
	-1.8	0.925594604526156\\
	-1.7	0.939496179330666\\
	-1.6	0.954229849856164\\
	-1.5	0.970217054309436\\
	-1.4	0.990873411257452\\
	-1.3	1.01006171366503\\
	-1.2	1.03102580526513\\
	-1.1	1.05153902795656\\
	-1	1.07839812209765\\
	-0.9	1.1114469970787\\
	-0.8	1.14349869731116\\
	-0.7	1.18343406409544\\
	-0.6	1.23324942801018\\
	-0.5	1.29702231085284\\
	-0.4	1.38782483680559\\
	-0.35	1.4518911591107\\
	-0.3	1.53747642672684\\
	-0.25	1.66226336221408\\
	-0.2	1.86589249825933\\
	-0.15	2.2593190397435\\
	-0.1	3.12433732716272\\
	-0.05	4.00094058115802\\
	0	2.84578325742006\\
	0.05	2.03853231582906\\
	0.1	1.66334198534656\\
	0.15	1.45797762521252\\
	0.2	1.3255710262024\\
	0.25	1.23516391906089\\
	0.3	1.16586242458337\\
	0.35	1.10884802626797\\
	0.4	1.06402182310826\\
	0.5	0.994214722381111\\
	0.6	0.94423641502416\\
	0.7	0.89933455944516\\
	0.8	0.867800939403477\\
	0.9	0.834996106825306\\
	1	0.812768402950485\\
	1.1	0.790146816626475\\
	1.2	0.769957282010221\\
	1.3	0.746384078400421\\
	1.4	0.726471637795559\\
	1.5	0.712395167487195\\
	1.6	0.696263052691475\\
	1.7	0.681552184003141\\
	1.8	0.667974345580825\\
	1.9	0.655483428267553\\
	2	0.641116208901405\\
};
\addlegendentry{$V_{\rm BG} = 0$V};
\addplot [color=mycolor3]
table[row sep=crcr, y expr=\thisrowno{1}*10]{%
	-2	0.332224718931403\\
	-1.9	0.333372223087102\\
	-1.8	0.335007398781272\\
	-1.7	0.337217037515935\\
	-1.6	0.338565125417558\\
	-1.5	0.34145979288476\\
	-1.4	0.343402167053931\\
	-1.3	0.346787969544824\\
	-1.2	0.350411274665952\\
	-1.1	0.354268867722977\\
	-1	0.359639386507308\\
	-0.9	0.366595810504317\\
	-0.8	0.375491146104334\\
	-0.7	0.387717220899506\\
	-0.6	0.406217878503009\\
	-0.5	0.436417554526223\\
	-0.4	0.492251939678483\\
	-0.35	0.54265781350148\\
	-0.3	0.626448093858013\\
	-0.25	0.783243807685799\\
	-0.2	1.11408312509586\\
	-0.15	1.84004099786854\\
	-0.1	2.32387143530808\\
	-0.05	1.34302187682469\\
	0	0.746939450753214\\
	0.05	0.49656427769039\\
	0.1	0.375132458745445\\
	0.15	0.306929592881777\\
	0.2	0.264345279336905\\
	0.25	0.235547585883477\\
	0.3	0.214912408238442\\
	0.35	0.19944461185335\\
	0.4	0.187442110938752\\
	0.5	0.170053581139118\\
	0.6	0.158074203629539\\
	0.7	0.14928621518106\\
	0.8	0.142534224399349\\
	0.9	0.137165886844502\\
	1	0.132774597891078\\
	1.1	0.129116525125077\\
	1.2	0.1259172555366\\
	1.3	0.123194357454728\\
	1.4	0.12087818130497\\
	1.5	0.118702333594077\\
	1.6	0.116894644337314\\
	1.7	0.115127380475133\\
	1.8	0.113594727861237\\
	1.9	0.112285745756088\\
	2	0.111007618809156\\
};
\addlegendentry{$V_{\rm BG} = 1$V};

\node[anchor=south, font=\footnotesize] at (rel axis cs: 0.5, 1) {R$_{\rm T}$};

\SubfigLetter{a)}{0.2}{0.9};

\draw (rel axis cs: 0.65, 0.38) ellipse (0.15cm and 0.15cm);
\draw[->] (rel axis cs: 0.7, 0.42) -- (rel axis cs: 0.82,0.42);

\draw (rel axis cs: 0.4, 0.1) ellipse (0.15cm and 0.2cm);
\draw[->] (rel axis cs: 0.32, 0.13) -- (rel axis cs: 0.2, 0.13);

\end{axis}

\begin{axis}[%
width=0.3\linewidth,
height=3cm,
at={(0,0)},
scale only axis,
xmin=-2,
xmax=2,
xticklabels={,,},
ymin=0,
ymax=4,
ytick={0,1,...,5},
minor y tick num=1,
axis y line*=right,
axis x line=none,
]

\addplot [color=mycolor1, dashed]
table[row sep=crcr, y expr=\thisrowno{1}*10]{%
	-2	0.0684093650881556\\
	-1.9	0.0694609534157627\\
	-1.8	0.0705611915541171\\
	-1.7	0.0717221463527891\\
	-1.6	0.0731299224191272\\
	-1.5	0.0745898549701305\\
	-1.4	0.0763601700362473\\
	-1.3	0.0782630429349725\\
	-1.2	0.0805806006628271\\
	-1.1	0.0831688309614338\\
	-1	0.0863059827597787\\
	-0.9	0.0901020929919077\\
	-0.8	0.094680543653632\\
	-0.7	0.10047043136316\\
	-0.6	0.108053799052018\\
	-0.5	0.118366180271374\\
	-0.4	0.13303228525529\\
	-0.35	0.142831524308804\\
	-0.3	0.154948680174125\\
	-0.25	0.170086125571254\\
	-0.2	0.188987889730973\\
	-0.15	0.21221704429978\\
	-0.1	0.238993950285457\\
	-0.05	0.264830946113177\\
	0	0.279509086799706\\
	0.05	0.274306869706932\\
	0.1	0.254354733253842\\
	0.15	0.230554517715521\\
	0.2	0.208764665258418\\
	0.25	0.190653958959609\\
	0.3	0.175996455297016\\
	0.35	0.16420451418515\\
	0.4	0.154653485460504\\
	0.5	0.140355281414399\\
	0.6	0.130302318127429\\
	0.7	0.122872891116528\\
	0.8	0.117178560706035\\
	0.9	0.112685600999717\\
	1	0.109051890133604\\
	1.1	0.10605070539319\\
	1.2	0.103515577461042\\
	1.3	0.101350576540739\\
	1.4	0.0994777304050328\\
	1.5	0.0978333440419492\\
	1.6	0.0963822511402647\\
	1.7	0.0950852385417358\\
	1.8	0.0939192784344847\\
	1.9	0.092870256586539\\
	2	0.0919422789476002\\
};

\addplot [color=mycolor2, dashed]
table[row sep=crcr, y expr=\thisrowno{1}*10]{%
	-2	0.1081979769967\\
	-1.9	0.109524374489376\\
	-1.8	0.110930510336236\\
	-1.7	0.112523170895281\\
	-1.6	0.114283718009704\\
	-1.5	0.116250804465066\\
	-1.4	0.118500293125134\\
	-1.3	0.121015967931311\\
	-1.2	0.123906248190533\\
	-1.1	0.127246414149359\\
	-1	0.13124506380179\\
	-0.9	0.136106181587804\\
	-0.8	0.142049104345544\\
	-0.7	0.149634889037317\\
	-0.6	0.159598668077156\\
	-0.5	0.173114535568881\\
	-0.4	0.19252347325736\\
	-0.35	0.205575323673035\\
	-0.3	0.221724925657578\\
	-0.25	0.241674057248038\\
	-0.2	0.265807938196392\\
	-0.15	0.292723499942474\\
	-0.1	0.316859613576182\\
	-0.05	0.326774561777032\\
	0	0.314913194201713\\
	0.05	0.28957048423002\\
	0.1	0.262277677027944\\
	0.15	0.238097704016028\\
	0.2	0.218107671196678\\
	0.25	0.201969191227312\\
	0.3	0.188886248702767\\
	0.35	0.178174249479807\\
	0.4	0.169363629501532\\
	0.5	0.155753777030467\\
	0.6	0.145825331233187\\
	0.7	0.138198423722049\\
	0.8	0.132249346997203\\
	0.9	0.127366272524325\\
	1	0.12338877747038\\
	1.1	0.119997081025486\\
	1.2	0.117086821369317\\
	1.3	0.11452124089641\\
	1.4	0.112294703220288\\
	1.5	0.110350296056052\\
	1.6	0.108576256408451\\
	1.7	0.106972280053194\\
	1.8	0.105514365370186\\
	1.9	0.104177252938851\\
	2	0.102916483309369\\
};

\addplot [color=mycolor3, dashed]
table[row sep=crcr, y expr=\thisrowno{1}*10]{%
	-2	0.0896924222359484\\
	-1.9	0.0907417577250208\\
	-1.8	0.0918597122334094\\
	-1.7	0.0931021345915911\\
	-1.6	0.0944985655425448\\
	-1.5	0.0960755107798504\\
	-1.4	0.0978911534947835\\
	-1.3	0.0999902269663433\\
	-1.2	0.102444725268977\\
	-1.1	0.105390755310072\\
	-1	0.108956490692626\\
	-0.9	0.113368788807973\\
	-0.8	0.119038307895309\\
	-0.7	0.126498668344513\\
	-0.6	0.136673198459123\\
	-0.5	0.151212496930425\\
	-0.4	0.172941443027282\\
	-0.35	0.187843336694416\\
	-0.3	0.206202256484804\\
	-0.25	0.228105396758036\\
	-0.2	0.251679338549325\\
	-0.15	0.27084037269019\\
	-0.1	0.274734659302736\\
	-0.05	0.258944137005241\\
	0	0.232622704955459\\
	0.05	0.205889122843224\\
	0.1	0.182896111373565\\
	0.15	0.164239450923272\\
	0.2	0.149353306113767\\
	0.25	0.137439425946981\\
	0.3	0.127804297204816\\
	0.35	0.119918060666034\\
	0.4	0.11337994069006\\
	0.5	0.103230172269721\\
	0.6	0.0957699868853388\\
	0.7	0.0900833614647051\\
	0.8	0.0856087970284794\\
	0.9	0.0819991766808309\\
	1	0.0790238893247543\\
	1.1	0.0765227955602368\\
	1.2	0.0743743898362844\\
	1.3	0.0725224390100205\\
	1.4	0.0709165012504551\\
	1.5	0.0694776001117364\\
	1.6	0.0682202637150236\\
	1.7	0.0670684165674304\\
	1.8	0.06603981389613\\
	1.9	0.0651277172732062\\
	2	0.0642784143186008\\
};

\end{axis}
}

{\begin{axis}[%
width=0.3\linewidth,
height=3cm,
at={(0.5\linewidth,0)},
scale only axis,
xmin=-2,
xmax=2,
xticklabels={,,},
ymin=0,
ymax=25,
ytick={0,5,...,25},
minor y tick num=1,
axis background/.style={fill=white},
axis x line*=bottom,
axis y line*=left,
xmajorgrids,
ymajorgrids,
clip mode=individual,
legend style={at={(0,-0.3)}, anchor=north west ,legend cell align=left, align=left, draw=white!15!black, legend columns=4, font=\footnotesize}
]

\addplot [color=mycolor1]
table[row sep=crcr, y expr=\thisrowno{1}*10]{%
	-2	0.00913868482784443\\
	-1.9	0.00975096120441273\\
	-1.8	0.0104437618776628\\
	-1.7	0.0112333719144071\\
	-1.6	0.0121427811007235\\
	-1.5	0.0131976731949362\\
	-1.4	0.0144354167678202\\
	-1.3	0.015905698540974\\
	-1.2	0.0176753495458256\\
	-1.1	0.019843808960996\\
	-1	0.0225484741433007\\
	-0.9	0.026007793498295\\
	-0.8	0.0305629559880753\\
	-0.7	0.0367759885711263\\
	-0.6	0.0456654616262697\\
	-0.5	0.0592176128649881\\
	-0.4	0.081799588566666\\
	-0.35	0.0994791209097789\\
	-0.3	0.124877381072401\\
	-0.25	0.163603725256094\\
	-0.2	0.2274799630961\\
	-0.15	0.34493779794157\\
	-0.1	0.594824320813138\\
	-0.05	1.2009827250099\\
	0	2.10823713818243\\
	0.05	1.4917201488246\\
	0.1	0.745118976333711\\
	0.15	0.410041366043158\\
	0.2	0.259574248032878\\
	0.25	0.181708204552024\\
	0.3	0.136190243774969\\
	0.35	0.107022171989759\\
	0.4	0.087069503788345\\
	0.5	0.0621263721050965\\
	0.6	0.0474580998173051\\
	0.7	0.0379729241005337\\
	0.8	0.0314056478240492\\
	0.9	0.0266317299762959\\
	1	0.0230265788360734\\
	1.1	0.0202198322776738\\
	1.2	0.0179781638131539\\
	1.3	0.0161538133858802\\
	1.4	0.0146433795034442\\
	1.5	0.0133728551449111\\
	1.6	0.0122933425873614\\
	1.7	0.0113633815739198\\
	1.8	0.0105570195120847\\
	1.9	0.00985077047199362\\
	2	0.00922708016391087\\
};

\addplot [color=mycolor2]
table[row sep=crcr, y expr=\thisrowno{1}*10]{%
	-2	0.0095030400664451\\
	-1.9	0.0101632561415483\\
	-1.8	0.0109129446764699\\
	-1.7	0.0117736262395219\\
	-1.6	0.012769100024045\\
	-1.5	0.0139333249832313\\
	-1.4	0.0153103884484032\\
	-1.3	0.0169592931276941\\
	-1.2	0.0189638959480874\\
	-1.1	0.0214483968023706\\
	-1	0.0245963328076252\\
	-0.9	0.028704723907887\\
	-0.8	0.0342206452620806\\
	-0.7	0.0419759169440991\\
	-0.6	0.0535068598210513\\
	-0.5	0.0720271302933395\\
	-0.4	0.105444369663042\\
	-0.35	0.133591310407718\\
	-0.3	0.176876355750485\\
	-0.25	0.250248845865365\\
	-0.2	0.388315634935112\\
	-0.15	0.696914131191403\\
	-0.1	1.47782085456081\\
	-0.05	2.36135474763794\\
	0	1.30717358676852\\
	0.05	0.624151455685058\\
	0.1	0.357853137777901\\
	0.15	0.234483108866774\\
	0.2	0.167707135221525\\
	0.25	0.127656086787737\\
	0.3	0.101298816282018\\
	0.35	0.0830424294155383\\
	0.4	0.069837321531607\\
	0.5	0.0521955759741577\\
	0.6	0.0411362336291155\\
	0.7	0.0336271512358546\\
	0.8	0.0282773807569287\\
	0.9	0.0242742704598859\\
	1	0.0212056999746096\\
	1.1	0.0187700100583023\\
	1.2	0.0168015303975054\\
	1.3	0.0151751154931692\\
	1.4	0.013818070380617\\
	1.5	0.0126725078141119\\
	1.6	0.011689159421528\\
	1.7	0.0108379715449664\\
	1.8	0.0100952918440422\\
	1.9	0.00944234141396283\\
	2	0.00886233452561202\\
};

\addplot [color=mycolor3]
table[row sep=crcr, y expr=\thisrowno{1}*10]{%
	-2	0.00988837250808473\\
	-1.9	0.0105997592490296\\
	-1.8	0.0114128880584204\\
	-1.7	0.0123503667089172\\
	-1.6	0.0134398600916213\\
	-1.5	0.0147220995007912\\
	-1.4	0.0162486471425665\\
	-1.3	0.0180930234223897\\
	-1.2	0.0203613999205091\\
	-1.1	0.0232080267241036\\
	-1	0.0268640746633644\\
	-0.9	0.0317116860940871\\
	-0.8	0.0383926344625801\\
	-0.7	0.0480795409279328\\
	-0.6	0.0631024446033957\\
	-0.5	0.0888238734119201\\
	-0.4	0.139760785952472\\
	-0.35	0.187482330994743\\
	-0.3	0.269776074190525\\
	-0.25	0.430204336280836\\
	-0.2	0.784343664620775\\
	-0.15	1.54821173305879\\
	-0.1	2.07391069060169\\
	-0.05	1.13753453830028\\
	0	0.568082710532792\\
	0.05	0.332776544634626\\
	0.1	0.221004480872261\\
	0.15	0.159723762143133\\
	0.2	0.122365378792632\\
	0.25	0.0977423411484728\\
	0.3	0.0805474783134387\\
	0.35	0.0679871812869464\\
	0.4	0.0584734238021077\\
	0.5	0.0451685376452295\\
	0.6	0.0364182477591979\\
	0.7	0.0302962709411221\\
	0.8	0.0258035622789069\\
	0.9	0.0223875363921715\\
	1	0.0197152154488755\\
	1.1	0.0175696887167412\\
	1.2	0.0158191564162962\\
	1.3	0.0143622474224094\\
	1.4	0.0131353135821923\\
	1.5	0.0120890303176162\\
	1.6	0.0111867202474274\\
	1.7	0.0104030889428739\\
	1.8	0.00971493000513579\\
	1.9	0.00910670698486418\\
	2	0.00856629267460917\\
};

\node[anchor=south, font=\footnotesize] at (rel axis cs: 0.5, 1) {R$_{\rm Chn}$};

\SubfigLetter{b)}{0.2}{0.9};

\draw (rel axis cs: 0.65, 0.15) ellipse (0.15cm and 0.15cm);
\draw[->] (rel axis cs: 0.7, 0.19) -- (rel axis cs: 0.82, 0.19);

\draw (rel axis cs: 0.42, 0.08) ellipse (0.1cm and 0.15cm);
\draw[->] (rel axis cs: 0.37, 0.05) -- (rel axis cs: 0.25, 0.05);

\end{axis}

\begin{axis}[%
width=0.3\linewidth,
height=3cm,
at={(0.5\linewidth,0)},
scale only axis,
xmin=-2,
xmax=2,
xticklabels={,,},
ymin=0,
ymax=2.5,
ytick={0, 0.5 , 1, 1.5, 2, 2.5},
\RemoveTickExp,
minor y tick num=1,
axis y line*=right,
axis x line=none,
]

\addplot [color=mycolor1, dashed]
table[row sep=crcr, y expr=\thisrowno{1}*10]{%
	-2	0.00875937065072184\\
	-1.9	0.00932001863614795\\
	-1.8	0.00995079476388735\\
	-1.7	0.0106651618008308\\
	-1.6	0.0114811292457228\\
	-1.5	0.0124198045798468\\
	-1.4	0.0135093450770877\\
	-1.3	0.0147886893452033\\
	-1.2	0.0163059347878545\\
	-1.1	0.0181338355391131\\
	-1	0.0203657687976874\\
	-0.9	0.023145235436\\
	-0.8	0.0266840601010213\\
	-0.7	0.0313035842193172\\
	-0.6	0.0375252876735924\\
	-0.5	0.0462240501479439\\
	-0.4	0.0589362227410112\\
	-0.35	0.0675919544716788\\
	-0.3	0.0784269772744154\\
	-0.25	0.0921187786484918\\
	-0.2	0.109390205594819\\
	-0.15	0.130758982681125\\
	-0.1	0.155448926342495\\
	-0.05	0.179001633583026\\
	0	0.191300133898721\\
	0.05	0.183843703719104\\
	0.1	0.162031439209705\\
	0.15	0.136959789326306\\
	0.2	0.114493323414226\\
	0.25	0.0961665909182841\\
	0.3	0.0815942072016459\\
	0.35	0.0700771506462238\\
	0.4	0.0609073578060811\\
	0.5	0.0475166963244545\\
	0.6	0.038415358212955\\
	0.7	0.0319411445528862\\
	0.8	0.0271559747382482\\
	0.9	0.0235068549165334\\
	1	0.0206499183814044\\
	1.1	0.0183610497431087\\
	1.2	0.0164920954584724\\
	1.3	0.0149433942927386\\
	1.4	0.013639707467842\\
	1.5	0.0125308079352774\\
	1.6	0.0115769850826698\\
	1.7	0.0107485224940382\\
	1.8	0.0100236529469748\\
	1.9	0.00938453376735903\\
	2	0.00881675144092703\\
};

\addplot [color=mycolor2, dashed]
table[row sep=crcr, y expr=\thisrowno{1}*10]{%
	-2	0.009077571276224\\
	-1.9	0.0096775548202516\\
	-1.8	0.0103555490223635\\
	-1.7	0.0111267886569402\\
	-1.6	0.0120109717549928\\
	-1.5	0.0130345202005572\\
	-1.4	0.0142292134675241\\
	-1.3	0.0156411614520505\\
	-1.2	0.0173295450783126\\
	-1.1	0.0193814089517812\\
	-1	0.0219131921776079\\
	-0.9	0.0251065692185917\\
	-0.8	0.0292306571464522\\
	-0.7	0.0347083410909778\\
	-0.6	0.0422224164194577\\
	-0.5	0.0529735585841938\\
	-0.4	0.0691601314359356\\
	-0.35	0.0803934896694736\\
	-0.3	0.0945888967306238\\
	-0.25	0.112487068637731\\
	-0.2	0.134570707531389\\
	-0.15	0.159649145099957\\
	-0.1	0.182541330925355\\
	-0.05	0.192179197915119\\
	0	0.18126066965505\\
	0.05	0.157504625251404\\
	0.1	0.132453108232251\\
	0.15	0.110694630027397\\
	0.2	0.0931100579034564\\
	0.25	0.0791957167873691\\
	0.3	0.0681796589590232\\
	0.35	0.0593875377532014\\
	0.4	0.0523051665869251\\
	0.5	0.0417292746244756\\
	0.6	0.0343379107782623\\
	0.7	0.0289507012599692\\
	0.8	0.0248917076335318\\
	0.9	0.0217423477818968\\
	1	0.0192445231452461\\
	1.1	0.0172179093914678\\
	1.2	0.0155479586866898\\
	1.3	0.0141506921165238\\
	1.4	0.012967467412877\\
	1.5	0.0119537440903061\\
	1.6	0.0110772003318061\\
	1.7	0.0103119190171701\\
	1.8	0.00963896700475698\\
	1.9	0.00904334690934497\\
	2	0.00851261611400393\\
};

\addplot [color=mycolor3, dashed]
table[row sep=crcr, y expr=\thisrowno{1}*10]{%
	-2	0.00941961115532268\\
	-1.9	0.0100633300046347\\
	-1.8	0.0107937892714205\\
	-1.7	0.0116291042892634\\
	-1.6	0.0125908620508367\\
	-1.5	0.0137101977147744\\
	-1.4	0.0150262396104495\\
	-1.3	0.0165917940355147\\
	-1.2	0.018481378457472\\
	-1.1	0.0207991552154708\\
	-1	0.0236929650361132\\
	-0.9	0.0273907939486988\\
	-0.8	0.0322478932898184\\
	-0.7	0.038831531959501\\
	-0.6	0.0481037351222892\\
	-0.5	0.0617747061834508\\
	-0.4	0.0828990213933537\\
	-0.35	0.0977443373714758\\
	-0.3	0.116357463894886\\
	-0.25	0.138954263156021\\
	-0.2	0.163780573704118\\
	-0.15	0.184777591975234\\
	-0.1	0.190914697495578\\
	-0.05	0.177525839612799\\
	0	0.15353227239084\\
	0.05	0.128924443677984\\
	0.1	0.107810484268793\\
	0.15	0.0908133804627457\\
	0.2	0.0773721709742086\\
	0.25	0.0667365734487357\\
	0.3	0.0582372725958281\\
	0.35	0.0513674437975778\\
	0.4	0.0457443746164882\\
	0.5	0.0371748147959986\\
	0.6	0.0310378542795673\\
	0.7	0.0264753999313186\\
	0.8	0.0229785746329401\\
	0.9	0.0202293621350009\\
	1	0.0180218078666616\\
	1.1	0.0162127042513825\\
	1.2	0.0147104385245673\\
	1.3	0.0134427514560618\\
	1.4	0.0123625751655366\\
	1.5	0.0114313835744562\\
	1.6	0.0106217450338156\\
	1.7	0.00991268290641927\\
	1.8	0.00928609092221224\\
	1.9	0.0087291194109623\\
	2	0.00823146249567683\\
};

\node[anchor=south, font=\footnotesize] at (rel axis cs: 0.5, 1) {R$_{\rm Chn}$};

\end{axis}

}

{\begin{axis}[%
width=0.3\linewidth,
height=3cm,
at={(0\linewidth,-3.8cm)},
scale only axis,
xmin=-2,
xmax=2,
xlabel={$V_{\rm FG}$ (V)},
ymin=0,
ymax=10,
minor y tick num=1,
ylabel={R ($\Omega\cdot$ cm)},
axis background/.style={fill=white},
axis x line*=bottom,
axis y line*=left,
xmajorgrids,
ymajorgrids,
clip mode=individual,
legend style={at={(0.4\linewidth,-1.6cm)}, anchor=north,legend cell align=left, align=left, draw=white!15!black, legend columns=4, font=\footnotesize}
]

\addplot [color=mycolor1]
table[row sep=crcr, y expr=\thisrowno{1}*10]{%
	-2	0.0593087298773239\\
	-1.9	0.0598430488581699\\
	-1.8	0.0603959708328404\\
	-1.7	0.0609641829028382\\
	-1.6	0.0616369304457395\\
	-1.5	0.062303048442403\\
	-1.4	0.0631032551802072\\
	-1.3	0.0639057668653403\\
	-1.2	0.0648823481624828\\
	-1.1	0.0658745574913975\\
	-1	0.0670386062085359\\
	-0.9	0.0683807222260335\\
	-0.8	0.0699655063405795\\
	-0.7	0.0718786420054206\\
	-0.6	0.0740366433820598\\
	-0.5	0.0767641101768847\\
	-0.4	0.0804057218949096\\
	-0.35	0.0827630970282313\\
	-0.3	0.0856650336835362\\
	-0.25	0.0893664700222367\\
	-0.2	0.0942911629557038\\
	-0.15	0.101395724586583\\
	-0.1	0.112664671586322\\
	-0.05	0.132604393453894\\
	0	0.165007083882006\\
	0.05	0.190970784429125\\
	0.1	0.200948400589867\\
	0.15	0.20432818033972\\
	0.2	0.204875828010283\\
	0.25	0.204584215900199\\
	0.3	0.204198350269337\\
	0.35	0.203629077978275\\
	0.4	0.202921939549142\\
	0.5	0.201986368191749\\
	0.6	0.200469577639067\\
	0.7	0.199536244814931\\
	0.8	0.198481169767992\\
	0.9	0.197351291943122\\
	1	0.196359320483376\\
	1.1	0.195922644828361\\
	1.2	0.195088333944618\\
	1.3	0.194109487835315\\
	1.4	0.193834051126015\\
	1.5	0.192863942852627\\
	1.6	0.192657533591846\\
	1.7	0.191837357999769\\
	1.8	0.191621538141597\\
	1.9	0.191243752400731\\
	2	0.190459426204788\\
};
\addlegendentry{No puddles};
\addplot [color=mycolor1, dashed]
table[row sep=crcr, y expr=\thisrowno{1}*10]{%
	-2	0.0593087298773239\\
	-1.9	0.0598430488581699\\
	-1.8	0.0603959708328404\\
	-1.7	0.0609641829028382\\
	-1.6	0.0616369304457395\\
	-1.5	0.062303048442403\\
	-1.4	0.0631032551802072\\
	-1.3	0.0639057668653403\\
	-1.2	0.0648823481624828\\
	-1.1	0.0658745574913975\\
	-1	0.0670386062085359\\
	-0.9	0.0683807222260335\\
	-0.8	0.0699655063405795\\
	-0.7	0.0718786420054206\\
	-0.6	0.0740366433820598\\
	-0.5	0.0767641101768847\\
	-0.4	0.0804057218949096\\
	-0.35	0.0827630970282313\\
	-0.3	0.0856650336835362\\
	-0.25	0.0893664700222367\\
	-0.2	0.0942911629557038\\
	-0.15	0.101395724586583\\
	-0.1	0.112664671586322\\
	-0.05	0.132604393453894\\
	0	0.165007083882006\\
	0.05	0.190970784429125\\
	0.1	0.200948400589867\\
	0.15	0.20432818033972\\
	0.2	0.204875828010283\\
	0.25	0.204584215900199\\
	0.3	0.204198350269337\\
	0.35	0.203629077978275\\
	0.4	0.202921939549142\\
	0.5	0.201986368191749\\
	0.6	0.200469577639067\\
	0.7	0.199536244814931\\
	0.8	0.198481169767992\\
	0.9	0.197351291943122\\
	1	0.196359320483376\\
	1.1	0.195922644828361\\
	1.2	0.195088333944618\\
	1.3	0.194109487835315\\
	1.4	0.193834051126015\\
	1.5	0.192863942852627\\
	1.6	0.192657533591846\\
	1.7	0.191837357999769\\
	1.8	0.191621538141597\\
	1.9	0.191243752400731\\
	2	0.190459426204788\\
};
\addlegendentry{$N_{\rm p} = 10^{12}$cm$^{-2}$};

\addplot [color=mycolor2]
table[row sep=crcr, y expr=\thisrowno{1}*10]{%
	-2	0.42517861601909\\
	-1.9	0.429768478700018\\
	-1.8	0.434774586343483\\
	-1.7	0.439976985623566\\
	-1.6	0.445476897615839\\
	-1.5	0.451444401626784\\
	-1.4	0.457770039500934\\
	-1.3	0.464815686783353\\
	-1.2	0.472548159999623\\
	-1.1	0.481052229715007\\
	-1	0.490578899642816\\
	-0.9	0.501310739175477\\
	-0.8	0.513674423745729\\
	-0.7	0.528039547804874\\
	-0.6	0.545094855298687\\
	-0.5	0.566585428380104\\
	-0.4	0.594880051850705\\
	-0.35	0.612717917688613\\
	-0.3	0.634147242301153\\
	-0.25	0.661824807991889\\
	-0.2	0.698604238500515\\
	-0.15	0.747579400775226\\
	-0.1	0.787437348056311\\
	-0.05	0.728884929800277\\
	0	0.676563170287457\\
	0.05	0.655549321528156\\
	0.1	0.626199879732631\\
	0.15	0.598968502875787\\
	0.2	0.573242613907786\\
	0.25	0.553221570761837\\
	0.3	0.534602913551266\\
	0.35	0.515966639793098\\
	0.4	0.501213045418527\\
	0.5	0.476247334102543\\
	0.6	0.458584348878166\\
	0.7	0.438494818771412\\
	0.8	0.426875072183247\\
	0.9	0.410416640125441\\
	1	0.402287839487728\\
	1.1	0.391753852611624\\
	1.2	0.382151093170198\\
	1.3	0.367546697957652\\
	1.4	0.355666504642404\\
	1.5	0.349298719693098\\
	1.6	0.339763691660302\\
	1.7	0.331038622475599\\
	1.8	0.322963571150596\\
	1.9	0.31550469454852\\
	2	0.304995122954978\\
};

\addplot [color=mycolor3]
table[row sep=crcr, y expr=\thisrowno{1}*10]{%
	-2	0.148099887522752\\
	-1.9	0.148363921737103\\
	-1.8	0.148748844230128\\
	-1.7	0.149514434088783\\
	-1.6	0.149605506645541\\
	-1.5	0.150488368837559\\
	-1.4	0.150689908430435\\
	-1.3	0.151628719870035\\
	-1.2	0.15215823440399\\
	-1.1	0.152752641457887\\
	-1	0.153674902175562\\
	-0.9	0.154757301001858\\
	-0.8	0.155774308283164\\
	-0.7	0.156949376386761\\
	-0.6	0.158651103285957\\
	-0.5	0.160684790367486\\
	-0.4	0.163051106780586\\
	-0.35	0.164285044391095\\
	-0.3	0.164282177829772\\
	-0.25	0.159680143564244\\
	-0.2	0.136727577307764\\
	-0.15	0.10539416300116\\
	-0.1	0.0891005174465976\\
	-0.05	0.0809272467592721\\
	0	0.0760243952761981\\
	0.05	0.072666732458473\\
	0.1	0.0701812518603259\\
	0.15	0.0682045834629152\\
	0.2	0.0665942359409336\\
	0.25	0.0652337859658146\\
	0.3	0.0640621177415209\\
	0.35	0.0630372530490764\\
	0.4	0.06212905386113\\
	0.5	0.0605692814515976\\
	0.6	0.0592508542995007\\
	0.7	0.0581100160414657\\
	0.8	0.0571064287457089\\
	0.9	0.0562112396107312\\
	1	0.0554076803229492\\
	1.1	0.0546612396206406\\
	1.2	0.0540034092837513\\
	1.3	0.0533635034704072\\
	1.4	0.0527342969632655\\
	1.5	0.0522255920676107\\
	1.6	0.051680375468529\\
	1.7	0.0512428127034425\\
	1.8	0.0507976051459398\\
	1.9	0.0503503441946526\\
	2	0.0499606484754695\\
};

\node[anchor=south, font=\footnotesize] at (rel axis cs: 0.5, 1) {R$_{\rm S, Acc}$};

\SubfigLetter{c)}{0.2}{0.9};
\draw (rel axis cs: 0.45, 0.38) ellipse (0.15cm and 0.3cm);
\draw[->] (rel axis cs: 0.48, 0.28) -- (rel axis cs: 0.6,0.28);

\draw (rel axis cs: 0.45, 0.13) ellipse (0.15cm and 0.25cm);
\draw[->] (rel axis cs: 0.4, 0.2) -- (rel axis cs: 0.28,0.2);

\end{axis}

\begin{axis}[%
width=0.3\linewidth,
height=3cm,
at={(0\linewidth,-3.8cm)},
scale only axis,
xmin=-2,
xmax=2,
xticklabels={,,},
ymin=0,
ymax=1,
\RemoveTickExp,
minor y tick num=1,
axis y line*=right,
axis x line=none,
]

\addplot [color=mycolor1, dashed]
table[row sep=crcr, y expr=\thisrowno{1}*10]{%
	-2	0.0302175296975595\\
	-1.9	0.030447190402836\\
	-1.8	0.0306816371284771\\
	-1.7	0.0309202173427429\\
	-1.6	0.0312032512568406\\
	-1.5	0.0314810379295038\\
	-1.4	0.0318126519985223\\
	-1.3	0.032143899452968\\
	-1.2	0.0325416144840342\\
	-1.1	0.0329464118566455\\
	-1	0.0334160802714016\\
	-0.9	0.0339488228410209\\
	-0.8	0.0345201147667989\\
	-0.7	0.0351714076507621\\
	-0.6	0.0359345740583589\\
	-0.5	0.0368462363938249\\
	-0.4	0.037959873093273\\
	-0.35	0.0386156181869357\\
	-0.3	0.0393533987625567\\
	-0.25	0.0401871508308483\\
	-0.2	0.0411240072500899\\
	-0.15	0.0421842071991568\\
	-0.1	0.0433562632501457\\
	-0.05	0.0445834530302973\\
	0	0.0457573433029989\\
	0.05	0.0467243929428811\\
	0.1	0.0474032739565256\\
	0.15	0.0477990775227047\\
	0.2	0.0479687534496041\\
	0.25	0.0479782773568162\\
	0.3	0.0478837800417826\\
	0.35	0.0477208670724969\\
	0.4	0.0475184455761992\\
	0.5	0.0470613841332659\\
	0.6	0.0465833557200973\\
	0.7	0.0461094682087568\\
	0.8	0.0456543469785165\\
	0.9	0.0452314627035056\\
	1	0.044844179498598\\
	1.1	0.0444871560932083\\
	1.2	0.0441546421276898\\
	1.3	0.0438498570430442\\
	1.4	0.0435638035400638\\
	1.5	0.0433001628320619\\
	1.6	0.0430506654347967\\
	1.7	0.0428212182320455\\
	1.8	0.0425998002399548\\
	1.9	0.0423970391627714\\
	2	0.0422165166209043\\
};

\addplot [color=mycolor2, dashed]
table[row sep=crcr, y expr=\thisrowno{1}*10]{%
	-2	0.0493787311124097\\
	-1.9	0.0497291082183848\\
	-1.8	0.0501066293672966\\
	-1.7	0.0505047066325052\\
	-1.6	0.0509300167593018\\
	-1.5	0.0513893430798031\\
	-1.4	0.0518786006196992\\
	-1.3	0.0524191733072635\\
	-1.2	0.0530097954808726\\
	-1.1	0.0536652975219784\\
	-1	0.0543884219717832\\
	-0.9	0.0551938893013566\\
	-0.8	0.0561170814109074\\
	-0.7	0.05716677845551\\
	-0.6	0.0583736328717692\\
	-0.5	0.0598062322527813\\
	-0.4	0.0615181055498722\\
	-0.35	0.0624810080043752\\
	-0.3	0.0635047105750857\\
	-0.25	0.0645548405219367\\
	-0.2	0.0655448239816984\\
	-0.15	0.066314074265116\\
	-0.1	0.0666030859267617\\
	-0.05	0.066286076839187\\
	0	0.0654377246697239\\
	0.05	0.0644872676257449\\
	0.1	0.063408485764347\\
	0.15	0.0623459335714357\\
	0.2	0.0613110120887495\\
	0.25	0.0603723003614788\\
	0.3	0.0594889562913925\\
	0.35	0.0586474415277496\\
	0.4	0.0578909605401054\\
	0.5	0.0565392289488896\\
	0.6	0.0554033199843403\\
	0.7	0.0543474618596701\\
	0.8	0.0534806877921638\\
	0.9	0.0526301903570286\\
	1	0.0519429845594345\\
	1.1	0.0512783952172215\\
	1.2	0.0506681744434947\\
	1.3	0.0500616954837328\\
	1.4	0.0495392614500129\\
	1.5	0.0491034978185659\\
	1.6	0.0486502685606589\\
	1.7	0.0482257304852703\\
	1.8	0.0478239361502989\\
	1.9	0.0474439912529649\\
	2	0.0470316081163651\\
};

\addplot [color=mycolor3, dashed]
table[row sep=crcr, y expr=\thisrowno{1}*10]{%
	-2	0.0394651653385623\\
	-1.9	0.03966025725366\\
	-1.8	0.0398528227714299\\
	-1.7	0.0400507413867802\\
	-1.6	0.0402674652741487\\
	-1.5	0.0404888880979544\\
	-1.4	0.040734752554621\\
	-1.3	0.0409901849594488\\
	-1.2	0.0412630705878576\\
	-1.1	0.0415594551525064\\
	-1	0.0418709101380846\\
	-0.9	0.0421908553465699\\
	-0.8	0.0425623568188538\\
	-0.7	0.0429521008953869\\
	-0.6	0.0433371746195463\\
	-0.5	0.0436693381006116\\
	-0.4	0.0437901204886636\\
	-0.35	0.0436720530621971\\
	-0.3	0.0433401727180397\\
	-0.25	0.0427203621318243\\
	-0.2	0.0417821910259498\\
	-0.15	0.0406155446743352\\
	-0.1	0.0394031682176475\\
	-0.05	0.0382877700613863\\
	0	0.0373149924863607\\
	0.05	0.0364745311458477\\
	0.1	0.0357530997281643\\
	0.15	0.0351213748519433\\
	0.2	0.0345685792995173\\
	0.25	0.0340762068686866\\
	0.3	0.0336344295255917\\
	0.35	0.0332349417206812\\
	0.4	0.032871801525943\\
	0.5	0.0322340914348261\\
	0.6	0.0316859277646217\\
	0.7	0.0312053330255886\\
	0.8	0.0307767208105161\\
	0.9	0.030390575369914\\
	1	0.0300401403709905\\
	1.1	0.0297184776499944\\
	1.2	0.0294222616505206\\
	1.3	0.0291448828606174\\
	1.4	0.0288824617869344\\
	1.5	0.0286464285269265\\
	1.6	0.0284166866505222\\
	1.7	0.0282102329069784\\
	1.8	0.0280114177623207\\
	1.9	0.027818446206854\\
	2	0.0276410186652332\\
};

\end{axis}

}

{\begin{axis}[%
width=0.3\linewidth,
height=3cm,
at={(0.5\linewidth,-3.8cm)},
scale only axis,
xmin=-2,
xmax=2,
xlabel={$V_{\rm FG}$ (V)},
ymin=0,
ymax=10,
minor y tick num=1,
axis background/.style={fill=white},
axis x line*=bottom,
axis y line*=left,
xmajorgrids,
ymajorgrids,
clip mode=individual,
legend style={at={(0.15,-0.3)}, anchor=north west ,legend cell align=left, align=left, draw=white!15!black, legend columns=4, font=\footnotesize}
]
\addplot [color=mycolor1]
table[row sep=crcr, y expr=\thisrowno{1}*10]{%
	-2	0.0562651891786298\\
	-1.9	0.0569840785810668\\
	-1.8	0.0575627640921058\\
	-1.7	0.057995848040871\\
	-1.6	0.0588384288356135\\
	-1.5	0.0593575142113263\\
	-1.4	0.0603034923311748\\
	-1.3	0.0609492181948269\\
	-1.2	0.0620351315950279\\
	-1.1	0.0628449454309013\\
	-1	0.0639420179167429\\
	-0.9	0.0651708026847403\\
	-0.8	0.0665858009736652\\
	-0.7	0.0682287811188045\\
	-0.6	0.0696453731171369\\
	-0.5	0.0713343722984182\\
	-0.4	0.0734490478687303\\
	-0.35	0.0747377822727861\\
	-0.3	0.0762449266621638\\
	-0.25	0.078049821127183\\
	-0.2	0.0802799418010096\\
	-0.15	0.083188523302506\\
	-0.1	0.0871588263928288\\
	-0.05	0.0932120368615743\\
	0	0.104022287478168\\
	0.05	0.126642650819434\\
	0.1	0.163588479496421\\
	0.15	0.17932735366148\\
	0.2	0.180396575058507\\
	0.25	0.179390768928549\\
	0.3	0.178155546411696\\
	0.35	0.177055866060167\\
	0.4	0.176030355145062\\
	0.5	0.174408554608283\\
	0.6	0.172915634067371\\
	0.7	0.171553597803071\\
	0.8	0.170383310036603\\
	0.9	0.169244113064079\\
	1	0.16818825659284\\
	1.1	0.167254962714197\\
	1.2	0.166578949981681\\
	1.3	0.165681778639688\\
	1.4	0.164956930965233\\
	1.5	0.164244006222143\\
	1.6	0.163672859189673\\
	1.7	0.163017090673164\\
	1.8	0.162574291084787\\
	1.9	0.161960387955778\\
	2	0.16154068057929\\
};

\addplot [color=mycolor2]
table[row sep=crcr, y expr=\thisrowno{1}*10]{%
	-2	0.46918098179774\\
	-1.9	0.476204187117878\\
	-1.8	0.479907073506203\\
	-1.7	0.487745567467578\\
	-1.6	0.49598385221628\\
	-1.5	0.504839327699421\\
	-1.4	0.517792983308115\\
	-1.3	0.528286733753985\\
	-1.2	0.539513749317416\\
	-1.1	0.549038401439184\\
	-1	0.563222889647204\\
	-0.9	0.581431533995336\\
	-0.8	0.59560362830335\\
	-0.7	0.613418599346464\\
	-0.6	0.63464771289044\\
	-0.5	0.658409752179397\\
	-0.4	0.687500415291846\\
	-0.35	0.705581931014373\\
	-0.3	0.726452828675198\\
	-0.25	0.750189708356824\\
	-0.2	0.7789726248237\\
	-0.15	0.814825507776869\\
	-0.1	0.859079124545597\\
	-0.05	0.910700903719801\\
	0	0.862046500364081\\
	0.05	0.758831538615849\\
	0.1	0.679288967836025\\
	0.15	0.624526013469963\\
	0.2	0.584621277073092\\
	0.25	0.554286261511316\\
	0.3	0.529960694750083\\
	0.35	0.509838957059337\\
	0.4	0.492971456158124\\
	0.5	0.46577181230441\\
	0.6	0.444515832516879\\
	0.7	0.427212589437893\\
	0.8	0.412648486463301\\
	0.9	0.400305196239979\\
	1	0.389274863488148\\
	1.1	0.379622953956548\\
	1.2	0.371004658442518\\
	1.3	0.3636622649496\\
	1.4	0.356987062772538\\
	1.5	0.350423939979985\\
	1.6	0.344810201609644\\
	1.7	0.339675589982576\\
	1.8	0.334915482586186\\
	1.9	0.33053639230507\\
	2	0.327258751420815\\
};

\addplot [color=mycolor3]
table[row sep=crcr, y expr=\thisrowno{1}*10]{%
	-2	0.174236458900567\\
	-1.9	0.17440854210097\\
	-1.8	0.174845666492723\\
	-1.7	0.175352236718234\\
	-1.6	0.175519758680396\\
	-1.5	0.17624932454641\\
	-1.4	0.176463611480929\\
	-1.3	0.177066226252399\\
	-1.2	0.177891640341453\\
	-1.1	0.178308199540987\\
	-1	0.179100409668382\\
	-0.9	0.180126823408372\\
	-0.8	0.18132420335859\\
	-0.7	0.182688303584813\\
	-0.6	0.184464330613656\\
	-0.5	0.186908890746818\\
	-0.4	0.189440046945426\\
	-0.35	0.190890438115643\\
	-0.3	0.192389841837716\\
	-0.25	0.19335932784072\\
	-0.2	0.19301188316732\\
	-0.15	0.186435101808594\\
	-0.1	0.160860227259791\\
	-0.05	0.124560091765137\\
	0	0.102832344944224\\
	0.05	0.0911210005972911\\
	0.1	0.0839467260128586\\
	0.15	0.0790012472757291\\
	0.2	0.0753856646033395\\
	0.25	0.0725714587691893\\
	0.3	0.0703028121834822\\
	0.35	0.0684201775173275\\
	0.4	0.0668396332755146\\
	0.5	0.0643157620422907\\
	0.6	0.0624051015708399\\
	0.7	0.0608799281984719\\
	0.8	0.0596242333747334\\
	0.9	0.0585671108415988\\
	1	0.0576517021192528\\
	1.1	0.056885596787695\\
	1.2	0.0560946898365526\\
	1.3	0.0554686065619116\\
	1.4	0.055008570759512\\
	1.5	0.0543877112088499\\
	1.6	0.0540275486213572\\
	1.7	0.0534814788288166\\
	1.8	0.0530821927101614\\
	1.9	0.052828694576571\\
	2	0.0524806776590776\\
};

\node[anchor=south, font=\footnotesize] at (rel axis cs: 0.5, 1) {R$_{\rm D, Acc}$};

\SubfigLetter{d)}{0.2}{0.9};

\draw (rel axis cs: 0.5, 0.38) ellipse (0.15cm and 0.3cm);
\draw[->] (rel axis cs: 0.53, 0.28) -- (rel axis cs: 0.65,0.28);

\draw (rel axis cs: 0.5, 0.13) ellipse (0.15cm and 0.25cm);
\draw[->] (rel axis cs: 0.44, 0.2) -- (rel axis cs: 0.32,0.2);

\end{axis}

\begin{axis}[%
width=0.3\linewidth,
height=3cm,
at={(0.5\linewidth,-3.8cm)},
scale only axis,
xmin=-2,
xmax=2,
xticklabels={,,},
ymin=0,
ymax=1,
\RemoveTickExp,
minor y tick num=1,
axis y line*=right,
axis x line=none,
]
\addplot [color=mycolor1, dashed]
table[row sep=crcr, y expr=\thisrowno{1}*10]{%
	-2	0.0294324647398743\\
	-1.9	0.0296937443767787\\
	-1.8	0.0299287596617527\\
	-1.7	0.0301367672092154\\
	-1.6	0.0304455419165639\\
	-1.5	0.0306890124607799\\
	-1.4	0.0310381729606373\\
	-1.3	0.0313304541368012\\
	-1.2	0.0317330513909383\\
	-1.1	0.0320885835656752\\
	-1	0.0325241336906897\\
	-0.9	0.0330080347148868\\
	-0.8	0.0334763687858118\\
	-0.7	0.0339954394930802\\
	-0.6	0.0345939373200668\\
	-0.5	0.0352958937296057\\
	-0.4	0.0361361894210056\\
	-0.35	0.0366239516501894\\
	-0.3	0.0371683041371524\\
	-0.25	0.0377801960919143\\
	-0.2	0.0384736768860644\\
	-0.15	0.0392738544194978\\
	-0.1	0.0401887606928158\\
	-0.05	0.0412458594998538\\
	0	0.0424516095979866\\
	0.05	0.0437387730449471\\
	0.1	0.0449200200876114\\
	0.15	0.045795650866511\\
	0.2	0.0463025883945877\\
	0.25	0.0465090906845084\\
	0.3	0.0465184680535872\\
	0.35	0.0464064964664288\\
	0.4	0.0462276820782234\\
	0.5	0.0457772009566781\\
	0.6	0.0453036041943772\\
	0.7	0.0448222783548848\\
	0.8	0.0443682389892705\\
	0.9	0.0439472833796781\\
	1	0.0435577922536015\\
	1.1	0.043202499556873\\
	1.2	0.0428688398748798\\
	1.3	0.0425573252049557\\
	1.4	0.0422742193971269\\
	1.5	0.0420023732746099\\
	1.6	0.0417546006227981\\
	1.7	0.0415154978156521\\
	1.8	0.0412958252475551\\
	1.9	0.0410886836564086\\
	2	0.0409090108857689\\
};

\addplot [color=mycolor2, dashed]
table[row sep=crcr, y expr=\thisrowno{1}*10]{%
	-2	0.0497416746080658\\
	-1.9	0.0501177114507396\\
	-1.8	0.0504683319465761\\
	-1.7	0.0508916756058359\\
	-1.6	0.051342729495409\\
	-1.5	0.0518269411847053\\
	-1.4	0.0523924790379107\\
	-1.3	0.0529556331719966\\
	-1.2	0.053566907631348\\
	-1.1	0.0541997076755995\\
	-1	0.0549434496523991\\
	-0.9	0.0558057230678561\\
	-0.8	0.056701365788184\\
	-0.7	0.057759769490829\\
	-0.6	0.0590026187859289\\
	-0.5	0.0603347447319059\\
	-0.4	0.0618452362715519\\
	-0.35	0.062700825999186\\
	-0.3	0.0636313183518683\\
	-0.25	0.0646321480883701\\
	-0.2	0.0656924066833043\\
	-0.15	0.0667602805774004\\
	-0.1	0.0677151967240644\\
	-0.05	0.0683092870227262\\
	0	0.0682147998769396\\
	0.05	0.067578591352871\\
	0.1	0.0664160830313468\\
	0.15	0.0650571404171951\\
	0.2	0.0636866012044721\\
	0.25	0.0624011740784639\\
	0.3	0.061217633452351\\
	0.35	0.0601392701988564\\
	0.4	0.0591675023745019\\
	0.5	0.0574852734571014\\
	0.6	0.0560841004705839\\
	0.7	0.0549002606024095\\
	0.8	0.0538769515715071\\
	0.9	0.0529937343853995\\
	1	0.0522012697656994\\
	1.1	0.0515007764167966\\
	1.2	0.0508706882391327\\
	1.3	0.050308853296153\\
	1.4	0.0497879743573981\\
	1.5	0.0492930541471796\\
	1.6	0.0488487875159865\\
	1.7	0.0484346305507536\\
	1.8	0.0480514622151303\\
	1.9	0.0476899147765416\\
	2	0.0473722590789998\\
};

\addplot [color=mycolor3, dashed]
table[row sep=crcr, y expr=\thisrowno{1}*10]{%
	-2	0.0408076457420634\\
	-1.9	0.0410181704667261\\
	-1.8	0.041213100190559\\
	-1.7	0.0414222889155475\\
	-1.6	0.0416402382175594\\
	-1.5	0.0418764249671215\\
	-1.4	0.042130161329713\\
	-1.3	0.0424082479713799\\
	-1.2	0.0427002762236476\\
	-1.1	0.0430321449420947\\
	-1	0.0433926155184279\\
	-0.9	0.0437871395127046\\
	-0.8	0.0442280577866366\\
	-0.7	0.0447150354896253\\
	-0.6	0.0452322887172874\\
	-0.5	0.0457684526463623\\
	-0.4	0.0462523011452647\\
	-0.35	0.0464269462607429\\
	-0.3	0.0465046198718788\\
	-0.25	0.0464307714701909\\
	-0.2	0.0461165738192571\\
	-0.15	0.0454472360406207\\
	-0.1	0.0444167935895114\\
	-0.05	0.0431305273310548\\
	0	0.0417754400782589\\
	0.05	0.0404901480193918\\
	0.1	0.0393325273766072\\
	0.15	0.0383046956085835\\
	0.2	0.0374125558400409\\
	0.25	0.0366266456295587\\
	0.3	0.0359325950833965\\
	0.35	0.0353156751477745\\
	0.4	0.0347637645476291\\
	0.5	0.0338212660388962\\
	0.6	0.0330462048411499\\
	0.7	0.0324026285077979\\
	0.8	0.0318535015850232\\
	0.9	0.031379239175916\\
	1	0.0309619410871022\\
	1.1	0.0305916136588598\\
	1.2	0.0302416896611964\\
	1.3	0.0299348046933413\\
	1.4	0.029671464297984\\
	1.5	0.0293997880103537\\
	1.6	0.0291818320306858\\
	1.7	0.0289455007540328\\
	1.8	0.0287423052115971\\
	1.9	0.02858015165539\\
	2	0.0284059331576908\\
};

\node[anchor=south, font=\footnotesize] at (rel axis cs: 0.5, 1) {R$_{\rm D, Acc}$};

\end{axis}

}

\end{tikzpicture}%

%% file: Figuras/GFET_Puddles_VBG_fT_Comp.tex
%
%
%

\begin{axis}[%
width=0.7\linewidth,
height=3.5cm,
at={(0\linewidth,0)},
scale only axis,
xmin=-1,
xmax=1,
minor x tick num=1,
xlabel={$V_{\rm FG}$ (V)},
ymode=log,
ymin=0.5,
ymax=2000,
ytick={0.1, 1, 10, 100, 1000, 1e4},
minor y tick num=1,
ylabel={f$_T$ (GHz)},
axis background/.style={fill=white},
axis x line*=bottom,
axis y line*=left,
xmajorgrids,
xminorgrids,
ymajorgrids,
clip mode=individual,
samples = 50,
]
\addplot [color=mycolor5, mark=square, mark options={scale=0.2mm}] 
(x, 1591.5494);

\addplot [color=mycolor2, dashdotted]
  table[row sep=crcr, x expr=\thisrowno{0}+5.7]{%
  	-7.7	3.79667903957228\\
  	-7.6	3.87810431122785\\
  	-7.5	4.16034851525463\\
  	-7.4	4.5340011136176\\
  	-7.3	4.90672071443425\\
  	-7.2	5.39001312981845\\
  	-7.1	5.89410832771394\\
  	-7	6.43975956360106\\
  	-6.9	7.11580888586245\\
  	-6.8	7.99300450232158\\
  	-6.7	9.15987333972572\\
  	-6.6	10.4762633754041\\
  	-6.5	12.1638291419646\\
  	-6.4	14.4921902538763\\
  	-6.3	17.5067667700062\\
  	-6.2	21.5818323366437\\
  	-6.1	25.6329116632607\\
  	-6.05	30.9805895062482\\
  	-6	35.4714889584669\\
  	-5.95	40.8491689901711\\
  	-5.9	47.2057912652366\\
  	-5.85	54.4638139550081\\
  	-5.8	61.9313900994026\\
  	-5.75	66.7738877422977\\
  	-5.7	56.3214884318106\\
  	-5.65	2.99468360307836\\
  	-5.6	53.1341381047352\\
  	-5.55	65.8178120202287\\
  	-5.5	61.4129933175523\\
  	-5.45	53.8674344249561\\
  	-5.4	46.4728958511232\\
  	-5.35	39.9879365122635\\
  	-5.3	32.2118165543142\\
  	-5.2	26.4152367800957\\
  	-5.1	20.7388312469177\\
  	-5	16.6837318214133\\
  	-4.9	13.7395475192669\\
  	-4.8	11.5279900872365\\
  	-4.7	9.82223070653905\\
  	-4.6	8.54665529865503\\
  	-4.5	7.55155295903345\\
  	-4.4	6.71862649403239\\
  	-4.3	5.9428816263998\\
  	-4.2	5.36758734598212\\
  	-4.1	4.9340418592699\\
  	-4	4.52770637829195\\
  	-3.9	4.17847454194327\\
  	-3.8	3.93562545845011\\
  	-3.7	3.85579037709444\\
  };
    
\addplot [color=mycolor3, dashdotted, forget plot]
  table[row sep=crcr, x expr=\thisrowno{0}+5.7]{%
  	-7.7	3.66877292496914\\
  	-7.6	3.87941063765569\\
  	-7.5	4.25985769628358\\
  	-7.4	4.63537458460961\\
  	-7.3	5.08806114799994\\
  	-7.2	5.63334129435885\\
  	-7.1	6.27566059461944\\
  	-7	7.04673434334077\\
  	-6.9	7.99924152423471\\
  	-6.8	9.15306004208155\\
  	-6.7	10.6085854241175\\
  	-6.6	12.4859286510096\\
  	-6.5	14.9196861110317\\
  	-6.4	18.157427626341\\
  	-6.3	22.5124984943397\\
  	-6.2	28.6276207667888\\
  	-6.1	34.7904472411687\\
  	-6.05	42.902478088076\\
  	-6	49.6648721626262\\
  	-5.95	57.4909422933575\\
  	-5.9	66.0492634457801\\
  	-5.85	73.8658752413386\\
  	-5.8	75.5278640122478\\
  	-5.75	51.5225741619656\\
  	-5.7	20.8341820642027\\
  	-5.65	80.7623581097501\\
  	-5.6	93.8280335819053\\
  	-5.55	89.4153932244467\\
  	-5.5	81.1064273529939\\
  	-5.45	72.3729575166374\\
  	-5.4	64.2553937625334\\
  	-5.35	57.018745165672\\
  	-5.3	47.942165826122\\
  	-5.2	40.7137940138891\\
  	-5.1	33.1349905573633\\
  	-5	27.4318592919594\\
  	-4.9	23.1217362706176\\
  	-4.8	20.0310748197032\\
  	-4.7	17.608516814934\\
  	-4.6	15.3224988831285\\
  	-4.5	13.5859318478763\\
  	-4.4	12.4668823568287\\
  	-4.3	11.0945536837249\\
  	-4.2	10.0860539903554\\
  	-4.1	9.21065784328963\\
  	-4	8.29496384529726\\
  	-3.9	8.00449772121172\\
  	-3.8	7.43536554880096\\
  	-3.7	6.98826263929142\\
  };
    
\addplot [color=mycolor1, dashdotted, forget plot]
  table[row sep=crcr, x expr=\thisrowno{0}+5.7]{%
  	-7.7	6.52470855753858\\
  	-7.6	6.86281483502839\\
  	-7.5	7.54586491955708\\
  	-7.4	8.09593360859801\\
  	-7.3	8.77921243485607\\
  	-7.2	9.58842884468203\\
  	-7.1	10.5028770354296\\
  	-7	11.6026312968762\\
  	-6.9	12.862265264073\\
  	-6.8	14.3994159730225\\
  	-6.7	16.2106187700392\\
  	-6.6	18.496672688192\\
  	-6.5	21.3358075689376\\
  	-6.4	24.9231543239764\\
  	-6.3	29.531899966021\\
  	-6.2	35.5216248772538\\
  	-6.1	41.1104812203599\\
  	-6.05	48.048670754979\\
  	-6	53.6122707455465\\
  	-5.95	60.0046365718961\\
  	-5.9	67.2826827182959\\
  	-5.85	75.4066916209247\\
  	-5.8	84.0829497687955\\
  	-5.75	92.3025078066649\\
  	-5.7	96.7922048117492\\
  	-5.65	84.7844751250204\\
  	-5.6	26.7894316831258\\
  	-5.55	49.1569646132495\\
  	-5.5	77.3264872703807\\
  	-5.45	77.1273601171047\\
  	-5.4	69.6227771344811\\
  	-5.35	61.0105045356901\\
  	-5.3	49.3117699989144\\
  	-5.2	40.0862929514464\\
  	-5.1	30.7840828977948\\
  	-5	24.1528845717125\\
  	-4.9	19.3779631365561\\
  	-4.8	15.8109677082048\\
  	-4.7	13.1440481771686\\
  	-4.6	11.1559181759816\\
  	-4.5	9.56977388061569\\
  	-4.4	8.30907089287518\\
  	-4.3	7.32482366673329\\
  	-4.2	6.48905901828139\\
  	-4.1	5.80178509703168\\
  	-4	5.23787132091262\\
  	-3.9	4.73507653329637\\
  	-3.8	4.39092141930761\\
  	-3.7	4.28259143661661\\
  };
  
\node[color=red, anchor=south, align=left, font=\scriptsize] at (axis cs: 1.1, 300) {\pgfuseplotmark{o}};
\draw[color=red, ->] (axis cs: 1.1, 300) -- (axis cs: 1, 300);
\node[color=blue, anchor=south, align=left, font=\scriptsize] at (axis cs: 1.1, 120) {\pgfuseplotmark{square}};
\draw[color=blue, ->] (axis cs: 1.1, 120) -- (axis cs: 1, 120);
\node[color=orange, anchor=north, align=left, font=\scriptsize] at (axis cs: 1.1, 105) {\pgfuseplotmark{triangle}};
\draw[color=orange, ->] (axis cs: 1.1, 105) -- (axis cs: 1, 105);

\end{axis}

\begin{axis}[%
	width=0.7\linewidth,
	height=3.5cm,
	at={(0\linewidth,0)},
	scale only axis,
	xmin=-1,
	xmax=1,
	xticklabels={,,},
	ymode=log,
	ymin=0.5,
	ymax=2000,
	yticklabels={,,},
	minor y tick num=1,
	xminorgrids,
	axis x line*=bottom,
	axis y line*=left,
	legend style={at={(0.5, -0.3)}, anchor=north,legend cell align=left, align=left, draw=white!15!black, legend columns=3, font=\scriptsize}
	]
	\addplot [color=mycolor4, dashed, forget plot]
  table[row sep=crcr, x expr=\thisrowno{0}+5.7]{%
  	-7.7	242.29115238895\\
  	-7.6	242.293963710252\\
  	-7.5	242.300935666258\\
  	-7.4	242.310094526906\\
  	-7.3	242.320192608052\\
  	-7.2	242.331945459325\\
  	-7.1	242.346299330451\\
  	-7	242.364026162833\\
  	-6.9	242.384568877884\\
  	-6.8	242.410707138533\\
  	-6.7	242.443840688453\\
  	-6.6	242.480575288015\\
  	-6.5	242.529784648655\\
  	-6.4	242.597869687363\\
  	-6.3	242.685855230699\\
  	-6.2	242.88399112436\\
  	-6.1	243.139111828647\\
  	-6.05	243.468532958027\\
  	-6	243.779535248124\\
  	-5.95	244.201199060436\\
  	-5.9	244.981891982653\\
  	-5.85	245.494816249455\\
  	-5.8	243.196479437646\\
  	-5.75	226.108896162237\\
  	-5.7	149.270267754623\\
  	-5.65	1.9731278706365\\
  	-5.6	152.537491049187\\
  	-5.55	228.293030841875\\
  	-5.5	243.798325412703\\
  	-5.45	244.900318754393\\
  	-5.4	244.583215827509\\
  	-5.35	244.077474125174\\
  	-5.3	243.396739820527\\
  	-5.2	243.071599269089\\
  	-5.1	242.886692874456\\
  	-5	242.74723220914\\
  	-4.9	242.641114253346\\
  	-4.8	242.557251454063\\
  	-4.7	242.496138330175\\
  	-4.6	242.451189220096\\
  	-4.5	242.413843757172\\
  	-4.4	242.385617355843\\
  	-4.3	242.363898029659\\
  	-4.2	242.345761245812\\
  	-4.1	242.331241247969\\
  	-4	242.319423725817\\
  	-3.9	242.309397202887\\
  	-3.8	242.300346312872\\
  	-3.7	242.296208873335\\
  };
	
	\addplot [color=mycolor1]
  table[row sep=crcr, x expr=\thisrowno{0}+5.7]{%
  	-7.7	3.16099970685965\\
  	-7.6	3.39509697041952\\
  	-7.5	3.87673612784948\\
  	-7.4	4.09053439635633\\
  	-7.3	4.43231987173912\\
  	-7.2	4.85922204591186\\
  	-7.1	5.32526265721771\\
  	-7	5.91303199162568\\
  	-6.9	6.57512088199229\\
  	-6.8	7.44705995208077\\
  	-6.7	8.42833493814882\\
  	-6.6	9.79062945468761\\
  	-6.5	11.5252427792746\\
  	-6.4	13.8527503190941\\
  	-6.3	17.0612039378343\\
  	-6.2	21.6684127245101\\
  	-6.1	26.3794693048715\\
  	-6.05	32.8736937459431\\
  	-6	38.7518279879324\\
  	-5.95	46.2796322822855\\
  	-5.9	56.09126754637\\
  	-5.85	69.03531281752\\
  	-5.8	86.0534473490767\\
  	-5.75	107.278141438234\\
  	-5.7	126.868486552786\\
  	-5.65	112.365965647132\\
  	-5.6	27.3158048515237\\
  	-5.55	64.137796194666\\
  	-5.5	85.9434374349636\\
  	-5.45	63.9241498531854\\
  	-5.4	42.3935236815394\\
  	-5.35	28.4693495596116\\
  	-5.3	17.079434758878\\
  	-5.2	11.1712388006539\\
  	-5.1	6.81201972969962\\
  	-5	4.45636223105747\\
  	-4.9	3.14050687010484\\
  	-4.8	2.38966442166577\\
  	-4.7	1.86352798052971\\
  	-4.6	1.40212063117571\\
  	-4.5	1.13347734926379\\
  	-4.4	1.06670056273\\
  	-4.3	0.813202825959635\\
  	-4.2	0.736566691794234\\
  	-4.1	0.644626766718626\\
  	-4	0.54366432791229\\
  	-3.9	0.570578440437121\\
  	-3.8	0.398427318741394\\
  	-3.7	0.331760660388826\\
  };
  \addlegendentry{$V_{\rm BG}=-1$V}

	\addplot [color=mycolor2]
table[row sep=crcr, x expr=\thisrowno{0}+5.7]{%
	-7.7	0.917781960232978\\
	-7.6	0.822205061323442\\
	-7.5	0.872918815733347\\
	-7.4	1.04557446347308\\
	-7.3	1.10850321585629\\
	-7.2	1.32269848887748\\
	-7.1	1.41309416697747\\
	-7	1.25752542751883\\
	-6.9	1.10103963109363\\
	-6.8	1.17169397398814\\
	-6.7	1.48583097208669\\
	-6.6	1.53080285059123\\
	-6.5	1.61225249799837\\
	-6.4	2.01791992743464\\
	-6.3	2.34521809050218\\
	-6.2	2.70613425523446\\
	-6.1	3.20768624446073\\
	-6.05	4.03607787301212\\
	-6	4.80314505837069\\
	-5.95	5.94637164639909\\
	-5.9	7.7367970264668\\
	-5.85	10.7501043074071\\
	-5.8	16.2531414558585\\
	-5.75	26.4137010600438\\
	-5.7	33.1432560041107\\
	-5.65	4.04067585416601\\
	-5.6	28.8357022404147\\
	-5.55	24.2445263725149\\
	-5.5	13.8063933365441\\
	-5.45	8.34519835660087\\
	-5.4	5.6706027120619\\
	-5.35	4.20121460820234\\
	-5.3	2.96718508772257\\
	-5.2	2.26008204167186\\
	-5.1	1.71798967052792\\
	-5	1.38080858710104\\
	-4.9	1.21427426543017\\
	-4.8	1.06287852970494\\
	-4.7	0.909467293310975\\
	-4.6	0.853624576697226\\
	-4.5	0.827411673212216\\
	-4.4	0.769676038125915\\
	-4.3	0.610147522134153\\
	-4.2	0.562228657692842\\
	-4.1	0.581093064173973\\
	-4	0.551852341872914\\
	-3.9	0.52401237690135\\
	-3.8	0.571325843587864\\
	-3.7	0.6326000376114\\
};
\addlegendentry{$V_{\rm BG}=0$V}

	\addplot [color=mycolor3]
  table[row sep=crcr, x expr=\thisrowno{0}+5.7]{%
  	-7.7	0.494662680943888\\
  	-7.6	0.470615604120709\\
  	-7.5	0.412061816650334\\
  	-7.4	0.495014866928926\\
  	-7.3	0.54336605198524\\
  	-7.2	0.608503836959459\\
  	-7.1	0.694999665799032\\
  	-7	0.791019041988614\\
  	-6.9	0.946547939363509\\
  	-6.8	1.00411883309328\\
  	-6.7	1.24075877969776\\
  	-6.6	1.60469651018884\\
  	-6.5	1.99610494642896\\
  	-6.4	2.57645945948233\\
  	-6.3	3.65024128601628\\
  	-6.2	5.44504803096105\\
  	-6.1	7.69072415352613\\
  	-6.05	11.839078661917\\
  	-6	16.5918408973562\\
  	-5.95	24.3883437419361\\
  	-5.9	37.3135253686971\\
  	-5.85	57.8178458205092\\
  	-5.8	81.1989129059267\\
  	-5.75	66.2159199824856\\
  	-5.7	20.2162971375129\\
  	-5.65	106.230795959633\\
  	-5.6	122.316032757416\\
  	-5.55	101.648535670869\\
  	-5.5	79.6916109694493\\
  	-5.45	62.5846655733642\\
  	-5.4	49.9388249017807\\
  	-5.35	40.5608634926905\\
  	-5.3	30.8504448567432\\
  	-5.2	24.2808084055698\\
  	-5.1	18.2465662075865\\
  	-5	14.7136507752456\\
  	-4.9	12.3548385683991\\
  	-4.8	10.3280597831862\\
  	-4.7	8.82307998889461\\
  	-4.6	7.49539341557282\\
  	-4.5	6.60558630898376\\
  	-4.4	6.28319344481888\\
  	-4.3	5.48046070127258\\
  	-4.2	4.99037334634242\\
  	-4.1	4.56474425066932\\
  	-4	4.00447397944082\\
  	-3.9	4.11976391312886\\
  	-3.8	3.84607687373217\\
  	-3.7	3.50724680702392\\
  };
  \addlegendentry{$V_{\rm BG}=1$V}

	\node[anchor=north, align=left, font=\scriptsize] at (axis cs: 0.75, 1500) {Physical \\ limit};	
	\node[anchor=west, align=left, font=\scriptsize] at (rel axis cs: 0.1, 0.6) {Intrinsic \\ Device};
	\draw[->] (rel axis cs: 0.2, 0.6) -- (rel axis cs: 0.3, 0.7);
\end{axis}
